\documentclass[final,5p,times,twocolumn,authoryear]{elsarticle}
\usepackage[utf8]{inputenc}

\usepackage{wrapfig}
\usepackage{graphicx}
\usepackage{subcaption}
\usepackage{amsmath}
\usepackage{amsthm}
\usepackage[dvipsnames]{xcolor}
\usepackage{listings}
\usepackage{booktabs}
\usepackage{array}
\usepackage{tabularx}
\usepackage{adjustbox}
\usepackage{multirow}  
\usepackage{makecell}
\usepackage{verbatim}
\usepackage{natbib}        
\usepackage{colortbl}
\usepackage{siunitx}
\usepackage{amsfonts}
\usepackage{amssymb}
\usepackage{soul}
\usepackage[inline]{enumitem}
\usepackage{mathtools}
\usepackage{algorithmicx}
\usepackage{algorithm}
\usepackage{algpseudocode}
\usepackage{listings}
\usepackage{xcolor}

\definecolor{codegreen}{rgb}{0,0.6,0}
\definecolor{codegray}{rgb}{0.5,0.5,0.5}
\definecolor{codepurple}{rgb}{0.58,0,0.82}
\definecolor{backcolour}{rgb}{0.95,0.95,0.92}
\lstdefinestyle{mystyle}{
  backgroundcolor=\color{backcolour}, commentstyle=\color{codegreen},
  keywordstyle=\color{magenta},
  numberstyle=\tiny\color{codegray},
  stringstyle=\color{codepurple},
  basicstyle=\ttfamily\footnotesize,
  breakatwhitespace=false,         
  breaklines=true,                 
  captionpos=b,                    
  keepspaces=true,                 
  numbers=left,                    
  numbersep=5pt,                  
  showspaces=false,                
  showstringspaces=false,
  showtabs=false,                  
  tabsize=2
}
\usepackage{url}
\usepackage[breaklinks=true, colorlinks, bookmarks=true]{hyperref}

\newtheorem{theorem}{Theorem}[section]

\newtheorem{definition}[theorem]{Definition}

\newtheorem{example}[theorem]{Example}

\newcommand{\frakq}{{\mathfrak{q}}}

\newcommand{\bbR}{\mathbb{R}}

\journal{Annual Reviews in Control}

\begin{document}

\begin{frontmatter}




\title{Control-Oriented System Identification: \\ Classical, Learning, and Physics-Informed Approaches}

\author[label1]{S. Sivaranjani \corref{cor1} \fnref{label_equal}}

\affiliation[label1]{organization={Edwardson School of Industrial Engineering, Purdue University},
            city={West Lafayette}, 
            state={IN},
            postcode={47907},
            country={USA}}
\fntext[label_equal]{These authors contributed equally to this work.}
\cortext[cor1]{sseetha@purdue.edu, yyshi@ucsd.edu, natanasov@ucsd.edu}

\author[label2]{Yuanyuan Shi \corref{cor1} \fnref{label_equal}}

\affiliation[label2]{organization={Electrical and Computer Engineering,  University of California San Diego},
            city={La Jolla},
            state={CA},
            postcode={92093}, 
            country={USA}}

\author[label2]{Nikolay Atanasov \corref{cor1} \fnref{label_equal}}


\author[label3]{Thai Duong \fnref{label_alphabet}}

\affiliation[label3]{organization={Department of Computer Science, Rice University},
            city={Houston}, 
            state={TX},
            postcode={77005},
            country={USA}}
\fntext[label_alphabet]{These authors contributed equally and are ordered alphabetically by last name.}

\author[label2]{Jie Feng \fnref{label_alphabet}}

\author[label4]{Tim Martin \fnref{label_alphabet}}

\affiliation[label4]{organization={University of Stuttgart, Institute for Systems Theory and Automatic Control},
            city={Stuttgart},
            country={Germany}}

\author[label1]{Yuezhu Xu \fnref{label_alphabet}}

\author[label5]{\\ Vijay Gupta}

\affiliation[label5]{organization={Elmore Family School of Electrical and Computer Engineering, Purdue University},
            city={West Lafayette}, 
            state={IN},
            postcode={47907},
            country={USA}}

\author[label4]{Frank Allgöwer}

\begin{abstract}
This article surveys classical, machine learning, and data-driven system identification approaches to learn control-relevant and physics-informed models of dynamical systems. In recent years, machine learning approaches have enabled system identification from noisy, high-dimensional, and complex data. However, their utility in control applications is limited by their ability to provide provable guarantees on control-relevant properties. Meanwhile, traditional control theory has identified several properties of  physical systems that are useful in analysis and control synthesis, such as  dissipativity, monotonicity, energy conservation, and symmetry-preserving structures. In this paper, we postulate that merging system identification algorithms with such control-relevant or physics-informed properties can provide useful inductive bias, enhance explainability, enable control synthesis with provable guarantees, and improve sample complexity. We formulate system identification as an  optimization problem where control-relevant properties can be enforced in three ways, namely, direct parameterization (constraining the model structure to satisfy a desired property by construction), soft constraints (encouraging control-relevant properties through regularization or penalty terms), and hard constraints (imposing control-relevant properties as constraints in the optimization problem). Through this lens, we survey methods to learn physics-informed and control-relevant models spanning classical linear and nonlinear system identification techniques, machine learning-based approaches, as well as direct identification through data-driven and behavioral representations. {Taken together, these perspectives suggest that control-oriented identification should be viewed not only as a problem of minimizing prediction error, but also as one of selecting model structures and learning methods that preserve the properties needed for downstream analysis and control synthesis.} Throughout the paper, we provide several expository examples that are accompanied by code and brief tutorials on a public Github repository. 
We also describe several challenging directions for future research in this area, including identification in networked, switched, and time-varying systems, experiment design, and bridging the gaps between data-driven, learning-based, and control-oriented system identification.

\end{abstract}




\end{frontmatter}



\section{Introduction}
\label{sec: intro}
Machine learning, at its core, is learning from data. Since control theory also seeks to utilize data to identify system models and estimate and regulate their states, it is natural to ask if the resurgent interest in machine learning can translate to new algorithms and guarantees useful in control. This survey focuses on techniques for system identification as a precursor to model-based control. System identification is a fundamental and widely studied topic in both the automatic control and signal processing communities. The basic problem considers a system that is excited using some `input' data and generates measurements or `output' data. Using the inputs and the outputs, we seek to generate a model of the system that not only explains the data that were used to generate it, but also generalizes to other unseen input data \citep{Ljung_dlsysid_2020}. One could interpret this as a function approximation problem, which is exactly the problem that machine learning seeks to solve with the potential ability to consider arbitrary noise distributions, high dimensional data, and complex loss functions. However, the central premise of this survey is that while machine learning can indeed yield exciting new algorithms for system identification, to realize this promise fully, `vanilla' machine learning algorithms will not do --- we need physics-informed and control-oriented learning algorithms that guarantee properties in the learned models useful for the subsequent control design.

What do we mean by control-oriented learning algorithms? Traditional control theory has identified several properties of dynamical system that are useful for analyzing dynamical systems or synthesizing control laws. These properties typically arise from first principles underlying the system physics or dynamics, e.g., symmetries or energy dissipation properties, and are useful for designing controllers to achieve desired closed-loop responses. By merging such physics-based or control-relevant properties with learning algorithms, we hope to identify system models that provide the best of both  philosophies.

The first reason to insist on such properties when learning a model is that they may provide a useful inductive bias to the learning algorithm. The possible choices of a model that can explain the given (or training) input-output data are many; the only test for utility among such models is the ability to generalize to unseen (or test) data. Constraining the learned model to display properties that we know hold from physics or system structure is potentially a strong and useful inductive bias to the learning algorithm. Even if the learned model yields some ground on the error criterion over the training data, a physics-constrained model will likely have better generalization ability.

The second reason for ensuring that such properties are satisfied by the learned model is that several decades of developments in control synthesis explicitly utilize these properties to provide provable guarantees. In control engineering applications, the aim is typically not to just learn an \emph{open-loop} model using the input-output data, but rather to utilize this model for control design with desired \emph{closed-loop} characteristics. Many methods for controller design utilize properties such as dissipativity and monotonicity to obtain guarantees on closed-loop system behavior. For instance, 
compositionality properties of dissipative and monotone systems have been widely leveraged for scalable control designs in networked systems. If the learned model also displays such properties that we expect to be satisfied from physics, similar controller synthesis procedures can be used and strong guarantees on the closed-loop system behavior can be obtained. 

Another reason to incorporate a priori information about physics is to obtain explainable system models. It is well-known that the outputs, and consequently the failure modes, of black-box models such as neural networks utilized in learning-based control cannot be readily intuited. Therefore, when control designs that utilize these models do not perform as intended, the root cause is often unclear, resulting in the designer having to resort to trial-and-error to achieve acceptable performance. Instead, if black-box system models like neural networks can be imbued with key physics properties, such as conservation laws, symmetries, or energy dissipation, their outputs or failure modes can be explained in terms of measures defined on these properties (e.g., dissipativity indices) or violation of these properties (e.g., violation of symmetries). Then, desired input-output behaviors can be translated into specifications on these properties and shaped by explainable control designs.

Finally, we also expect that enforcing such properties could improve the sample complexity of the learning algorithms. The intuition here is that these properties constrain the degrees of the freedom for the learned model and hence the model can be learned in a more efficient fashion. {Some empirical evidence for this can be observed in Miller et al. (2020); Cuomo et al. (2022); Son et al. (2023), where physics-informed and structure-preserving neural-network models are compared with unstructured neural networks in limited-data settings and are shown to achieve comparable or improved prediction accuracy using fewer training samples.}

Due to these reasons, extending the powerful methods developed in the learning literature to identify dynamical system models that are physics-informed and control-oriented has emerged as a key direction of work in multiple research communities. It has the potential to move the identification methods in automatic control and signal processing beyond the limiting assumptions classically made, while at the same time, making the learning algorithms more relevant for analysis and control of physical systems. The purpose of this survey is to provide students, researchers, and practitioners with an overview of this field. We aim to categorize the types of methods that have been proposed so that the reader obtains a map of the area, given the often bewildering mix of modeling philosophies and approaches that have been proposed. We also identify how classical system identification techniques rooted in the control literature form an important pillar in this area and must be mastered by anybody seeking to make meaningful contributions in this domain. At the same time, we present the reader with a succinct understanding of where these methods can be complemented and extended by newer approaches and tools from learning, as well as new ideas in data-driven representations stemming from notions like behavioral systems theory. 

While there are many excellent surveys on various facets at the intersection of learning and control, our aim and presentation differ from them. Specifically considering the problem of system modeling and identification, there are two lines of research closely aligned with the focus of this paper. The first is \emph{physics-informed or scientific machine learning}, surveyed in works such as \cite{Yu_dldynsyssurvey_2021,nature_sysid,phyrev_sysid,jsc_pinn,willard2020integrating,yu2024learning}, where the idea is to leverage prior knowledge of system physics to enhance the generalizability and interpretability of deep learning models, with applications to a wide array of problems including forecasting, inference, and solution of ordinary or partial differential equations.  However, these works do not typically focus on control-relevant models, with the exception of recent tutorials \citep{nghiem2023physics,drgona2025safe}, some parts of which discuss physics-informed learning for system identification for specific model classes and structural priors.  The second closely related line of research is \emph{identification for control}, which has a long rich history, surveyed in works such as \citet{gevers2005identification}, which typically focus on classical system identification tools. An exception here is \citet{pillonetto2025deep} that focuses on deep learning architectures for system identification, primarily from the viewpoint of optimization algorithms and kernel-based methods.  The renewed interest in data-driven control \citep{DBLSCT2025} in recent years has also led to surveys on data-driven system representations based on behavioral systems theory \citep{behavioral_system_theory} and operator dynamical models such as Koopman operators \citep{bevanda2021koopman}. Even within these domains, there are no comprehensive surveys focusing on the problem of learning representations or models capturing specific control-relevant properties or physics-based constraints. This paper bridges the gap between these two lines of research and existing review papers, surveying both \emph{classical and learning-based approaches to address the specific problem of identifying physics-constrained and control-relevant models for dynamical systems and control synthesis.} 

This paper is organized as follows. In Section \ref{sec:properties}, we begin by introducing system modeling techniques and describing the control-relevant properties considered in this work, namely, dissipativity, monotonicity, and symmetry-preserving structures (such as Lagrangian and Hamiltonian dynamics). We then formally pose the problem of identifying a model satisfying these properties as a constrained optimization problem, and outline various perspectives on how to solve this problem in Section \ref{sec:perspectives}. Then, we survey classical (Section \ref{sec:classical}) and deep-learning (Section \ref{sec:deep}) system identification techniques capturing control-relevant properties, as well as direct data-driven identification of these properties through, e.g., behavioral representations (Section \ref{sec:direct}). The code for all the numerical examples in this paper is available in our Github repository.\footnote{\scriptsize\url{https://github.com/ExistentialRobotics/physics_based_sysid}} We conclude by providing several directions for future exploration in Section \ref{sec: future}.

\textbf{Notation:} Throughout the manuscript, scalars and vectors are denoted by lower-case letters. The set of real numbers is denoted by $\mathbb{R}$, and the set of $n$-dimensional real vectors by $\mathbb{R}^n$. For a vector $x \in \mathbb{R}^n$, $\|x\|$ denotes its Euclidean norm. 
The operator $\nabla_x$ denotes the gradient with respect to the variable $x$. For a function $t \mapsto x(t)$, the notation $\dot{x}$ and $\ddot{x}$ denotes the first and second time derivatives of $x(t)$, respectively. The set of complex numbers is denoted by $\mathbb{C}$. The variable $s \in \mathbb{C}$ denotes the Laplace variable (complex frequency), and $\omega \in \mathbb{R}$ denotes the real frequency, with $s = i\omega$ in the Fourier domain. Matrices are denoted by upper-case letters. For a matrix $A$, its transpose is denoted by $A^\top$, and its inverse (if it exists) is denoted by $A^{-1}$. Inequalities between matrices are defined in the standard positive (semi)definite sense: ${A} \succ 0$ indicates that ${A}$ is positive definite, while ${A} \succeq 0$ indicates positive semidefiniteness. Similarly, ${A} \prec {0}$ and ${A} \preceq {0}$ denote negative definite and negative semidefinite matrices, respectively.

\section{Control-Related Properties and Constraints}
\label{sec:properties}
\subsection{System Modeling}\label{sec: sys_modeling}
Continuous-time dynamical systems are commonly modeled using ordinary differential equations (ODEs):
\begin{equation} \label{eq:ode}
\frac{d^n}{dt^n} y(t) = \phi\left(y(t), \frac{d}{dt} y(t), \ldots,  \frac{d^{n-1}}{dt^{n-1}} y(t), u(t) \right),
\end{equation}
where $u(t)$ is the system input at time $t$, $y(t)$ is the system output, and $\phi$ is a function capturing the relationship among the system input and the time-derivatives of its output. 
A state-space model converts the $n$-th order ODE in \eqref{eq:ode} to a system of $n$ first-order ODEs by defining state variables $x(t) = (x_1(t),\ldots,x_n(t))$ corresponding to the output time derivatives:
\begin{equation} \label{eq:states}
x_1(t) = y(t), \quad x_2(t) = \frac{d}{dt} y(t),\quad \ldots, \quad x_n(t) = \frac{d^{n-1}}{dt^{n-1}} y(t)
\end{equation}
and observing that one can write  $\dot{x}_1(t) = x_2(t)$, $\dot{x}_2(t) = x_3(t)$, $\ldots$, $\dot{x}_{n-1}(t) = x_{n}(t)$, and $\dot{x}_n(t) = \phi(x_1(t),\ldots,x_n(t), u(t))$, is shorthand notation for the first time derivative of state $x_i$.

In general, a state-space model describes a continuous-time dynamical system as:
\begin{equation} \label{eq:control_system}
\begin{aligned}
\dot{x}(t) &= f(x(t), u(t)),\\
y(t) &= h(x(t), u(t)),
\end{aligned}
\end{equation}
where $x(t) \in \mathcal{X} \subseteq \mathbb{R}^{n_x}$ is the state, $u(t) \in  \mathcal{U} \subseteq \mathbb{R}^{n_u}$ is the input, $y(t) \in \mathcal{Y} \subseteq \mathbb{R}^{n_y}$ is the output, $f: \mathbb{R}^{n_x} \times \mathbb{R}^{n_u} \rightarrow \mathbb{R}^{n_x}$ is the system dynamics model, $h : \mathbb{R}^{n_x} \times \mathbb{R}^{n_u} \rightarrow \mathbb{R}^{n_y}$ is the output model, and the initial condition $x(0)$ is given.

It is often sufficient from a system modeling perspective and beneficial from a control design perspective to consider models that are affine in the control input $u(t)$:
\begin{equation} \label{eq:affine_system}
\begin{aligned}
\dot{x}(t)  &= A(x(t)) + B(x(t))u(t),\\
y(t) &= C(x(t)) + D(x(t))u(t),
\end{aligned}
\end{equation}
or even in both the state $x(t)$ and the input $u(t)$:
\begin{equation} \label{eq:linear_system}
\begin{aligned}
\dot{x}(t)  &= A x(t) + B u(t),\\
y(t) &= Cx(t) + Du(t).
\end{aligned}
\end{equation}

The Laplace transform converts a linear time-invariant (LTI) system from a linear ODE in the time domain, given by \eqref{eq:linear_system}, to a linear algebraic equation in the complex domain. The input-output relationship of an LTI system in the complex domain with zero initial conditions is described by its transfer function $G : \mathbb{C} \mapsto \mathbb{C}^{n_y \times n_u}$:
\begin{equation} \label{eq:transfer_function}
Y(s) = G(s)U(s) = (C(sI - A)^{-1}B + D) U(s),
\end{equation}
where $U(s)$ and $Y(s)$ are the Laplace transforms of the input $u(t)$ and output $y(t)$, respectively. In the time domain, the relationship in \eqref{eq:transfer_function} shows that, with zero initial conditions, the system output is the result of convolution:
\begin{equation} \label{eq:convolution}
 y(t) = g(t) * u(t) = \int_0^t g(t - \tau) u(\tau) d\tau
\end{equation}
of the system impulse response $g(t) = Ce^{At}B + D\delta(t)$ (inverse Laplace transform of $G(s)$) with the input signal $u(t)$.

A \emph{control law}, also referred to as \emph{control policy}, is a function $\pi: \mathcal{X} \times [0,\infty) \to \mathcal{U}$ that determines an input $u(t) = \pi(x(t),t)$ for a dynamical system based on the system's current state $x(t)$, time $t$, and desired behavior. { Note that $\pi(x(t),t)$ describes a memoryless, or static, state-feedback control law. More general dynamic controllers also arise frequently in control design \citep[Ch.~12]{Khalil}. These controllers introduce internal controller states, for example to store accumulated tracking error in integral control, estimate unmeasured states in observers, filter measured signals, or implement dynamic compensation \citep{krstic1995nonlinear}.   In the following discussion, for simplicity of exposition, we use the static feedback form  to introduce the closed-loop dynamics and notation, while noting that the interconnection of such controllers with the plant can be analyzed in a manner similar to that of memoryless controllers by augmenting the plant state with the controller state.  For the static feedback case, the composition of the system dynamics model $f$ in \eqref{eq:control_system} with a control policy $\pi$ creates a \emph{closed-loop system} with dynamics model: }
\begin{equation} \label{eq:closed_loop_system}
\dot{x}(t) = F(x(t),t) := f(x(t), \pi(x(t),t)).
\end{equation}
To guarantee existence and uniqueness of solutions of the ODE in \eqref{eq:closed_loop_system}, the function $F$ needs to be Lipschitz continuous in $x$ and continuous in $t$ \citep[Thm.~3.1]{Khalil}. A special case of \eqref{eq:closed_loop_system} arises when $F$ does not explicitly depend on $t$; that is 
\begin{equation} 
\label{eq:closed_loop_autonoumoussystem}
\dot{x} = F(x)\,,
\end{equation}
in which case the closed-loop system is said to be \emph{autonomous}.

If the system state $x(t)$ is not available to the control policy, the input $u(t)$ must be determined by an \emph{output-feedback} control policy, relying on the output $y(t)$ instead of $x(t)$. In this case, the closed-loop dynamics $F$ is obtained by the composition of the open-loop dynamics model $f$, the control policy $\pi$, and the output model $h$ in \eqref{eq:control_system}.

Instead of representing a system using a state-space model \eqref{eq:control_system} or transfer function \eqref{eq:transfer_function}, behavioral system theory \citep{behavioral_system_theory,willems_part1} considers a representation-free perspective. In this framework, a system is described by its behavior \(\mathcal{B}\), defined as the set of all trajectories compatible with the system laws: 
\begin{equation}\label{eq:behavioral_system_theory}
\left\{ \omega : [0,\infty) \rightarrow \mathbb{R}^q  \right\}.
\end{equation}
{  In other words,} the behavioral framework describes a system as the span of all possible input-output trajectories generated by its dynamics.
{  
Thus, the system is represented by its behavior rather than by a particular state-space realization or transfer function.}
A trajectory $\omega(t)$ may be related to an input-output representation via a permutation matrix $\Pi \in \mathbb{R}^{q \times q}$ with $q = n_u+n_y$ and $\omega(t) = \Pi \begin{bmatrix} u(t) \\ y(t) \end{bmatrix}$. {   Note that the permutation matrix does not encode the system dynamics; {rather,} it only specifies the ordering of the input and output variables in the trajectory vector. The dynamics are encoded by which trajectories { are considered to be compatible with} the behavior { of the system}.}
The complexity of a system is defined by its structure indices \citep{willems_part1}: the number of inputs $n_u$, the order $n_x$, and the lag $\ell$ (the smallest integer $\ell$ for which the matrix $\begin{bmatrix} C & CA & \cdots & CA^{\ell-1} \end{bmatrix}$ has rank $n_x$).

The system models described so far can also be formulated in discrete time $k \in \mathbb{N}$ with input $u_k \in\mathbb{R}^{n_u}$, state $x_k \in\mathbb{R}^{n_x}$, and output $y_k \in\mathbb{R}^{n_y}$. Discrete-time versions of the nonlinear model in \eqref{eq:control_system}, control-affine model in \eqref{eq:affine_system}, and LTI model in \eqref{eq:linear_system} are defined using difference equations with equivalent notation. Specifically, a discrete-time nonlinear system can be written as:
\begin{equation}\label{Sys_IO}
\begin{aligned}
     x_{k+1}&=f(x_k,u_k),\\
    y_k&=h(x_k,u_k),
    \end{aligned}
\end{equation} and a discrete-time 
LTI model can be formulated as:
\begin{equation}\label{LTI_DT}
\begin{aligned}
    x_{k+1}&=Ax_k+Bu_k,\\
    y_k&=Cx_k+Du_k,
\end{aligned}
\end{equation}
with given initial conditions $x_{0}$.
Instead of using the Laplace transform, discrete-time LTI systems are converted to the complex domain using the Z transform and are associated with a discrete-time transfer function $G(z)$ such that $Y(z) = G(z)U(z)$, where $U(z)$ and $Y(z)$ are the Z transforms of the input $u_k$ and output $y_k$.

In the discrete-time setting\footnote{Continuous time analogues of Willems' fundamental lemma have also been recently proposed \citep{lopez2022continuous}.}, within the behavioral framework, \textit{Willems' fundamental lemma} \citep{Willems} provides a representation of an LTI system using a single sufficiently informative and long trajectory. Consider a trajectory $\{\tilde{u}_k,\tilde{y}_k\}_{k=0}^{N-1}$ from an LTI system \eqref{LTI_DT} that satisfies a condition called persistent excitation. Then, we can verify whether any arbitrary trajectory $\{u_k,y_k\}_{k=0}^{L-1}$ of shorter length $L$ comes from the same LTI system as long as $N\geq(n_u+1)(n_x+L)-1$. In detail, let $\tilde{\mathbf{u}}=\begin{bmatrix}\tilde{u}_0^\top & \cdots & \tilde{u}^\top_{N-1}\end{bmatrix}^\top$, $\tilde{\mathbf{y}}=\begin{bmatrix}\tilde{y}_0^\top & \cdots & \tilde{y}^\top_{N-1}\end{bmatrix}^\top$, $\mathbf{u} = \begin{bmatrix}u_0^\top & \cdots & u^\top_{L-1}\end{bmatrix}^\top$, and $\mathbf{y}=\begin{bmatrix}y_0^\top & \cdots & y^\top_{L-1}\end{bmatrix}^\top$. The input sequence $\tilde{\mathbf{u}}$ is \emph{persistently exciting} of order $k \in [1,N]$ if the Hankel matrix $\mathcal{H}_k(\tilde{\mathbf{u}})$ of depth $k$ constructed from $\tilde{\mathbf{u}}$ has full row rank \citep{Willems}. {  The Hankel matrix $\mathcal{H}_k(\tilde{\mathbf{u}})$ is constructed by stacking $k$ consecutive shifted copies of the sequence $\tilde{\mathbf{u}}$ to obtain a $(n_uk) \times (N-k+1)$ matrix with block elements $[\mathcal{H}_k(\tilde{\mathbf{u}})]_{ij} = \tilde{u}_{i-1+j-1} \in \bbR^{n_u}$:
\begin{equation}
    \mathcal{H}_k(\tilde{\mathbf{u}}) = \begin{bmatrix}
        \tilde{u}_0 & \tilde{u}_1 & \cdots & \tilde{u}_{N-k}\\
        \tilde{u}_1 & \tilde{u}_2 & \cdots & \tilde{u}_{N-k+1}\\
        \vdots & \vdots & \ddots & \vdots\\
        \tilde{u}_{k-1} & \tilde{u}_k & \cdots & \tilde{u}_N
    \end{bmatrix}
\end{equation}}
If $\tilde{\mathbf{u}}$ is persistently exciting of order $n_x+L$, then $\mathbf{u}$, $\mathbf{y}$ is a trajectory of \eqref{LTI_DT} if there exists a vector $\alpha\in\mathbb{R}^{N-L+1}$ such that
\begin{equation}\label{WFL}
    \begin{bmatrix}
	\mathcal{H}_L(\tilde{\mathbf{u}})\\ \mathcal{H}_L(\tilde{\mathbf{y}})\end{bmatrix}\alpha=\begin{bmatrix}\mathbf{u}\\ \mathbf{y}\end{bmatrix}.
\end{equation}
{  
Each column of the block Hankel matrix corresponds to a finite-length segment of the measured input sequence. Similarly, the columns of the stacked Hankel matrix in \eqref{WFL} collect input-output trajectory segments from the measured data. Intuitively, the algebraic condition in \eqref{WFL} states that, if the measured input \(\tilde{u}\) is persistently exciting of order \(n_x+L\), then these measured trajectory segments span the set of length-\(L\) input-output trajectories generated by the same discrete-time LTI system. In this sense, the stacked Hankel matrix provides a behavioral representation of the system over horizon \(L\), without requiring explicit identification of a state-space realization.}

\subsection{System Identification}
Given a set $\mathcal{W} = \{(u_i(t),y_i(t))\}_i$ of input-output trajectories from a continuous-time dynamical system, \emph{system identification} is the problem of constructing a model of the system. This involves choosing a model order (state dimension $n_x$) and mathematical structure (e.g., nonlinear, control-affine, or linear state-space model), and specifying associated model parameters (e.g., matrices $A,B,C,D$ for the linear state-space model in \eqref{eq:linear_system},  parametrization of the nonlinear functions $f,h$ in \eqref{eq:control_system}, or parameterization of the transfer function $G(s)$ in \eqref{eq:transfer_function}). The parameters of the chosen model are determined by minimizing a suitably chosen difference between the predicted model output and the actual system output observed in the data $\mathcal{W}$. For example, considering a nonlinear state-space model with dynamics $f_\theta$ and output model $h_\theta$, the model parameters $\theta$ may be obtained from an optimization problem:
\begin{equation}\label{eq:general_sysid_optimizaiton}
\begin{aligned}
    \min_\theta\;&\; c(\theta; \mathcal{W}) := \sum_i \int_t \|y_i(t) - \hat{y}_i(t) \|^2 dt\\
    \text{s.t.}\;&\; \dot{\hat{x}}_i(t) = f_\theta(\hat{x}_i(t), u_i(t)),\\
    \;&\; \hat{y}_i(t) = h_\theta(\hat{x}_i(t), u_i(t)),
\end{aligned}
\end{equation}
{  where \(\{(\hat{x}_i(t),\hat{y}_i(t))\}_i\) are state and output trajectories predicted by the model \(f_\theta,h_\theta\). The objective in (14) should not be interpreted as requiring the optimal cost to be zero. A zero value can only be expected in idealized settings where the measured data are noise-free, the chosen model class contains the true system, the initial conditions and inputs are exactly known, and there is no numerical or discretization error. In practical identification problems, measurement noise, process disturbances, unmodeled dynamics, finite data, limited excitation, and model mismatch typically lead to a nonzero residual even at the optimum. Therefore, successful identification is usually assessed not by vanishing training error, but by the model's ability to explain the data, generalize to validation trajectories, and satisfy relevant structural or control-oriented properties. 
}

The above formulation \eqref{eq:general_sysid_optimizaiton} focuses on parametric system identification, where $f_\theta$, $h_\theta$ are the models to be identified. Motivated by Willems's work \citep{willems_part1,willems_part2,willems_part3}, system identification can also be carried out using non-parametric techniques, such as total least-squares \citep{total_least_squares}, deterministic subspace \citep{vanoverschee_subspaceidentification}, and structured low-rank approximation \citep{low_rank_approximation} approaches. 

After the identification of a system model, model validation and model analysis are crucial next steps to guarantee the usefulness of the model for prediction, optimization, and control design. Model validation refers to assessing the accuracy of the identified model by comparing its predictions with additional (validation) data that was not used during the parameter estimation phase. This helps to verify that the model's performance on the validation data and the training data are comparable, i.e., the model is not overfitting. Model analysis refers to asserting that the identified model satisfies properties and performance characteristics expected to be true for the considered system. This is essential to ensure that the model accurately represents the real-world system, facilitating reliable prediction, robust control design, and safe operation.

\subsection{System Properties}
\label{sec: properties}
Real-world dynamical systems display important properties. From a control perspective, desirable properties that may be considered include  stability, dissipativity, monotonicity, symmetries, to name a few. Ensuring that, when a system satisfies these properties, they are reflected in the system model is important for model accuracy and control design. 
Capturing these properties in the system model can often lead to simplified mathematical treatment, accurate reflection of a real system's behavior, and convenient control design.

\subsubsection{Lyapunov Stability}
Lyapunov stability is a fundamental concept for understanding the behavior of a dynamical system. Stable systems behave predictably in the sense that they return to an equilibrium (or a set of equilibria) after a disturbance and do not exhibit unbounded behavior. 
Suppose $x_e \in D, D \subset \mathbb{R}^n$, is an equilibrium of the closed-loop autonomous system \eqref{eq:closed_loop_autonoumoussystem}, that is $F(x_e) = 0$. For convenience, we state all definitions and theorems in this section for the case when the equilibrium point is at the origin of $\mathbb{R}^n$, that is $x_e = 0$. There is no loss of generality in doing so because any equilibrium point can be shifted to the origin via a change of variables \citep{Khalil}.

\begin{theorem}[Lyapunov Stability]
\label{thm: Lyapunov_stability}
Let $x = 0$ be an equilibrium point for the closed-loop autonomous system in \eqref{eq:closed_loop_autonoumoussystem} and $D \subset \mathbb{R}^n$ be a domain containing $x = 0$. Let $V: D \rightarrow \mathbb{R}$ be a continuously differentiable function such that
\begin{gather*}
    V(0) = 0 \text{ and } V(x)>0, \; \forall x \in D \setminus \{0\},\\
    \dot{V}(x) := \frac{\partial V}{\partial x} F(x) \leq 0\,, \forall x \in D.
\end{gather*}
{  Then, \(x=0\) is stable in the sense of Lyapunov, that is, for every \(\epsilon>0\), there exists a \(\delta>0\) such that, whenever \(\|x(0)\|<\delta\), the solution satisfies \(\|x(t)\|<\epsilon\) for all \(t\geq 0\). Moreover, if \[
\dot{V}(x) < 0\,, \forall x \in D \setminus \{0\},
\] then \(x=0\) is asymptotically stable, that is, \(x=0\) is stable and there exists \(c>0\) such that, whenever \(\|x(0)\|<c\), the solution satisfies \(\lim_{t\to\infty}x(t)=0\).}
\end{theorem}

The condition that $\dot{V}(x)$ is negative in Thm.~\ref{thm: Lyapunov_stability} suggests that for any initial condition $x(0) \in D$, the system trajectories $x(t)$ move to smaller and smaller
values of $V(x)$, eventually approach or converge to $x=0$. 

LaSalle’s invariance principle \citep[Theorem 4.4]{Khalil} generalizes the above notion of stability from an equilibrium point to stability with respect to a set. To state LaSalle's invariance theorem, we need to introduce a few definitions. A set $M$ is said to be an invariant set with respect to the closed-loop autonomous system \eqref{eq:closed_loop_autonoumoussystem} if $x(0) \in M \Rightarrow x(t) \in M \,, \forall t \in \mathbb{R}$. That is, if a solution belongs to $M$ at some time instant, then it belongs to $M$ for
all future and past time. A set $M$ is said to be a positively invariant set with respect to the closed-loop autonomous system \eqref{eq:closed_loop_autonoumoussystem} if $x(0) \in M \Rightarrow x(t) \in M \,, \forall t \geq 0$.

\begin{theorem}[LaSalle's Invariance Principle]
\label{thm: LaSalle_invariance}
Let $\Omega \subset D$ be a compact set that is positively invariant with respect to \eqref{eq:closed_loop_autonoumoussystem} and $D \subset \mathbb{R}^n$ be a domain containing $x = 0$. Let $V: D \rightarrow \mathbb{R}$ be a continuously differentiable function such that $\dot{V}(x) \leq 0$ in $\Omega$. Let $E$ be the set of all points in $\Omega$ where $\dot{V} = 0$. Let $M$ be the largest invariant set in $E$. Then every solution starting in $\Omega$ approaches $M$ as $t$ approaches infinity. That is, for any $\epsilon > 0$, there exists a $T > 0$ such that $\text{dist}(x(t), M) < \epsilon$ for all $t > T$, where $\text{dist}(x(t), M) = \inf_{p \in M} \|x(t)-p\|$.
\end{theorem}

In Lyapunov stability analysis and its generalization via the LaSalle’s invariance principle, one first needs to identify a candidate Lyapunov function $V(x)$. However, finding a Lyapunov function for a nonlinear system is challenging, especially when the system dynamics are unknown. For dynamical systems with a special structure, such as networked systems or mechanical systems, there are some natural choices of Lyapunov or generalized energy functions, which we will introduce in the following subsections.

\subsubsection{Dissipativity and Passivity} \label{subsubsec:dissipativity_passivity}
Dissipativity extends the concept of passive circuit elements from circuit theory to a  generalized notion of energy (not necessarily corresponding to a physical quantity) that is applicable to nonlinear dynamical systems. Dissipativity theory offers a powerful framework for analyzing dynamical systems and designing controllers, as it can be used to establish crucial properties such as stability, passivity, and sector-boundedness (see Definition \ref{def:qsr_dissipativity}). One of the key advantages of dissipativity is its preservation across various network interconnection structures, making it particularly valuable for compositional and distributed control synthesis \citep{antsaklis2013control,agarwalsequential2,agarwal2020compositional, agarwal2019sequential,vidyasagar1979new}. Owing to these attractive properties, dissipativity has been widely studied in the control literature, and finds application in several domains including, but not limited to, robotics \citep{hatanaka2015passivity}, electromechanical systems \citep{ortega2013passivity}, aerospace systems \citep{forbes2011extensions},  process control \citep{bao2007process}, and power and energy systems \citep{sivaranjani2020distributed,sivaranjani2020mixed}. Concretely, we  define dissipativity for a nonlinear system as follows.

\begin{definition}[Dissipativity]\label{def:dissipativity}
The dynamical system \eqref{eq:control_system} is said to be dissipative with respect to a \textit{supply rate} $s:\mathcal{U}\times \mathcal{Y}\to \mathbb{R}$, where $\int\limits_{0}^{T}|s(u,y) dt| <\infty, \forall 0,T \in \mathbb{R}_{+}$, if there exists a positive definite function $V:\mathcal{X}\to \mathbb{R}_{+}$, such that, $\forall t\in[0,T], x(0)\in \mathcal{X}$, and $y\in \mathcal{Y}$, 
\begin{equation}
    \int\limits_{0}^{T} s(u,y) dt \geq V(x(T))-V(x(0)),
\end{equation}
where $x(T)$ is the state of the dynamical system \eqref{eq:control_system} at time $T$, resulting from the initial condition $x(0)$ and input $u(\cdot)$.
\end{definition}

For nonlinear systems, dissipativity from Definition~\ref{def:dissipativity} often holds only in the neighbourhood of an equilibrium. Incremental dissipativity \citep{DifferentialSys, IncrementalRoland} extends this notion by considering the relationship between two arbitrary system trajectories. In general, incremental dissipativity implies dissipativity but not vice versa. For linear systems, incremental and non-incremental properties are equivalent. The analysis of incremental dissipativity can be particularly useful when operating a system in a transient mode, between operating points, or in a reference tracking scenario where stability and performance with respect to a certain trajectory are desired. Specifically, incremental dissipativity can be obtained by replacing the supply rate $s(u,y)$ in Definition \ref{def:dissipativity} by $s(u_1-u_2,y_1-y_2), u_1,u_2 \in \mathcal{U}, y_1, y_2 \in \mathcal{Y},$ and the storage functions by $V(x_1(\cdot),x_2(\cdot)).$ Analogous discrete-time notions of dissipativity and incremental dissipativity can also be appropriately defined, but are omitted here for brevity of exposition.

A particularly useful special case of dissipativity is the property of quadratic dissipativity, commonly referred to as $QSR$-dissipativity, defined as follows. 
   \begin{definition}[QSR-Dissipativity]\label{def:qsr_dissipativity}
The nonlinear system \eqref{eq:control_system} is said to be \textit{$QSR$-dissipative} with dissipativity matrices $Q=Q^\top $, $S$ and $R=R^\top $ of appropriate dimensions, if $\forall x \in \mathcal{X}$ and control inputs $u \in \mathcal{U}$, Definition \ref{def:dissipativity} holds with $s(u,y)=y^\top Qy+u^\top Ru+2y^\top Su$. By selecting appropriate dissipativity matrices $Q, S,$ and $R$, this formalism can capture several control-theoretic properties of importance, including but not limited to:
		\begin{enumerate}
			\item[(i)] \textit{passivity}, with $Q = 0$, $S = \frac{1}{2}I$ and $R =0$,
			\item[(ii)] \textit{strict passivity}, with $Q=-a I$, $S= \frac{1}{2}I$ and $R =-b I$, where $a,b \in \mathbb{R}^+\backslash\{0\}$,
			\item[(iii)] \textit{$\mathcal{L}_2$ stability}, with $Q=-\frac{1}{\gamma} I$, $S= 0$ and $R =\gamma I$ where $\gamma\in \mathbb{R}^+$ is an $\mathcal{L}_2$ gain of the system, 
			\item[(iv)] \textit{conicity}, with $Q=-I$, $S = c I$ and $R = (r^2-c^2)I$, where $c\in\mathbb{R}$ and $r \in \mathbb{R}^+\backslash\{0\}$, and,
			\item[(v)] \textit{sector-boundedness}, with $Q=-I$, $S = (a+b)I$ and $R = -ab I$, where $a,b\in \mathbb{R}$.
		\end{enumerate}
Note that the same choices of supply rates can also be used in the appropriate definitions to establish discrete-time or incremental counterparts of these properties. 
\end{definition}

{  Dissipativity is useful for stabilization because the storage function can often be used as a Lyapunov function once the supply rate is made nonpositive by feedback. For example, {if the system is passive, so that}  \(s(u,y)=u^\top y\), negative output feedback \(u=-Ky\), with \(K\succeq 0\), gives \(\dot{V}(x)\leq u^\top y=-y^\top K y\). Thus, the stored energy is nonincreasing along closed-loop trajectories; under additional strictness or detectability conditions, this yields asymptotic stability. More generally, passivity-based control exploits the fact that passive systems remain passive under suitable interconnections, and that a passive plant interconnected with a passive or strictly passive controller admits a natural closed-loop Lyapunov function obtained from the sum of the corresponding storage functions. As mentioned in Section 3.1, these dissipativity-based interconnection properties also support compositional and distributed controller synthesis in networked systems \citep{agarwalsequential2}; see the accompanying Github repository for an illustrative example.}

For LTI systems, dissipativity is guaranteed if there exists matrix $P\succ 0$ satisfying the linear matrix inequality (LMI)
{\small
\begin{equation}\label{eq: dissipativity}
\begin{bmatrix}
A^\top P + PA - C^\top QC & PB -C^\top S- C^\top QD \\
(PB -C^\top S- C^\top QD)^\top  & -D^\top QD-(D^\top S+S^\top D)-R
\end{bmatrix}
\preceq 0,
\end{equation}}
\noindent where appropriate choices of the $Q,S,$ and $R$ matrices can be made to reflect special cases like passivity, as described in Definition \ref{def:qsr_dissipativity}.

For transfer function models $G_\theta(s)$, 
where the parameters $\theta$ represent the coefficients of the numerator and denominator polynomials of the transfer function, properties like passivity, positive realness, and bounded realness, which are special cases of dissipativity, can be directly imposed by enforcing  their corresponding frequency domain conditions. For example, strict passivity (positive realness) can be enforced by ensuring that $G_\theta(s)$ is stable, while satisfying the hard constraint 
\begin{equation}\label{eq: hard_stability}
g(\theta)=-G_\theta(i\omega)-G_\theta(-i\omega)<0, \forall \omega \in \mathbb{R}.\end{equation}
Similarly, bounded realness can be enforced by requiring that $G_\theta(s)$ is stable and satisfies the hard constraint \begin{equation}\label{eq: hard_bounded_real}g(\theta)=-I+G_\theta^\top (-i\omega)G_\theta(i\omega)<0, \forall \omega \in \mathbb{R}.\end{equation} 
Analogous definitions of these properties for discrete-time models and LMI conditions for state-space models (similar to \eqref{eq: dissipativity}) are also available \citep{kottenstette2010relationships}.

There are also related properties such as differential dissipativity \citep{DifferentialSys,van2013differential} and equilibrium-independent dissipativity \citep{simpson2018equilibrium,hines2011equilibrium}; however, these notions have not yet been explored as much in the context of system identification. Similarly, integral quadratic constraints (IQCs) establish an attractive framework to achieve tighter descriptions of input-output properties, with a number of uncertainties that can be characterized by IQCs in \cite{IQC_Rantzer}. However, in the context of physics-constrained system identification, IQCs have not been explored much thus far.

\subsubsection{Monotonicity}
Monotone systems have outputs that change in a consistent direction in response to changes in inputs. Similar to dissipativity, this property simplifies the model analysis and control design by providing guarantees about the direction of the system responses.
A special case of monotonicity is positivity of the dynamics, where the state variables are constrained to be positive. This corresponds to the physics of several systems such as reaction networks where the concentration of a chemical species cannot be negative. 

Monotone systems feature in several control applications of interest, including chemical reaction networks \citep{leenheer2007monotone}, biological systems \citep{angeli2021monotone},  epidemiological processes \citep{brauer2012mathematical}, and traffic flow networks \citep{lovisari2014stability,coogan2015compartmental}. In fact, most compartmental flow networks, which describe the flow of a commodity (such as traffic or a reactant) from one compartment to another under a conservation law, follow monotone dynamics \citep{haddad2010nonnegative}. Cooperation and competition dynamics in populations and teams can also be modeled using monotone systems \citep{hirsch2005monotone,smith2017monotone}.  From a control standpoint, monotonicity is an important property that can be exploited for distributed stability analysis  \citep{coogan2019contractive}, reachability analysis \citep{althoff2021set}, and control design \citep{shiromoto2018distributed,rantzer2014control, cui2024structured} in large-scale networks. This is due to the fact that linear monotone systems admit Lyapunov functions that are separable in terms of the Lyapunov functions corresponding to individual state variables, allowing for compositional stability verification \citep{haddad2010nonnegative}. These compositional properties also extend to certain classes of nonlinear monotone systems \citep{dirr2015separable}. Mathematically, monotonicity of dynamical systems is defined below, following~\citet{angeli2003monotone}.
\begin{definition}[Monotone Systems]
\label{def: monotone}
    Consider a controlled dynamical system $f$ as defined in \eqref{eq:control_system} $\dot{x} = f(x, u)$. Assume that solution $x(t) = \phi(t, x_0, u)$ with initial condition $x(0) = x_0$. 
    The dynamics $f$ is said to be \emph{monotone} if the following implication holds for all $t \geq 0$:
    $$u_1 \succeq u_2, x_1 \succeq x_2 \Rightarrow \phi(t, x_1, u_1) \succeq \phi(t, x_2, u_2),$$
    for a suitably defined ordering $\succeq.$ 
    The typical case for the ordering $y \succeq z$ for $y, z \in \mathbb{R}^n$ means each coordinate of $y$ is bigger or equal than the corresponding coordinate of $z$.  
    However, other orderings could be induced by choosing other orthants in $\mathbb{R}^n$ other than the positive orthants $K = \mathbb{R}^n_{\geq 0}$.
\end{definition}

\subsubsection{Symmetries of Physical Systems}
Symmetric systems have responses that are invariant to certain continuous transformations. System models that capture symmetries reduce the model complexity and improve the computational efficiency and accuracy of system dynamics prediction.  A continuous transformation $\gamma \in \Gamma$ is typically described by the action of a compact Lie group $\Gamma$ on the state space $\mathcal{X}$ of the system.

\begin{definition}[Group Action] \label{def:group_action}
Let $\Gamma$ be a group with operation $\cdot$ and identity element $e$. An action of $\Gamma$ on set $\mathcal{X}$ is a function $\rho: \Gamma \times \mathcal{X} \rightarrow \mathcal{X}$ that satisfies an identity property, $\rho(e,x)=x$, and a compatibility property, $\rho(a,\rho(b,x)) = \rho(a\cdot b,x)$ for $a,b \in \Gamma$. The notation is commonly simplified as $\gamma x := \rho(\gamma,x)$.  
\end{definition}

A group element $\gamma \in \Gamma$ is a \emph{symmetry} of the dynamics $F$ of a closed-loop system \eqref{eq:closed_loop_system} if, for every solution $x(t)$, $\gamma x(t)$ is also a solution. This leads to a useful condition for $\gamma$ to be a symmetry of $F$ \citep{golubitsky2002symmetry}.

\begin{definition}[Equivariant System] \label{def:ode_symmetry}
A system $\dot{x} = F(x,t)$ with $x\in \mathcal{X}$ is equivariant under an action of group $\Gamma$ on $\mathcal{X}$ if $F(\gamma x,t) = \gamma F(x,t)$ for all $t$, $\gamma \in \Gamma$, $x \in \mathcal{X}$.
\end{definition}

Symmetry in physical systems \citep{marsden2013introduction} has been widely studied in physics, from Lagrangian and Hamiltonian mechanics \citep{lurie2013analytical, HolmBook} to quantum mechanics \citep{greiner2012quantum}. One of the most well-known results is Noether's theorem \citep{noether1983invariante}, which states that there is a conservation law for every continuous symmetry of a system. The dynamics of a physical system can be described using a Lagrangian function $\mathcal{L}$. The action of the system, defined as the integral over time of the Lagrangian, governs the system behavior via the principle of least action \citep{HolmBook}. If the Lagrangian is invariant under a group of transformations $\Gamma$, i.e., $\mathcal{L}(\gamma x) = \mathcal{L}(x)$, $\forall \gamma \in \Gamma$, Noether's theorem guarantees the existence of conserved quantities associated with the generators of that group. For example, time-invariant Lagrangians, i.e., $\mathcal{L}(t + \delta t) = \mathcal{L}(t)$, lead to the conservation of the total energy of the system. Translation-invariant Lagrangians, i.e., $\mathcal{L}(x + \delta x) = \mathcal{L}(x)$, lead to the conservation of linear momentum, while rotational invariant Lagrangians, i.e., $\mathcal{L}(Rx) = \mathcal{L}(x)$ where $R$ is a rotation matrix, lead to the conservation of angular momentum. 

In reality, the assumption on conservative forces rarely holds as there are often external non-conservative forces, such as frictions or resistance, applied on the system. Such forces dissipate energy from the system and require a generalization of the mechanics model to account for energy dissipation. For example, a port-Hamiltonian model \citep{van2014port} combines concepts from Hamiltonian mechanics (which uses energy to describe motion) with port-based modeling (which views systems as interconnected elements exchanging power).

\section{Physics-Informed System Identification: Motivation, Problem, and Approaches}\label{sec:perspectives}

\subsection{Why Preserve Control-Relevant Properties?}
In general, preserving control-relevant and other physics-informed system properties, such as those described in Section \ref{sec: properties}, provides several advantages. Firstly, 
incorporating a priori information regarding system physics often improves the sample efficiency of the identification algorithms and generalizability of the identified models.
Secondly, control-relevant properties such as dissipativity and monotonicity can be exploited to facilitate scalable and distributed control designs with provable guarantees. Finally, capturing physics-based properties such as symmetries allows for explainability of the identified model, which is crucial when utilized in high-stake control applications. Here, we demonstrate through two concrete examples how control-relevant and physics-informed properties can directly aid in control design.

We start by considering the example of conservation laws - specifically, energy conservation. 
Energy conservation, if preserved in the dynamics model, has been shown to be beneficial for data efficiency, control design and stability analysis \citep{greydanus2019hamiltonian, zhong2020symplectic, wangincorporating, duong2021hamiltonian}. While the law of energy conservation universally holds for physical systems, a black-box machine learning model may struggle to learn this knowledge, even from a large amount of data. 
By enforcing the law of energy conservation in the machine learning model, e.g., using Lagrangian or Hamiltonian formulations of physical systems, energy conservation is guaranteed by design, thus improving data efficiency and prediction accuracy \citep{miller2020scaling}. Preserving the law of energy conservation also facilitates control design with the learned model from an energy perspective. Given a desired state $x^*$, the total energy of the system is generally not minimized at $x^*$, i.e., $x^*$ is not an equilibrium point of the system. By decomposing the system into energy storing and dissipating elements in a Hamiltonian formulation, one can inject energy through the control input to shape the total energy of the system to a desired total energy function, minimized at $x^*$, and thus, move the equilibrium point to $x^*$ \citep{van2014port}. 
We now illustrate how preserving the Hamiltonian structure in neural dynamical models can help to achieve superior training performance compared to  black box models.

\begin{example}[Hamiltonian neural ODE]\label{ex:hamiltonian_pendulum}
Consider a pendulum system with dynamics:
\begin{equation}
\label{eq:pend_gt_q_dyn}
\ddot{\varphi} = -\frac{g}{l}\sin{\varphi} + u,
\end{equation}
where $\varphi$ is the angle of the pendulum with respect to its downward position,$u$ is a scalar control input, $g$ is the gravitational acceleration and $l$ is the length of the pendulum. As a physical system, the pendulum dynamics can be formulated using Hamiltonian formulation, either on $\bbR^n$, i.e., using the angle $\varphi$ as the state \citep{zhong2020symplectic}, or on a Lie group, i.e. using the rotation matrix, generated from the Euler angle $[0, 0, \varphi]$, as a state $R$ on the $SO(3)$ manifold:
\begin{equation}
R = \begin{bmatrix}
\cos{\varphi} & - \sin{\varphi} & 0 \\
\sin{\varphi} & \cos{\varphi} & 0 \\
0 & 0 & 1 
\end{bmatrix}.
\end{equation}

\begin{figure}[t]
\begin{subfigure}[t]{0.23\textwidth}
        \centering
        \includegraphics[width=\textwidth]{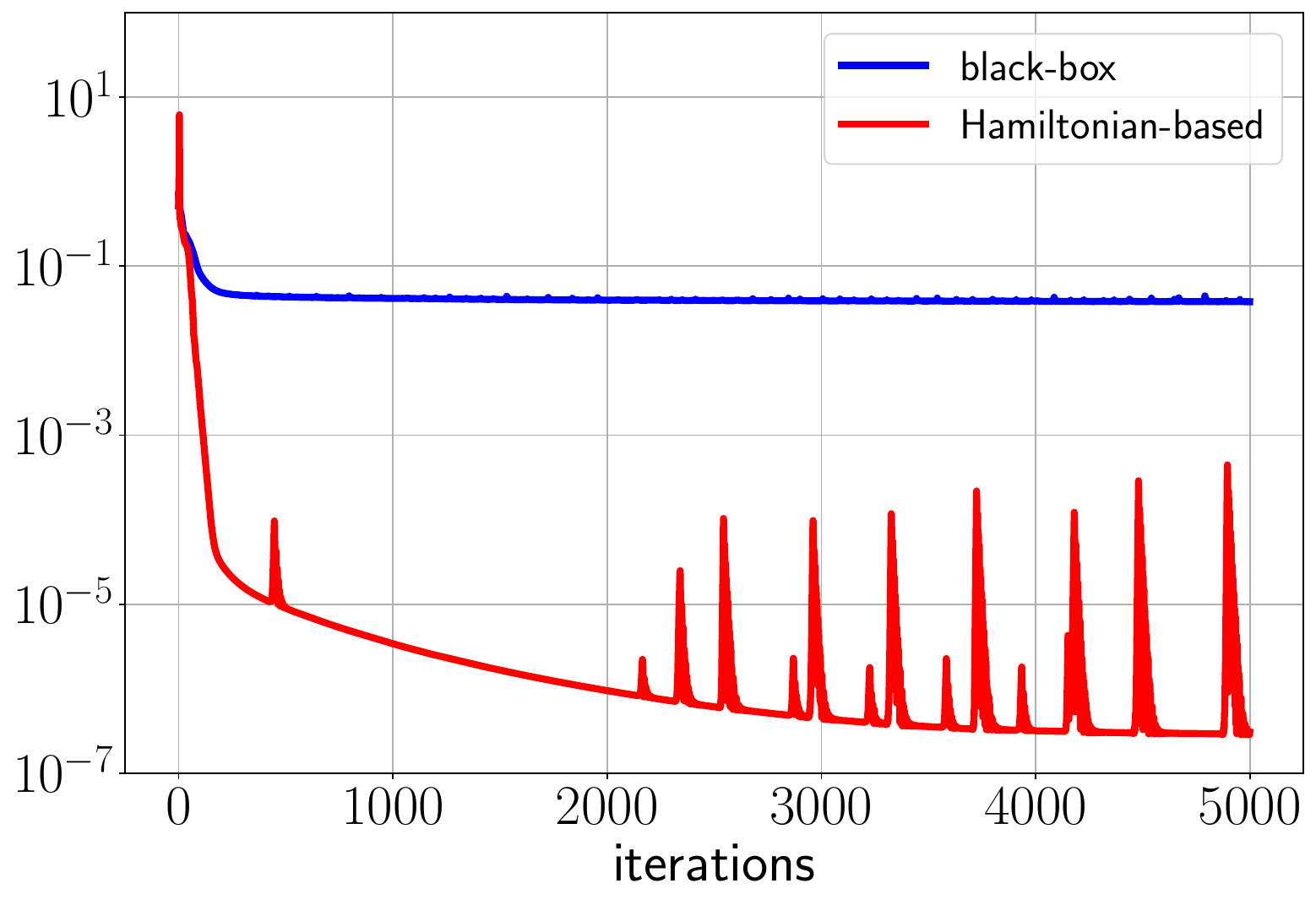}
        \caption{Training and test loss.}
        \label{fig:pend_loss}
\end{subfigure}%
\hfill%
\begin{subfigure}[t]{0.23\textwidth}
        \centering
        \includegraphics[width=\textwidth]{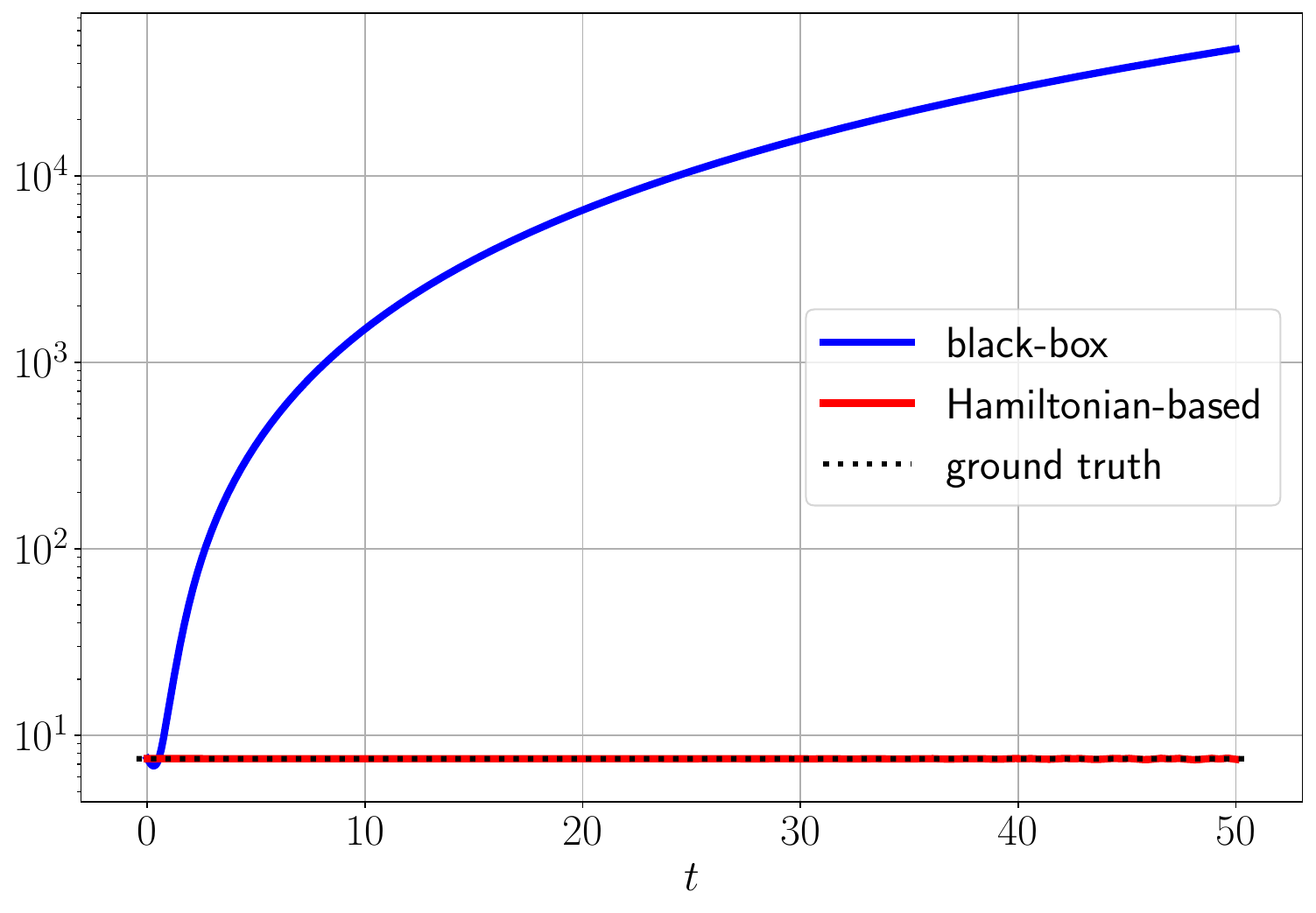}
        \caption{Total energy (Hamiltonian).}
        \label{fig:pend_ham}
\end{subfigure}%

\begin{subfigure}[t]{0.23\textwidth}
        \centering
\includegraphics[width=\textwidth]{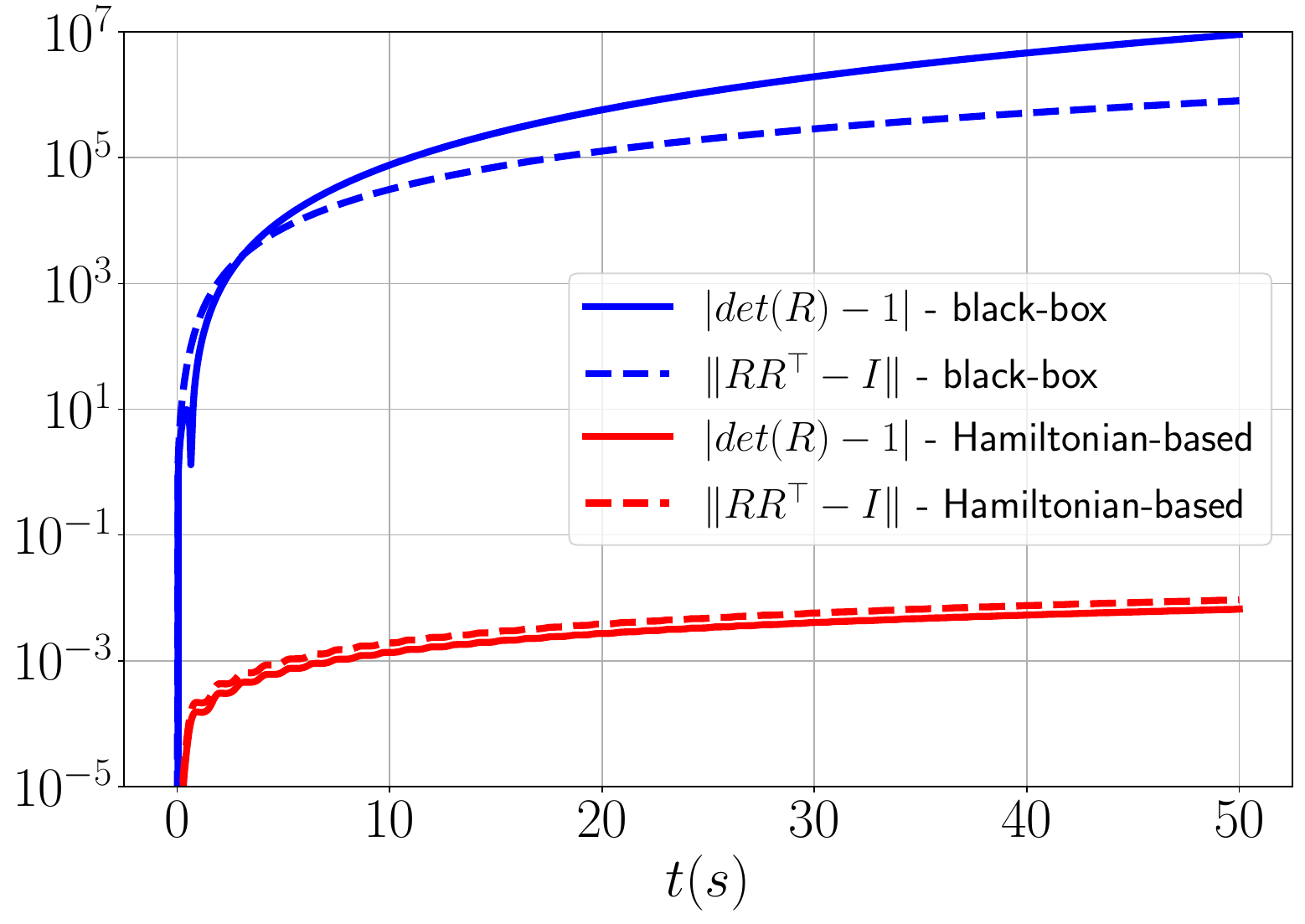}%
        \caption{Manifold constraints.}
        \label{fig:pend_constraints}
\end{subfigure}%
\hfill
\begin{subfigure}[t]{0.245\textwidth}
        \centering
        \includegraphics[width=\textwidth]{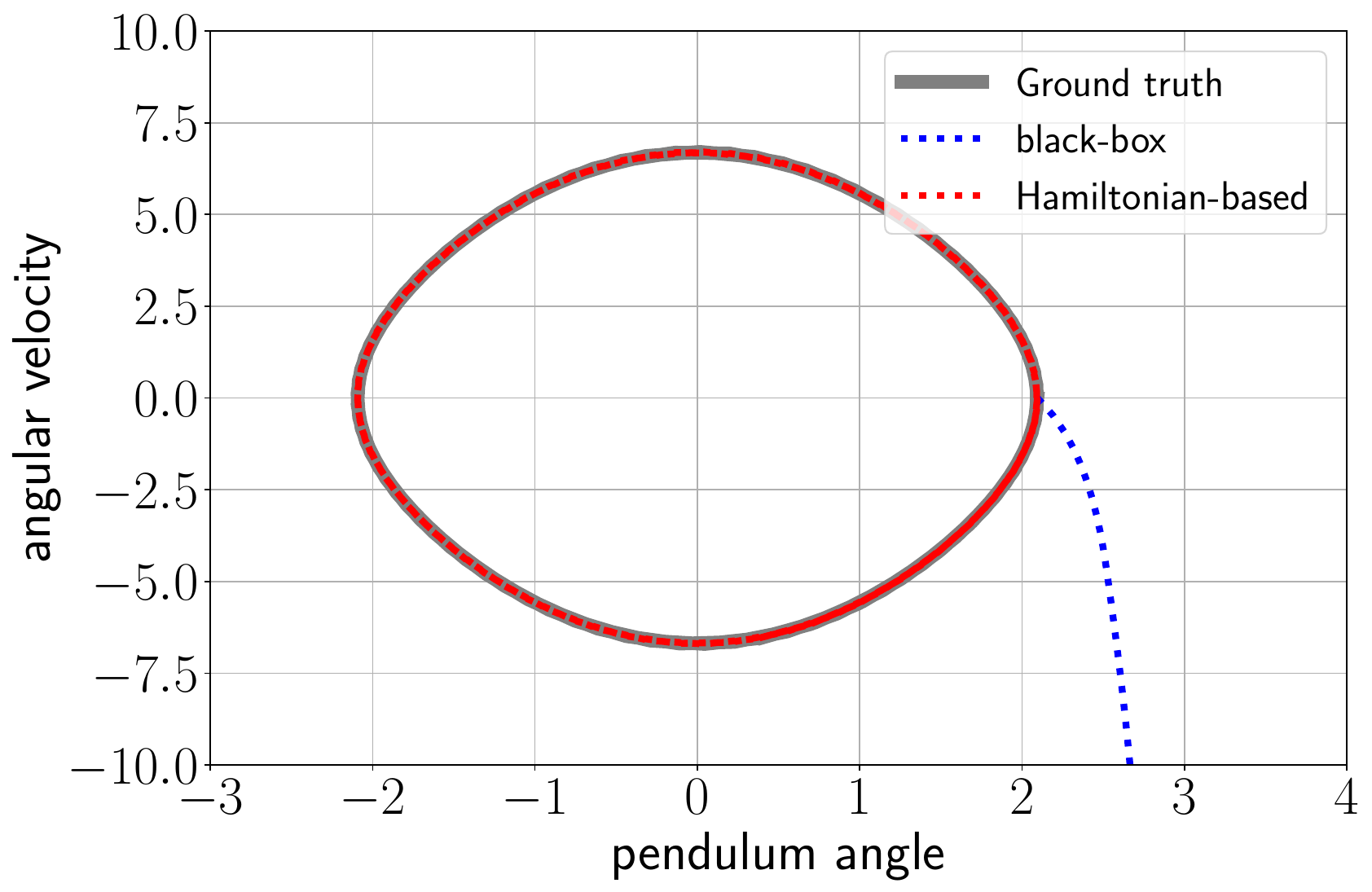}
        \caption{Phase portraits of state predictions.}
        \label{fig:pend_phase_portraits}
\end{subfigure}%

\caption{Pendulum dynamics identification using an $SO(3)$ port-Hamiltonian neural ODE network \citep{Duong_HNODE_2024}.}
\label{fig:pend_exp}
\end{figure}
The matrix $R$ satisfies the $SO(3)$ constraints: $R^\top R = I$, $\det(R) = 1$. The control input $u$ is randomly sampled and applied to the pendulum to collect a dataset of state-control trajectories. This dataset trains a $SO(3)$ Hamiltonian-based neural ODE network \citep{Duong_HNODE_2024}, which encodes both energy conservation and $SO(3)$ constraints via a Hamiltonian formulation. We discuss such physics-informed architectures in Section \ref{sec:deep}. {The dataset is also used to train a black-box neural ODE network for comparison, where the dynamics function $f$ in \eqref{eq:control_system} is represented by a multilayer perceptron network}. Fig. \ref{fig:pend_loss} shows that enforcing physics constraints improves training with faster convergence. Figs. \ref{fig:pend_ham} and \ref{fig:pend_constraints} show the learned model preserves energy and $SO(3)$ constraints when rolled out from an initial angle $\varphi = \frac{2\pi}{3}$ without control. The predicted angle and velocity also follow the pendulum's phase portraits (Fig. \ref{fig:pend_phase_portraits}).
\end{example}

The above example demonstrates how  properties such as energy conservation  can be leveraged to fit models with better sample efficiency and interpretability. Incorporating properties like dissipativity can also facilitate scalable analysis and control in large-scale networked systems \citep{arcak2016networks,arcak2022compositional,agarwalsequential2,agarwal2020compositional,jena2021distributed,sivaranjani2020distributed}. We provide an additional example demonstrating how dissipativity can be leveraged for distributed synthesis of controllers in our Github repository. Overall, these benefits motivate the idea of preserving control-relevant properties in system identification. 

\subsection{Does Classical System Identification Preserve Control-relevant Properties?}
Since it is desirable to preserve properties such as energy conservation and dissipativity that can inform or simplify control designs for complex systems, a natural question is: \emph{If we are able to identify models that approximate the dynamics very closely, will these system properties be naturally inherited in the identified model?} Unfortunately, this is not the case, as we illustrate with a simple counter-example. We show that even for a  linear system, there is no guarantee that a learned model that approximates the system very closely will preserve the  dissipativity property possessed by the original system. 

\begin{figure}[t]
    \centering
    \includegraphics[scale=0.55,trim=0cm 0.5cm 0cm 0cm]{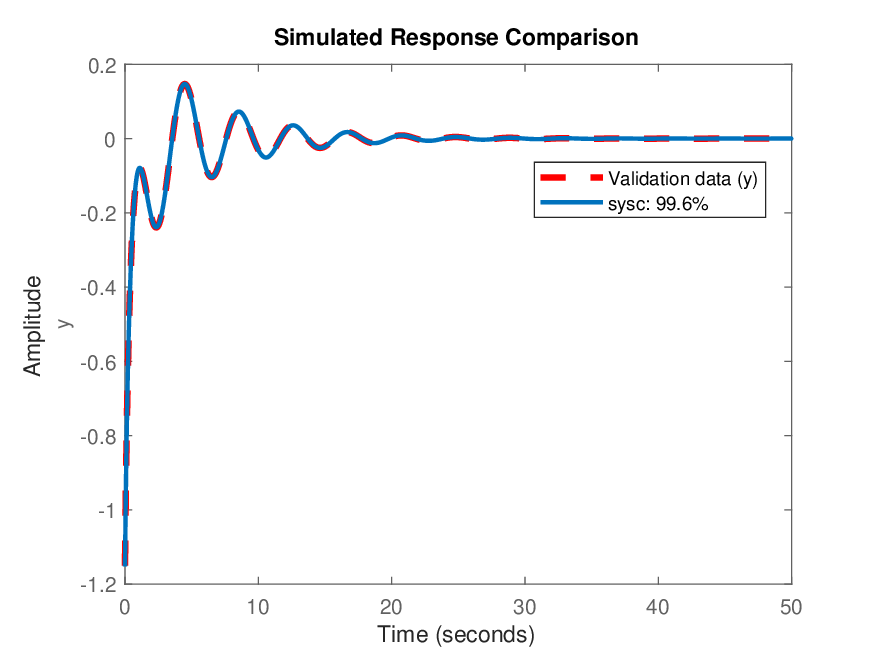}
    \caption{Linear system vs model trajectories for Example \ref{ex:counter-example}. The model closely approximates the system test data  (with a fit of 99.6\% in this case) but does not preserve the passivity of the system.}
    \label{fig:id-counter-example}
\end{figure}

\begin{example}[Linear system counter-example]\label{ex:counter-example}
We consider a linear system as follows.
\begin{align*}
    \dot{x}(t) &= Ax(t)+Bu(t) \\
    y(t) &= Cx(t)+Du(t),
\end{align*}
where $$A=\begin{bmatrix}
    -2 & -3 & -4\\
    1 & 0 & 0\\
    0 & 1 & 0
\end{bmatrix}, \quad B=\begin{bmatrix}
    1 \\ 0 \\ 0
\end{bmatrix}, \quad C=\begin{bmatrix}
    -1 & 0 & 2
\end{bmatrix}, \quad D = 1.$$ 
This system possesses the property of passivity, that is, it is dissipative with $Q=0, S=\frac{1}{2}, R=0$, which we can verify, e.g., using the MATLAB {\tt{isPassive}} function. For system identification, we generate training data using the input  $u(t)=te^{-5t}$. We show that even if we can measure the exact output, which is the most desirable case, and identify a linear model that closely approximates the system behavior, we still lose the passivity property in the identified model. 

In Fig. \ref{fig:id-counter-example}, we compare the trajectory of the ground truth and the one generated by the identified linear model using the MATLAB System Identification Toolbox. The identified learned model has matrices 
$$A_{id} = \begin{bmatrix}
    -0.033 & -1.325 & -0.001\\
    1.176 & -0.146 & -1.693\\
    -0.5 & 0.654 & -1.842
\end{bmatrix} \quad B_{id} =\begin{bmatrix}
    -4.68 \\
    14.2 \\
    -6.28
\end{bmatrix},$$ 
$C_{id} = \begin{bmatrix}
    3.368 & -0.023 & 0\\
\end{bmatrix}$ and $D_{id}=0$.
We can observe that the model manages to approximate the input-output behavior of the system extremely well, as seen from the validation/test data, with goodness of fit being 99.6\%.  However, when we test the passivity of the identified model $(A_{id},B_{id},C_{id},D_{id})$ using the MATLAB function {\tt isPassive}, we find that this model is, in fact, \textit{not} passive! This demonstrates that system identification does not guarantee to preserve the underlying dynamical system properties, even when the identified model behavior is extremely close to the system input-output behavior.
\end{example}

One intuition behind this observation is that there are an \emph{infinite} number of models that can achieve a desired level of fit for a set of system input-output trajectories. These models may differ significantly in their parameters (such as the $(A,B,C,D)$ matrices in the above example), but be close in terms of overall input-output behavior in the region of the state-space in which they are trained to approximate the unknown system dynamics.

From the preceding discussion, given that system properties like passivity are not inherited in system identification even for linear models, it is clear that we cannot provide any guarantees on more complex models like neural network models--even if the model accurately approximates the system behavior. Therefore, it necessitates specific approaches to impose control-relevant properties during identification. 

\subsection{Physics-Informed Identification as an Optimization Problem: Parametrization, Constraints, or Regularization?}
\label{sec:ways_incorporating_physics}
\begin{figure*}[t]
    \centering
    \includegraphics[width=0.95\linewidth]{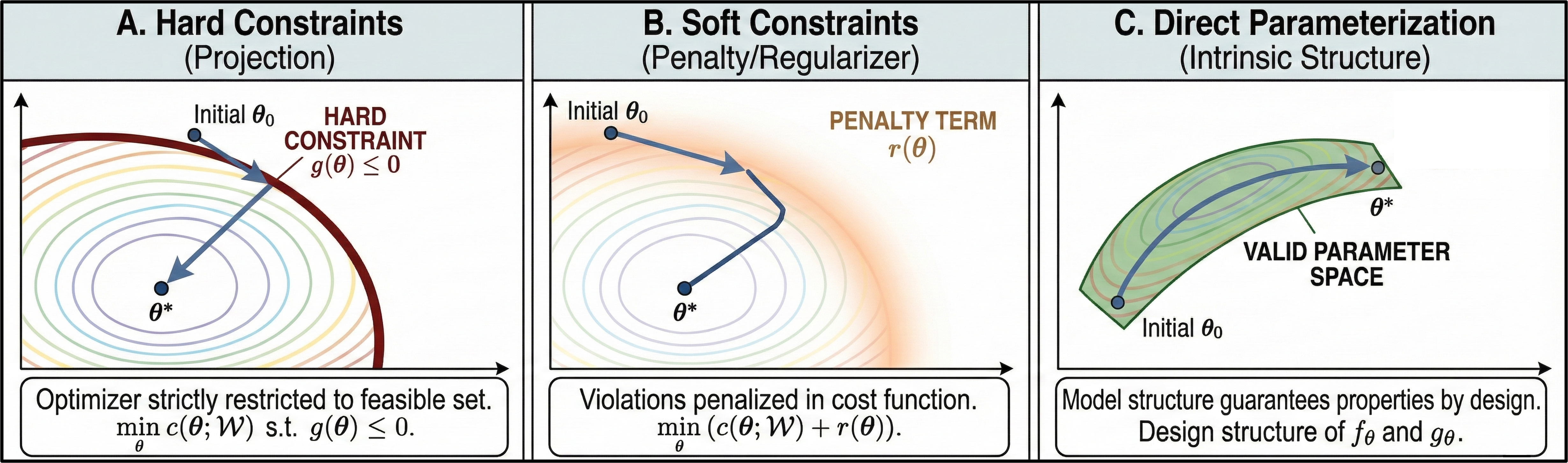}
    \caption{Approaches to impose control-oriented properties in optimization-based formulation of system identification.}
    \label{fig:sys_id_optimization}
\end{figure*}
Considering the identification problem for a nonlinear system \eqref{eq:control_system}, there are, in general, \textit{three ways} of imposing structure on the obtained system model: \textit{{direct parameterization}, {hard constraints}, and {soft constraints}}. 
In direct parametrization, we may use a particular parametrization of $f_\theta$ and $h_\theta$ to impose system properties, such as assuming linear structure and identifying $A, B, C, D$ matrices rather than a general neural network parameterized dynamics $f_{\theta}, h_{\theta}$. Alternately, we may impose hard constraints on the parameters $\theta$ in the optimization problem in \eqref{eq:general_sysid_optimizaiton}, such as dissipativity imposed as an LMI \citep{MartinIQC}. {Finally}, we may impose a soft constraint on $\theta$ in the form of a regularization term in the cost function in \eqref{eq:general_sysid_optimizaiton}, which promotes a property of interest such as stability \citep{lechner2020gershgorin} and energy conservation \citep{greydanus2019hamiltonian}. These three ways of imposing model structure are summarized in Fig.~\ref{fig:sys_id_optimization}. They can be introduced in the general system identification optimization in \eqref{eq:general_sysid_optimizaiton} as:
\begin{equation}
\label{eq:constrained_sysid_optimizaiton}
\begin{aligned}
    \min_\theta\;&\; c(\theta; \mathcal{W}) := \sum_i \int_t \|y_i(t) - \hat{y}_i(t) \|^2 dt + r(\theta)\\
    \text{s.t.}\;&\; \dot{\hat{x}}_i(t) = f_\theta(\hat{x}_i(t), u_i(t)),\\
    \;&\; \hat{y}_i(t) = h_\theta(\hat{x}_i(t), u_i(t)),\\
    \;&\; g(\theta) \leq 0, 
\end{aligned}
\end{equation}
where $\mathcal{W} = \{(u_i(t),y_i(t))\}_i$ 
represents the set of input-output trajectories from a continuous-time dynamical system, $f_\theta$, $h_\theta$ encode model structure through their parametrization, $g(\theta)$ captures hard constraints on the model parameters $\theta$, and $r(\theta)$ is a regularization term capturing soft constraints.

{  These three approaches differ in the type of guarantee they provide, as well as in their computational scalability. Direct parameterization can enforce a desired property by construction, provided that an appropriate parametrization is available. Hard constraints provide the most explicit mechanism for enforcing a property during identification, but the resulting optimization problem may be computationally expensive, especially when the constraints involve LMIs, semidefinite programs, nonlinear inequalities, or sampled conditions over large domains. Soft constraints are often the easiest to incorporate into large-scale learning problems, but generally encourage rather than guarantee satisfaction of the desired property. These tradeoffs are summarized in Table~\ref{tab:enforcement_tradeoffs}.
}

\begin{table*}[t]
\centering
\small
 {
\caption{Qualitative comparison of approaches for imposing control-relevant properties in system identification.}
\vspace{1mm}
\label{tab:enforcement_tradeoffs}
\begin{tabular}{p{0.15\linewidth}p{0.25\linewidth}p{0.24\linewidth}p{0.24\linewidth}}
\hline
Approach & Nature of guarantee & Advantages & Scalability and Limitations \\
\hline
Direct\newline parameterization &
Property satisfied by construction, guarantee tied to the chosen parametrization &
No post-processing, often compatible with unconstrained or simply constrained training, interpretable model structure &
Scalable after parametrization is derived, property- and model-class-specific, can limit model expressiveness \\
Hard constraints &
Explicit constraint satisfaction, certificate through feasible LMIs, SDPs, algebraic inequalities, or sampled conditions &
Direct enforcement, explicit feasibility conditions, compatibility with control-theoretic certificates &
Potentially expensive for large state dimension, model order, network size, or horizon length, possible conservatism or model bias \\
Soft constraints &
Penalty-based encouragement of the desired property, no general certificate without additional verification &
Easy integration into standard loss functions, compatible with large-scale neural-network training, flexible model classes &
Property satisfaction not guaranteed,  penalty tuning required, requires post-training verification \\
\hline
\end{tabular}}
\end{table*}

We provide concise illustrations of each method. An example of imposing structure through \emph{direct parameterization} is a Lagrangian neural network \citep{cranmer2020lagrangian}, which imposes the Euler-Lagrange equation in the design of $f_\theta$ as a physics prior. Given a physical system with state $x = (q,\dot{q})$, consisting of the system configuration $q$ and velocity $\dot{q}$, and torque input $u$, the Euler-Lagrange equation describes the system's motion as:
\begin{equation}
    \frac{d}{dt}\nabla_{\dot{q}} L(t,q,\dot{q}) - \nabla_q L(t,q,\dot{q}) = u
\end{equation}
where $L$ is the Lagrangian. Using the chain rule to expand the time derivative through the gradient of the Lagrangian and a neural network to parameterize $L_\theta$ leads to a parametric model of the system dynamics that encodes the Lagrangian structure:
\begin{align}
&\begin{bmatrix}
    \dot{q}\\\ddot{q}
\end{bmatrix} = f_\theta(q,\dot{q},u)\\
&\;\;=\begin{bmatrix} 0 & 1 \\ 0 & -(\nabla_{\dot{q}} \nabla_{\dot{q}}^\top L_\theta)^{-1}\nabla_{q} \nabla_{\dot{q}}^\top L_\theta\end{bmatrix} \begin{bmatrix}
    q\\\dot{q}
\end{bmatrix} + \begin{bmatrix}
    0 \\ (\nabla_{\dot{q}} \nabla_{\dot{q}}^\top L_\theta)^{-1}
\end{bmatrix}(\nabla_q L_\theta + u). \notag
\end{align}

As an example of imposing a \emph{hard constraint} on the model parameters $\theta$, we consider requiring that the identified system model is internally stable in the Lyapunov sense. \citet{lyapunov_net} propose a candidate Lyapunov function of the form:
\begin{equation}
    V_\theta(x) = \|\phi_\theta(x) - \phi_\theta(0)\|^2 + \alpha \|x\|^2,
\end{equation}
where $\alpha>0$ and $\phi_\theta(x)$ is a fully connected neural network with parameters $\theta$ and $\tanh$ activations. This construction ensures that $V_\theta(0) = 0$ and $V_\theta(x) \geq \alpha \|x\|^2 > 0$ for $x \neq 0$ as required by Theorem~\ref{thm: Lyapunov_stability}. The last condition in that result for $V_\theta(x)$ to be a valid Lyapunov function can be ensured by imposing a hard constraint in \eqref{eq:constrained_sysid_optimizaiton} of the form: 
\begin{equation}\label{eq:lyapunov_net_constraint}
    g(\theta) = \int \max\{\dot{V}_\theta(x) + \gamma\|x\|,0\} \ dx \leq 0,
\end{equation}
where $\gamma > 0$ and the time derivative of $V_\theta(x)$ is computed along the system trajectory. In practice, the integral in \eqref{eq:lyapunov_net_constraint} is approximated with a finite number of samples.

Finally, an example of imposing structure through \textit{soft constraints (parameter {regularization})} is the use of Gershgorin circle theorem \citep{lechner2020gershgorin} to encourage the eigenvalues of $A_{\theta}$ in a linear system model \eqref{eq:linear_system} to be negative. This can be achieved by adding a regularization term $r(\theta)$ to the cost function in \eqref{eq:constrained_sysid_optimizaiton} of the form:
\begin{equation}
    r(\theta) = \sum_{i=1}^n \max\{0,[A_\theta]_{i,i}+\sum_{j\neq i}|[A_\theta]_{i,j}|+\epsilon\},
\end{equation}
where $\epsilon > 0$ is a small constant.

With the above formulation in hand, we can now survey and classify approaches for imposing 
control-relevant and physics-informed structures into system identification proposed in the literature, in the context of both classical system identification and learning-based methods.

\section{Classical System Identification}
\label{sec:classical}
We now focus on imposing physics-informed and control-relevant properties in classical system identification for linear and nonlinear dynamical systems. 

\subsection{Linear System Identification}
Linear system identification techniques are concerned with learning either time-domain representations in the form of state space models like \eqref{eq:linear_system}, frequency domain representations like transfer function models  \eqref{eq:transfer_function}, or their discrete-time counterparts. Typical approaches for unconstrained identification such as least squares regression and subspace methods are surveyed in classic texts such as \cite{ljung1987theory}.{   System identification objectives can also be distinguished by how the prediction error is computed. In equation-error formulations, the residual is evaluated directly in the state or output equations using the measured data; for example, in discrete time, one may penalize \(x_{k+1}-Ax_k-Bu_k\), or the corresponding residual in the output equation. Such formulations often lead to least-squares or convex optimization problems when the model is linear in the unknown parameters, and can therefore be convenient when algebraic constraints on the model parameters pertaining to control-relevant properties  are imposed. In output-error or simulation-error formulations, the candidate model is simulated using the measured input sequence, and the predicted output is compared with the measured output. These criteria more directly measure the input-output behavior of the identified model over a time horizon, but the dependence of the predicted trajectory on the model parameters can make the optimization problem nonconvex. Consequently, hard constraints such as stability, passivity, or dissipativity may be easier to impose in equation-error formulations, while output-error and simulation-error formulations may better reflect the behavior of the identified model when used for prediction and control, at the cost of more challenging constrained optimization problems.}
\subsubsection{Stable System Identification}
First, consider the fundamental property of stability. In linear system identification, stability can be imposed using any of the three approaches described in Section \ref{sec:ways_incorporating_physics}. For example, in subspace identification of state-space models of the form \eqref{eq:linear_system} or \eqref{LTI_DT} parameterized by $\theta=(A,B,C,D)$, stability can be imposed through suitable choice of the structure of the  $A$ matrix as in \cite{maciejowski1995guaranteed}. Alternatively, the Lyapunov equation can be introduced as a hard constraint as in \cite{lacy2003subspace}; for example, we have the constraint $g(\theta)= A^\top PA-P$ with $P\succ 0$ to enforce stability in the discrete-time model \eqref{LTI_DT}. Another approach is to impose stability through soft constraints as in \cite{van2001identification}, by introducing a regularization term $r(\theta)=Tr(AWA^\top )$ with positive semi-definite weighting matrix $ W\succeq 0$, where the magnitude of the regularization is determined by solving a generalized eigenvalue problem. 

Similarly, in identifying stable frequency domain models, stable parameterizations of transfer functions including certain canonical forms can be chosen; alternatively, hard or soft constraints on the pole locations can be imposed to ensure stability.  Since stability-preserving system identification is a vast and well-studied topic, we direct the reader to pedagogical resources such as \cite{pintelon2012system,verhaegen2007filtering,van2012subspace}. The remainder of this section will then focus on identification approaches (direct parameterization, hard constraint, and soft constraint based) to preserve other control-relevant properties like dissipativity, monotonicity, and symmetries, following the framework  described in Section \ref{sec:ways_incorporating_physics}. 

\subsubsection{Hard Constraints}
\label{sec:linear_hard_constraints}
Consider the problem of identifying a linear system \eqref{eq:linear_system} with parameters $(A,B,C,D).$ 
Convex formulations of constraints for special cases like positive real systems and passive systems as defined in Section \ref{subsubsec:dissipativity_passivity}  are considered in works such as \cite{hoagg2004first,coelho2004convex}. For example, \eqref{eq: dissipativity} can be directly imposed as hard constraints while minimizing traditional identification objectives such as error between the actual system outputs and the predicted outputs from the identified model.

However, directly imposing hard constraints during identification is often not desirable since it may introduce uncontrollable model bias \citep{pintelon2012system}. 
To understand this, consider the identification problem posed in \eqref{eq:general_sysid_optimizaiton}. This problem can be viewed as filtering out the noise in the observed input-output data to obtain the best model fit. Thus, the cost function should ideally be weighted by the inverse of the noise variance. Adding hard constraints to this problem introduces a fundamental tradeoff since it does not hold a priori that  model achieving the best noise variance will satisfy these hard constraints.   For example, in the frequency domain, the model errors corresponding to this constraint and those corresponding to noise filtering may be distributed in different ranges in the frequency spectrum. 
In order to address this challenge, hard constraints such as dissipativity are often imposed through a two-stage identification process. First, the unconstrained optimization problem \eqref{eq:general_sysid_optimizaiton} is solved to identify a model that achieves the best fit to the observed system behavior. Then, a small perturbation is introduced in the parameters to ensure that the perturbed values satisfy the hard constraints while minimizing deviation from the original model. Let $\hat \theta$ be the solution to the unconstrained problem \eqref{eq:general_sysid_optimizaiton}, and let $G_{\theta_p}(s)$ be the perturbed model with parameter values $\theta_p$ that satisfy the desired constraints. Then, in frequency domain identification, the constrained optimization problem in \eqref{eq:constrained_sysid_optimizaiton} is modified to minimize the cost function $c(\theta,\mathcal{W})=||G_{\theta}(s)-G_{\theta_p}(s)||^2_F$ where $||\cdot||_F$ denotes the Frobenius norm, with hard constraints $g(\theta_p)$ on model $G_{\theta_p}(s)$ corresponding to the desired property, such as \eqref{eq: hard_stability} or \eqref{eq: hard_bounded_real}. In essence, this approach attempts to find the minimal deviation from the best unconstrained model that satisfies the desired hard constraints. A similar perturbation approach can be employed in the state space setting to enforce hard constraints corresponding to the desired property, such as the dissipativity constraint in \eqref{eq: dissipativity}. Here, we demonstrate through an example how passivity constraints can be enforced through such a perturbation approach.

\begin{example}[Passive linear model identification for an RLC circuit]   
\label{ex: linear passive LMI}
We consider a RLC circuit where resistor, inductor and capacitor are in series. We set the state variable $x_1(t)$ to be the current in the circuit and $x_2(t)$ to be the voltage of the capacitor. We let input $u(t)$ to be source voltage and output $y(t)$ to be current (i.e. $x_1(t)$). With $R=L=C=1$, we write the linear state-space model as 
\begin{equation}
\begin{aligned}
\begin{bmatrix}
{\dot x_1(t)}\\
{\dot x_2(t)}
\end{bmatrix}
&=\begin{bmatrix}
-1 & -1\\
1 & 0
\end{bmatrix}\begin{bmatrix}
x_1(t)\\
x_2(t)
\end{bmatrix}
+\begin{bmatrix}
1\\
0
\end{bmatrix}u(t)\\
y(t) &= \begin{bmatrix}
1 &
0
\end{bmatrix}\begin{bmatrix}
x_1(t)\\
x_2(t)
\end{bmatrix}
\end{aligned}
\end{equation}
In other words, the real coefficient matrices are
$A=\begin{bmatrix}
-1 & -1\\
1 & 0
\end{bmatrix}$, $B=\begin{bmatrix}
1\\
0
\end{bmatrix}$, $C = \begin{bmatrix}
1 & 0
\end{bmatrix}$, and $D=0$. Let $P=I_2>0$. Then we have 
$$
\begin{bmatrix}
A^\top P+PA & PB-C^\top \\
B^\top P-C & -(D+D^\top )
\end{bmatrix} = \begin{bmatrix}
-2 & 0 & 0\\
0 & 0 & 0 \\
0 & 0 & 0
\end{bmatrix}\preceq 0.
$$
According to the Positive Real Lemma, the real system is passive.
Now we generate trajectory data for input $u(t)=e^{-0.1t}sin(\pi t)$ with initial state being $[0.1, 0.2]^\top $. We also add additional Gaussian noise $\sim \mathcal{N}(0, 0.005^2)$ to make it more realistic. With the data, we use function {\tt{ssest}} from the System Identification Toolbox in MATLAB and identify a baseline system model where the fit goodness reaches 95.81\%. The identified model has coefficients $A_{id}=\begin{bmatrix}
 -0.6661  & -0.2035\\
    3.7286 &  -0.2876
\end{bmatrix}$, $B_{id}=\begin{bmatrix}
-4.3142 \\
   -1.7815
\end{bmatrix}$, $C_{id} = \begin{bmatrix}
-0.2338  &  0.0144
\end{bmatrix}$, and $D_{id}=0$. However, the following problem is not feasible, 
\begin{equation}
\begin{aligned}
    P_1\succeq 0 \quad 
    \begin{bmatrix}
A_{id}^\top P_1+P_1A_{id} & P_1B_{id}-C_{id}^\top \\
B_{id}^\top P_1-C_{id} & -(D_{id}+D_{id}^\top )
\end{bmatrix} \preceq 0,
\end{aligned}
\end{equation}
where $P_1$ is a symmetric matrix of size 2, which indicates that the identified model is not passive. Now we will design a perturbation $\Delta C$, where the perturbed $C_p = C_{id}+\Delta C$ is as small as possible to impose passivity. This is formulated as an optimization problem as below, where $P_2$ is a symmetric matrix of size 2 and the objective is the Frobenius norm of the perturbation $\Delta C$. 
\begin{equation}
\begin{aligned}
& \min \quad \|\Delta C\|_F
\\
    & s.t. \quad  P_2\succeq 0 \quad 
    \begin{bmatrix}
A_{id}^\top P_2+P_2A_{id} & P_2B_{id}-C_p^\top \\
B_{id}^\top P_2-C_p & -(D_{id}+D_{id}^\top )
\end{bmatrix} \preceq 0
\end{aligned}
\end{equation}
We have the resulting $\Delta C = \begin{bmatrix}
-0.0001  &  -0.0025
\end{bmatrix}$ with $P_2 = \begin{bmatrix}
0.0559  &  -0.0042 \\ -0.0042 & 0.0119
\end{bmatrix}\succeq0$. The corresponding $C_p= \begin{bmatrix}
-0.2339  &  -0.0119
\end{bmatrix}$. 
{
Finally, we obtain a linear model with system matrices to be $ A_{p} = A_{id}, B_{p} = B_{id}, C_{p} = C_{id} + \Delta C, D_{p}=D_{id}$ that is passive.}
We demonstrate the fit of the perturbed system, which is certified to be passive in Fig. \ref{fig:linear passive LMI}. The goodness of fit score is still high (95.23\%) and we can see that the final model closely matches the ground-truth.

\begin{figure}[t]
    \centering
    \includegraphics[scale=0.55,trim=0cm 0.5cm 0cm 0cm]{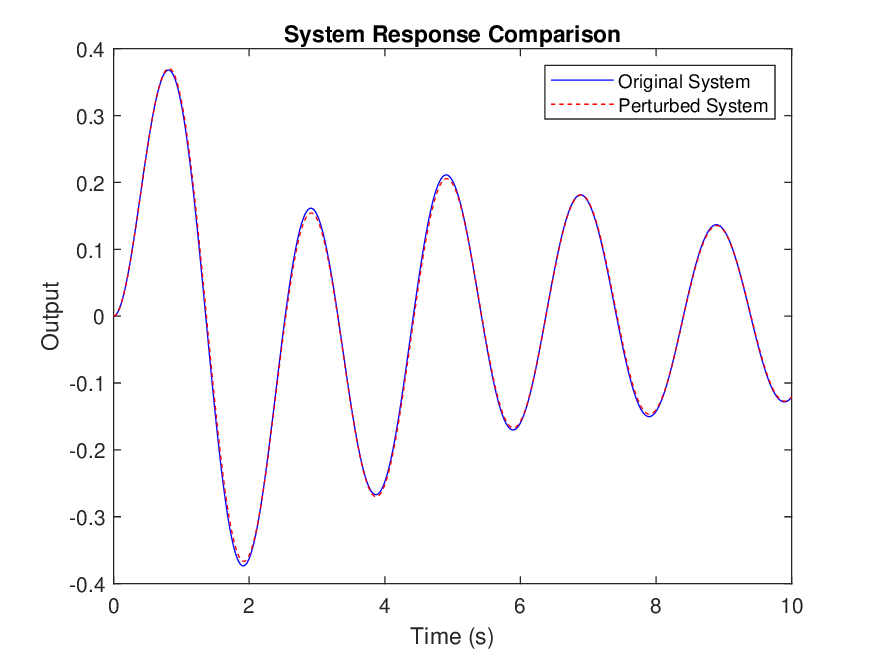}
    \caption{Passive linear model vs system trajectories for  for Example \ref{ex: linear passive LMI}. The model closely approximates the system test data and is certified to be  passive.}
    \label{fig:linear passive LMI}
\end{figure}
 \end{example}
There are multiple choices for deciding  which parameters to perturb. Perturbing the system matrix $A$ is generally avoided, as it contains key dynamical features like dominant modes or eigenvalues. Alternatively, perturbation of the feedthrough matrix $D$ can be leveraged in cases where $D>0$ is necessary to enforce strict dissipativity or other similar properties though it only affects direct input-output coupling, not internal dynamics. From a transfer function view, this is an asymptotic perturbation affecting behavior as $s \to \infty$. Since passivity is an input-output property, perturbing the input matrix $B$ or output matrix $C$ is also natural, as in Example~\ref{ex: linear passive LMI}. Such approaches for identifying dissipative (including $QSR$-dissipative) models have been explored in \cite{sivaranjani2022data,bradde2020bounded}.


Another approach for passivity enforcement in identified models is to consider the Hamiltonian matrix. Consider a parameterized state-space realization $(A,B,C,D)$ of order $n$ with corresponding transfer function $G(s).$ The associated Hamiltonian matrix $M_H \in \mathbb{R}^{2n\times2n}$ is defined as:
\begin{equation}\label{eq:dissipation_matrix_hamiltonian}
M_H = \begin{bmatrix}
A - BR_1^{-1}D^\top C & -BR_1^{-1}B^\top  \\
C^\top S_1^{-1}C & -A^\top  + C^\top  DR_1^{-1}B^\top ,
\end{bmatrix}
\end{equation}
where $R_1 = I - D^\top D$, and $S_1 = I - DD^\top $. Assuming that $||D||=\sigma_{max}(D)<1$, passivity violations occur when $M$ has purely imaginary eigenvalues $\mu_k=j\omega_k$, or correspondingly, at frequencies where the largest singular value of $G(j\omega)$ exceeds unity. 
When only the $C$ matrix of the system is perturbed to $C+\Delta C$ as discussed above, the corresponding  perturbation on the Hamiltonian can be written using the first-order approximation:
\begin{equation}
\Delta M_H \approx \begin{bmatrix}
BR^{-1}_1D^\top \Delta C & 0 \\
-C^\top S^{-1}_1\Delta C - \Delta C^\top S^{-1}_1C & -\Delta C^\top DR^{-1}_1B^\top .
\end{bmatrix}
\end{equation} The perturbation affects the eigenvalues of the Hamiltonian matrix according to
\begin{equation} \label{eq:hamiltonian_svd}
\Delta\lambda_i = \frac{v_i^H\Delta M_H w_i}{v_i^Hw_i}
\end{equation}
where $v_i$ and $w_i$ are the left and right eigenvectors corresponding to the imaginary eigenvalues of the original Hamiltonian.

Using this property, a general passivity enforcement approach can be developed as follows. The  eigenvalues of $M_H$ are computed for the baseline (unperturbed) model to identify frequencies where $\sigma_{\text{max}}(G(j\omega)) > 1$. Then, the singular value decomposition is computed at these violation frequencies to obtain left and right singular vectors $u_i, v_i$  of $M_H$ corresponding to the imaginary eigenvalues that represent the directions of passivity violation in the frequency response. Finally, the optimization problem to determine the minimal passivity-enforcing perturbation is formulated as:
\begin{equation}
\begin{aligned}
\min \quad & ||\Delta C||_F \\
\text{s.t.} \quad & \Delta\lambda_i < 0, \quad \text{Re}\{u_i^H\Delta 
G(j\omega_k)v_i\} < 0.
\end{aligned}
\end{equation}
Such Hamiltonian perturbation approaches were originally proposed in  \cite{grivet2004passivity}, with subsequent works extending this approach to large-scale systems \citep{grivet2006generation}, descriptor systems where $D+D^\top $ is singular \citep{wang2010peds}, negative imaginary systems \citep{mabrok2011enforcing}, and proposing improved algorithms to decrease the dispersion of the perturbed model from the original dynamics \citep{gustavsen2008passivity}. We further direct the reader to the monograph by \cite{grivet2015passive} for an in-depth discussion of these approaches. 

In the frequency domain, baseline rational transfer function models are often obtained through vector fitting  approaches \citep{gustavsen1999rational}. When the resulting models do not satisfy desired  dissipativity properties, iterative pole-residue perturbation of the identified transfer function is employed to correct for violations at specific frequency ranges through heuristic or linear and quadratic programming approaches \cite{gustavsen2001enforcing,gustavsen2007computer,saraswat2005global,saraswat2004fast,chen2003enforcing}. Here, LMIs directly corresponding to properties like dissipativity or tools the Hamiltonian matrix in \eqref{eq:dissipation_matrix_hamiltonian} may be employed for property verification. 

\subsubsection{Soft Constraints}
In several system identification problems, hard constraints as described in Section \ref{sec:linear_hard_constraints}, while desirable in terms of provably guaranteeing control-relevant properties, are challenging to impose. For instance, efficiently solving such formulations requires convex parameterizations of the constraints which may not be possible. Even if such parameterizations are available, direct imposition during learning is rarely adopted, and a two-stage perturbation or projection approach is often employed \citep{pintelon2012system} due to issue of model bias as described in Section \ref{sec:linear_hard_constraints}. 

Instead, it may be easier to obtain system models satisfying desired properties like stability or passivity by imposing soft constraints in the form of regularization terms as in the identification problem \eqref{eq:constrained_sysid_optimizaiton}. While such approaches may not possess  theoretical guarantees of yielding an identified model that satisfies the desired property, they nonetheless often yield property-preserving models in practice. These soft constraint based identification approaches may be particularly attractive in online system identification applications, where minimal violations of desired properties may be tolerated in favor of decreased computational complexity. As an example of such an approach, we now illustrate how soft constraints can help to enforce dissipativity-type properties in system identification.

Consider the problem of identifying a positive real discrete-time linear state-space model of the form \eqref{LTI_DT}. 
Then, as described in \cite{goethals2003identifying}, the cost function in \eqref{eq:constrained_sysid_optimizaiton} can be modified to obtain a positive real model parameterized by $\theta=(A,C)$ as 
$$c(\theta,\mathcal{W})=c_1(\theta)+r(\theta).$$ Here, the term
\begin{equation}
    c_1(\theta)=\left\Vert\begin{bmatrix} \hat{X}_{k+1} \\ Y_k \end{bmatrix} - \begin{bmatrix} A \\ C \end{bmatrix}\hat{X}_k\right\Vert^2_F
\end{equation}
is employed to minimize model error with respect to the input-output data. Additionally, the regularization term
\begin{equation} 
    r(\theta)= \text{Tr}\left(\begin{bmatrix} A \\ C \end{bmatrix}^\top  W \begin{bmatrix} A \\ C \end{bmatrix}\right), \quad W\succ 0
\end{equation}
is employed to promote positive realness of the identified model. Note that \cite{goethals2003identifying} considers identification with noise in the system dynamics and outputs, and additional constraints on $W$ are imposed based on the covariance matrices obtained from data.  
Similar regularization approaches have been employed in \cite{peternell1995identification} and \cite{vaccaro1993solution} where the output covariance matrix is regularized to ensure positive realness. However, such soft constraint based approaches as a whole have received limited attention in classical system identification. This is in stark contrast to machine learning based approaches where soft constraint based approaches are, in fact, dominant. We will describe such learning-based models in Section \ref{sec:PINN_systemID}.

\subsubsection{Direct Parameterization}
Considering the identification problem in \eqref{eq:constrained_sysid_optimizaiton}, one approach to imposing physics-informed or control-relevant properties is to directly parameterize the functions $f_\theta$ and $g_\theta$, representing the system dynamics, to preserve the desired property of interest, while choosing a parameterization that facilitates tractable solutions of \eqref{eq:constrained_sysid_optimizaiton}. For example, consider the property of positive realness. Given a continuous‐time SISO transfer function structure
$$
H(s)=d+\frac{c(s)}{a(s)}, 
\quad
a(s)=s^n + a_1 s^{n-1} + \cdots + a_n,
$$
where $a(s)$ is monic and Hurwitz, following \cite{dumitrescu2002parameterization},  every positive‐real $H(s)$ can be parameterized using a single symmetric positive semidefinite matrix $Z\in\mathbb{R}^{n\times n}$ as follows: 
$$
d = \mathrm{tr}(\Lambda_0 Z), 
\qquad
c_k = \mathrm{tr}(\Lambda_k Z)\quad(k=1,\dots,n),
$$
where  \(\{\Lambda_k\}_{k=0}^n\) are constant symmetric matrices computed as explicit linear functions of \((a_1,\dots,a_n)\).
This construction guarantees positive realness of \(H(s)\)  for all \(Z\succeq 0\). Parameterizations of the numerator polynomial of the transfer function have also been developed to capture properties like passivity and positive realness \citep{marquez1995design}. 

Similarly, in discrete-time, any stable LTI system \(G(z)\) can be uniquely represented as a series expansion 
$
G(z) \;=\;\sum_{i=1}^\infty \theta_i\,\psi_i(z^{-1}),
$
where \(\{\psi_i(z^{-1})\}_{i=1}^\infty\) is a prechosen orthonormal basis (e.g., Laguerre or Kautz) and 
\(\theta_i = \langle G,\psi_i\rangle_{\mathcal H_2}\) are model parameters. In practice, we can obtain an approximate finite-dimensional representation:
$$G_N(z) \;=\;\sum_{i=1}^N \theta_i\,\psi_i(z^{-1}),
$$
where hard constraints can be enforced via linear inequalities on \(\theta\) similar to Section \ref{sec:linear_hard_constraints} to enforce properties like positive realness, passivity, or dissipativity. For example, \cite{prakash2022data} formulate a convex optimization problem 
\begin{equation}\label{eq:direct_param}
\begin{aligned}
\min_{\theta \in \mathbb R^N}\quad & \sum_{k=1}^p \biggl|\,Y\bigl(e^{j\omega_k}\bigr) - \sum_{i=1}^N \theta_i\,\psi_i\bigl(e^{-j\omega_k}\bigr)\biggr|^2,\\
\text{s.t.}\quad & \Re\!\Bigl\{\sum_{i=1}^N \theta_i\,\psi_i(e^{-j\omega_k})\Bigr\}\;\ge\;\epsilon>0,
\quad k=1,\dots,p,
\end{aligned}
\end{equation}
where \(Y(e^{j\omega_k})\) are measured frequency‐response samples on the grid of frequencies \(\{\omega_k\}\), the constraint enforces strict positive‐realness.  
In multivariable settings, balanced state-space realizations (with equal and diagonal reachability and observability Gramians) can be utilized to develop stable minimal parameterizations capturing a variety of useful properties including minimum phase and positive realness \citep{chou1997system}. Such parameterizations have the advantage of providing injective mappings from the parameter space to input–output behavior, thus ensuring the existence of a unique
parameter vector corresponding to each transfer function. Further,  contraction mappings have been leveraged to preserve positive realness under reduction with explicit \(\mathcal L_\infty\) bounds~\citep{opdenacker1988contraction}, and reproducing kernel Hilbert space (RKHS) methods have been adopted to identify nonnegative input-output operators that preserve passivity by solving semidefinite programs (SDPs) \citep{shali2024towards}. In summary, some control‐relevant properties such as stability and passivity can be directly embedded through parameterization of transfer functions or state-space models. However, capturing more complex physics-informed or control-relevant properties through identifiable parameterizations remains an open challenge. We will further discuss this in the context of learning-based models in Section \ref{sec:deep}. 

\subsection{Nonlinear System Identification}
Nonlinear system identification, focused on learning representations of the form \eqref{eq:control_system}, is a classical topic \citep{sjoberg1995nonlinear,schon2011system,ljung2010perspectives}, with special cases like the identification of linearized models and Lure-type systems receiving especially significant attention \citep{schoukens2017identification}.  
However, identification in the nonlinear systems context remains a challenging problem due to local minima in model fitting, high long-term prediction errors, and difficulty in selecting stable model classes or parameterizations that adequately capture the rich and complex spectrum of dynamical behavior that such systems can exhibit.  We will now describe specific model classes that have been extensively utilized in preserving physics-informed and control-relevant properties during nonlinear system identification. 

\subsubsection{Implicit Representations}
Implicit representations of nonlinear systems that allow for a convex parameterization of all stable models have been introduced to facilitate robust nonlinear system identification minimizing simulation or prediction errors \citep{megretski2008convex,tobenkin2010convex,manchester2012stable,umenberger2018specialized}. For example, considering an implicit representation of the form 
\begin{equation}\label{eq:implicit_nonlinear}
e(x_{k+1}) = f(x_k, u_k), 
\quad
y_k = g(x_k, u_k),
\end{equation}
where \(e, f, g\) are drawn from a linearly parameterized polynomial basis with parameters $\rho$, stability can be enforced through the convex inequalities
{\small
\begin{align}
&F(x,u)\,P^{-1}F(x,u)^{T}
- E(x)\,P^{-1}E(x)^{T}+ \mu I + G(x,u)\,P^{-1}G(x,u)^{T}>0,
\nonumber\\ &P\succ0, \quad \mu>0,
\end{align}}
where 
$E(x) = \nabla_x e(x),$  
$F(x,u) = \nabla_x f(x,u),$  
$G(x,u) = \nabla_x g(x,u),$
which can be solved using sum-of-squares relaxations \citep{umenberger2018specialized}. Such implicit representations have also been extended to obtain convex parametrizations 
guaranteeing monotonicity, positivity, and contraction properties \citep{revay2021distributed}. 

\subsubsection{Koopman Operator Models} \label{sec:hard constraint Koopman}
The Koopman operator provides an approach to model nonlinear dynamical systems by approximating them through high-dimensional linear models. { This is accomplished by introducing a lifting map $\Phi$ that stacks scalar-valued functions of the system states and inputs, termed as \textit{observables}. On this observable space, the time evolution is linear under the  typically infinite-dimensional Koopman operator, which acts by composition with the system dynamics.}
In practical implementations, this infinite-dimensional operator is approximated using a finite set of basis functions, yielding a high-dimensional albeit linear representation of the nonlinear dynamics.

{
Specifically, considering a nonlinear dynamical system in continuous-time as in \eqref{eq:control_system} or discrete-time as in \eqref{Sys_IO}, 
denote $\mathbb{F}$ to be the forward state-input nonlinear operator such that $\mathbb{F}(x,u)\triangleq \begin{bmatrix}
    f(x,u) \\ \mathcal{S}(u)
\end{bmatrix}$, where $\mathcal{S}$ is the forward time-shift forward operator. } 
{
The infinite dimensional lifting function is expressed as 
$
    \Phi_{\text{inf}}(x,u)=\begin{bmatrix}
        \phi_{1,\text{inf}}(x,u),
        \phi_{2,\text{inf}}(x,u),
        \dots
    \end{bmatrix}^\top .
$
The Koopman operator $\mathcal{K}$ is then defined as 
\begin{equation}
    \mathcal{K} [\Phi_{\text{inf}}(x,u)] \triangleq \Phi_{\text{inf}}(\mathbb{F}(x, u)),
\end{equation}
that is, $\mathcal{K}$ takes $\Phi_{\mathrm{inf}}$ as the input and produces a new function whose value at $(x,u)$ equals the value of the original observable evaluated at the state-input pair obtained by evolving $(x,u)$ forward in time.
}

In practice, finite-dimensional approximations of the Koopman operator, as defined below, are often employed. Consider the class of lifting functions defined as \begin{equation}
    \phi_i(x,u)\triangleq\begin{bmatrix}
    \phi_i(x) \\ u
\end{bmatrix} \quad i=1,2,....,n_z,
\end{equation} 
{
where each $\phi_i(x)$ is a scalar function of the state, and $u$ is included in its original form to preserve control-affine structure.  Typical examples of $\phi_i(x)$ include monomials (for polynomial lifting), radial basis functions, Fourier modes, or features learned by a neural network.}
Then, the high-dimensional observable is defined as $z\triangleq\Phi(x)\triangleq \begin{bmatrix}
    \phi_1(x), \phi_2(x),\dots,\phi_{n_z}(x) 
\end{bmatrix}^\top \in\mathbb{R}^{n_z}$, and the Koopman linear model of the nonlinear dynamics \eqref{eq:control_system} can be written as 
\begin{equation} \label{eq: Koopman model}
    z^+ = A z + B u  \quad y = Cz+Du.
\end{equation}
{ This finite-dimensional representation is exact only when the chosen observables span a Koopman-invariant subspace; otherwise, \eqref{eq: Koopman model} should be interpreted as a finite-dimensional approximation of the infinite-dimensional Koopman representation.} 
{In discrete time, $z^+$ denotes the lifted state at the next time step, $z_{k+1}$, while in continuous time it denotes the time derivative $\dot{z}$.}

Given trajectory data, data-driven methods such as Dynamic Mode Decomposition (DMD) \citep{rowley2009spectral}, Extended DMD (EDMD) \citep{williams2015data}, and Sparse Identification of Nonlinear Dynamics (SINDy) \citep{brunton2016discovering} are commonly used to learn the Koopman operators and to design controllers \cite{korda2018linear}. \cite{mauroy2016linear,haseli2023modeling} improve the methods addressing accuracy and efficiency issues and \cite{StrasserKoopman} provides closed-loop guarantees. Neural networks can also be employed to learn the lifting functions in the Koopman operator framework \citep{yeung2019learning,zinage2023neural,hao2024deep}.

While Koopman operator models can be trained to accurately approximate system behavior, control-relevant properties are not inherently guaranteed. To address this, efforts have been made to impose physics-informed and control-relevant constraints on Koopman operator models. Stability in learning the Koopman operator can be achieved by limiting the parameterization to Hurwitz matrices \citep{bevanda2022diffeomorphically}, or by establishing an equivalence between stability conditions on the Koopman model and contraction criteria, which can be embedded into constrained optimal control formulations \citep{fan2022learning,fan2024learning,wang2024monotone}. Alternatively, stability can be enforced by applying a suitable control barrier function or control Lyapunov function-based gradient descent during the iterative learning process \citep{mitjans2024learning, huang2018feedback}.

Other control-relevant properties, such as quadratic dissipativity \citep{hara2020learning} and special structures like negative imaginary properties \citep{mabrok2023koopman}, are typically embedded by deriving convex matrix inequality constraints that can be imposed during the Koopman identification. For example, consider identifying a nonlinear discrete-time dissipative system with no direct relationship between the lifted states and outputs (i.e., $D=0$). Define the input, output, and lifted states sampled from dynamical trajectory data as
\begin{equation} \label{eqn:Koopman data}
    \begin{aligned}
        & U_k \triangleq [u(k), u(k+1), \dots, u(k+M-1)]\in \mathbb{R}^{n_u\times M}\\
        & Y_k \triangleq [y(k), y(k+1), \dots, y(k+M-1)]\in \mathbb{R}^{n_y\times M}\\
        & Z_k \triangleq [z(k), z(k+1), \dots, z(k+M-1)]\in \mathbb{R}^{n_z\times M}\\
        & Z_{k+1} \triangleq [z(k+1), z(k+2), \dots, z(k+M)]\in \mathbb{R}^{n_z\times M},
    \end{aligned}
\end{equation}
where $n_u=n_y$ under the quadratic dissipativity constraints. In order to identify a Koopman operator model \eqref{eq: Koopman model} parameterized by $\theta=(A,B,C,D)$, the cost function  in \eqref{eq:constrained_sysid_optimizaiton} takes the form $$c(\theta,\mathcal{W})=c_1(A,B)+c_2(C),$$ where 
\begin{equation}
    c_1(A,B)\triangleq \left\|Z_{k+1}-[A\ B ]
    \begin{bmatrix}
        Z_k\\U_k
    \end{bmatrix}\right\|_F^2, \ \ c_2(C)\triangleq \left\|Y_k-CZ_k\right\|_F^2.
\end{equation}
Further, we can ensure that this model preserves dissipativity by solving \eqref{eq:constrained_sysid_optimizaiton} with hard constraints 
\begin{equation}
  g_1(\theta)=  P \succ 0,
\end{equation}
\begin{equation}
g_2(\theta)=    \begin{bmatrix}
        A & B \\ 
        I & 0
    \end{bmatrix}^{T}
    \begin{bmatrix}
        P & 0 \\ 
        0 & -P
    \end{bmatrix}
    \begin{bmatrix}
        A & B \\ 
        I & 0
    \end{bmatrix} + \Theta \prec 0,
\end{equation}
where
\begin{equation}
    \Theta =
    \begin{bmatrix}
        C & 0 \\ 
        0 & I
    \end{bmatrix}^{T}
    \Xi
    \begin{bmatrix}
        C & 0 \\ 
        0 & I
    \end{bmatrix}, \quad \Xi = \begin{bmatrix}
        Q & S \\S^\top  & R
    \end{bmatrix},
\end{equation}
and $Q,S,R$ are the dissipativity indices as in Definition \ref{def:qsr_dissipativity}. 
This approach of imposing hard constraints to preserve properties like dissipativity is similar to the methods described in Section \ref{sec:linear_hard_constraints}. Another common practice is to use a soft constraint that penalizes the violation of the constraints as a part of the objective. We demonstrate such a soft-constraint based approach to enforce stability in Koopman operator identification in the following example.

\begin{example}[Stable Koopman linear model identification for Duffing oscillator]\label{ex:Koopman soft constraint}
Consider a Duffing oscillator with state $x(t)=[x_1(t), x_2(t)]^\top \in\mathbb{R}^2$, input $u(t)\in\mathbb{R}$, and dynamics:
\begin{equation}
\dot{x}(t)
=
\begin{bmatrix}
x_2(t) \\
-2 x_2(t) - x_1(t)\cos\bigl(x_1(t)+x_2(t)\bigr)+u(t)
\end{bmatrix}.
\end{equation}
While the Duffing oscillator is defined in continuous time, we identify the Koopman model from data sampled every $\Delta t=0.01$\,s under zero‐order hold on $u(t)$. The exact flow map is approximated using a fourth‐order Runge–Kutta step. 
To build a Koopman model, we define a lifting function  $\psi:\mathbb{R}^2\to\mathbb{R}^{2+N_{\text{rbf}}}$ and observables as follows:
\begin{equation} \label{eqn:Koopman lifting}
z\triangleq\psi(x)
=
\begin{bmatrix}
x \\[4pt]
\phi(x)
\end{bmatrix},
\quad
\phi_j(x)=\mathrm{rbf}\bigl(x,c_j\bigr),
\quad j=1,\dots,N_{\text{rbf}},
\end{equation}
where \texttt{rbf} is the thin‐plate kernel with $\{c_j\}$ being centers randomly chosen from $[-1,1]^2$. Here, we choose $N_{\text{rbf}}=8$.
With the discretized system, we generate $N_{\text{traj}}$ trajectories of length $M$ each, with initial states $x_0\sim\mathcal{U}([-1,1]^2)$ and control inputs $u_k\sim\mathcal{U}([-1,1])$. Then we calculate the lifted states according to \eqref{eqn:Koopman lifting} to form $U_k$, $Z_k$ and $Z_{k+1}$ as in \eqref{eqn:Koopman data}. 
{
With the Koopman model structure in \eqref{eq: Koopman model}, we define the reconstruction error to be $\|\mathcal{R}\|_F^2$, where $\mathcal{R}$ is the long stacked vector of all $\mathcal{R}_k$ from every time step $k$ along each trajectory, with all trajectories concatenated, that is, there  are $N_{traj}(M-1)$ of $\mathcal{R}_k$ in total. Each $\mathcal{R}_k$ is formed from the quadruple $(Z_k, U_k, Z_{k+1}, Y_k)$ at time step $k$ as
\begin{equation}
\mathcal{R}_k=
\begin{bmatrix}
Z_{k+1} \\[4pt]
Y_k
\end{bmatrix}
-
\begin{bmatrix}
A\,Z_k + B\,U_k \\[4pt]
C\,Z_k + D\,U_k
\end{bmatrix}.
\end{equation}
}

To enforce stability of the learned model, we desire the spectral radius of $A$ to be smaller than 1. 
However, directly penalizing the spectral radius \(\rho(A)\) yields a non‐smooth, non‐convex objective.  As a practical convex surrogate for the upper bounds of \(\rho(A)\), we instead impose a penalty on the operator 2-norm \(\|A\|_2\).
We enforce this penalty in the form of a soft constraint by solving the unconstrained optimization problem
\begin{equation}
    \min_{A,B,C,D}\quad  c(\theta,\mathcal{W}) := \|\mathcal{R}\|_F^2 +\lambda \|A\|_2^2,
\end{equation}
where $\lambda>0$ is a hyper-parameter that balances model accuracy and stability. To make the problem more computationally tractable, we introduce slack variables $\tau$ and $\gamma$ to transform the problem into a convex problem
\begin{equation}\label{eq:koopman_convex}
\begin{aligned}
( A_1^*,B_1^*,C_1^*,D_1^*, \gamma^*,t^*)&=argmin_{A,B,C,D,\gamma,t}\quad \|\mathcal{R}\|_F^2 \;+\;\lambda\,\tau^2,\\
\mathrm{s.t.}\quad
&\begin{bmatrix}
\gamma\,I & A \\[3pt]
A^\top   & \gamma\,I
\end{bmatrix}
\succeq 0,
\quad
\tau \ge \gamma,
\quad
\tau\ge 0,
\end{aligned}
\end{equation}
We choose $\lambda=200$ and solve \eqref{eq:koopman_convex} using the Mosek  solver \citep{mosek}.
For comparison, we also obtain the Koopman model without soft constraints, we solve the problem 
\begin{equation}
   (A_2^*,B_2^*,C_2^*,D_2^*)= argmin_{A,B,C,D} \|\mathcal{R}\|_F^2
\end{equation}
To assess the performance of the identified Koopman model, we simulate the system dynamics and that of the learned model with  initial condition $[-0.6,1.4]^\top $ on the horizon $T=[0,10]$ with time step $\Delta t =0.01$s (that is, total number of steps  $M=1000$). We set the test input to be \begin{equation}
    u(k) = 0.8\sin(0.2\,k),\quad k=0,\dots,M-1.
\end{equation}
From Fig. \ref{fig:Koopman soft constraints}, we can observe that the Koopman model with soft constraints is stable and closely approximates the true system trajectory, with the optimal solution $\rho(A_1^*)=0.9958$, satisfying the stability condition. On the other hand, it is obvious that the Koopman model learned without the soft constraints diverges from the true dynamics and is in fact unstable with $\rho(A_2^*) = 1.0035$.

\begin{figure}[t]
    \centering
    \includegraphics[scale=0.55,trim=0cm 0.5cm 0cm 0cm]{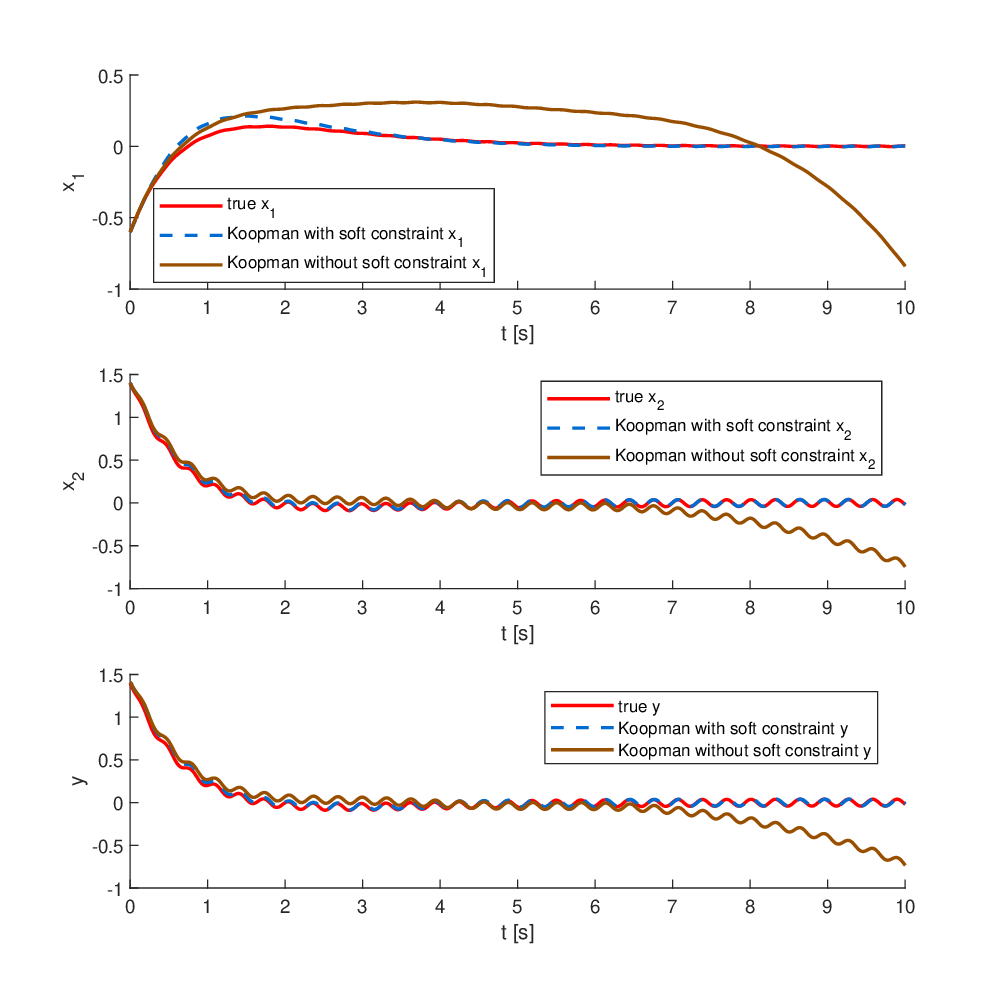}
    \caption{Koopman operator model with soft constraints vs Koopman Model without soft constraints vs ground truth for Example \ref{ex:Koopman soft constraint}.}
    \label{fig:Koopman soft constraints}
\end{figure}

\end{example}

\subsubsection{Sparse Identification of Nonlinear Dynamics (SINDy)}
Recently, Sparse Identification of Nonlinear Dynamical systems (SINDy)  \citep{brunton2016discovering} has emerged as a widely-adopted data-driven method for nonlinear system identification. In many physical systems, the dynamics can be parameterized by a small number of nonlinear function terms, making the governing equations naturally sparse in the space of candidate nonlinear functions. The key idea in SINDy is to identify a sparse set of active terms from the candidate nonlinear function space by transforming the system identification problem into a sparse regression problem. Specifically, for a dynamical system for the state $x(t)$
\begin{equation} \label{eq:autonomous dynamics}
    \dot{x}(t)=f(x(t)),
\end{equation} 
the first step is to construct a library of candidate functions $\Theta(x)=[\theta_1(x), \dots, \theta_p(x)]$, where $p$ is the number of candidates. The next step is to collect trajectory data of length $M$, defined as $X\triangleq [x(t_1),x(t_2),\dots,x(t_M)]^\top \in \mathbb{R}^{M\times n}$, $\Theta(X)\triangleq\left[\theta_1(X),\dots,\theta_p(X)\right]\in\mathbb{R}^{M\times np}$, and $\dot{X}(t)\triangleq [\dot{x}(t_1),\dot{x}
(t_2),\dots,\dot{x}(t_M)]^\top \in \mathbb{R}^{M\times n}$. The autonomous dynamics \eqref{eq:autonomous dynamics} can then be modeled in terms of the candidate functions as 
\begin{equation}\label{eq:sindy_model}
    \dot{X}(t) \approx \Theta(X(t))\Xi,
\end{equation}
where $\Xi=(\xi_1,\xi_2,\dots,\xi_n)$, and each $\xi_k$, $k\in\{1,2,,\dots,n\}$ is a sparse vector of coefficients that indicate the active terms. Sparsity is promoted by solving the following optimization problem 
\begin{equation}
    \min\limits_{\Xi} \frac{1}{2}\|\dot{X}-\Theta(X)\Xi\|^2+\lambda R(\Xi),
\end{equation}
where $\lambda>0$ and $R(\cdot)$ is a regularization term that promotes sparsity. The standard choices for $R$ include the $l_0$ norm or the $l_1$ norm, which is the convex relaxation of the $l_0$ norm. To solve the sparse regression problem, algorithms such as sequential thresholded least squares (STLSQ) \citep{brunton2016discovering}, LASSO \citep{hastie2015statistical}, and the sparse relaxed regularized regression (SR3) \citep{champion2020unified} have been  developed. There are also several variants that extend SINDy to other control-relevant settings. For example, implicit-SINDy \citep{mangan2016inferring} is designed for rational dynamics, while integral-SINDy \citep{schaeffer2017sparse} deals with noise robustness using integrated rather than differentiated data. Further, SINDy with control (SINDyc) \citep{brunton2016discovering} extends the framework beyond autonomous systems to enable identification with control inputs.

Recent work focuses on embedding prior knowledge to promote physics-informed properties such as Hamiltonian structures \citep{dipietro2020sparse,lee2022structure}, energy conservation \citep{holmsen2023pseudo}, and stability \citep{kaptanoglu2021promoting}. Due to the challenges associated with imposing the properties as hard constraints in the SINDy framework, the common practice is to use soft constraints, augmenting the objective function with a penalty term to encourage the desired physical behavior. As a representative example, we consider the `trapping SINDy' framework \citep{kaptanoglu2021promoting} that preserves system stability during identification. Trapping SINDy is particularly applicable to fluid and plasma flows with an energy-preserving, quadratic nonlinear structure. {  Using the standard Lyapunov function \(V(x)=\frac{1}{2}\|x\|^2\), a sufficient condition for Lyapunov stability of the dynamics \eqref{eq:autonomous dynamics} is \(x^\top f(x)\leq 0\). In trapping SINDy, the quadratic nonlinear component $q(x)$ of $f(x)$ is constrained to be energy-preserving, meaning that its contribution to the Lyapunov derivative satisfies $x^\top q(x)=0$ for all $x$. Since the quadratic component of $f(x)$ contributes cubic terms to $x^\top f(x)$, this condition can be achieved by eliminating self-coupled cubic terms such as $x_i^3$ and imposing an antisymmetric structure on cross terms of the form $x_i x_j^2$, so that their contributions cancel.} 
Such conditions, termed as energy-preserving quadratic nonlinearity constraints, ensure that only quadratic components of $x^\top f(x)$ remain. These conditions can be collectively expressed as a set of linear equality constraints, compactly denoted as $r(\Xi)\triangleq \mathcal R vec(\Xi)=0$, where $\mathcal R$ is an appropriate constant matrix and $vec(\Xi)$ is the vectorized form of $\Xi$. Then, we can obtain a stable nonlinear system model of the form \eqref{eq:sindy_model} by solving the following optimization problem:
\begin{equation} \label{eq:soft_constraint_formulation} 
\min_{\Xi}\, \frac{1}{2}\|\dot{X}-\Theta(X)\Xi\|_F^2+\lambda \|\Xi\|_1+\alpha\|r(\Xi)\|_2^2, 
\end{equation}
where $\lambda>0$  and
$\alpha>0$ weight the sparsity of the model and the stability preservation conditions respectively, and can be suitably adjusted to  balance accuracy, sparsity, and physically consistent behavior. Several simulation examples illustrating this approach for concrete examples such as  the atmospheric oscillator model and the Lorenz system are available in \cite{pysindy_trapping2024}.

\subsubsection{Port-Hamiltonian Models}
\label{sec:phmodels}

Port-Hamiltonian (PH) dynamics have been increasingly used in recent years as a unified formulation for energy-based modeling of systems \citep{ortega2001putting,vanderSchaft2004port,vanderSchaft2014PHBook,beattie2018linear,beattie2019robust,Mehrmann_Unger_2023}. PH systems theory observes that any physical system can be modeled by energy-storing elements (capacitors, inductors, masses, springs, etc.) and energy-dissipating elements (resistors, dampers, etc.), linked to each other by power-conserving connections, called a Dirac structure. The Dirac structure links flow variables, which quantify the rate of energy transfer (e.g., current, velocity), and effort variables, which quantify the potential for energy transfer (e.g., voltage, force), such that the total power (product of flow and effort variables) is equal to zero. 

A standard example of a Dirac structure is the graph of a skew-symmetric map from the effort to the flow variables. PH models used for control have input $u$ representing external effort variables, output $y$ representing external flow variables, and skew-symmetric relationship between the storage, dissipation, and external port variables (see \cite{vanderschaft2024pHnonlinearsystems} for details), leading to an input-state-output PH system:
\begin{equation} \label{eq:phs}
\begin{aligned}
\dot{x} &= (J(x) - R(x)) \nabla H(x) + G(x)u,\\
y &= G^\top(x)\nabla H(x),
\end{aligned}
\end{equation}
where $H(x)$ is the Hamiltonian representing the total energy of the system, $J(x)$ is a skew-symmetric interconnection matrix, $R(x)$ is a symmetric positive semidefinite dissipation matrix, and $G(x)$ is an input gain matrix.

A key property is that any power-conserving interconnection of PH systems is again a PH system. More precisely, if PH systems are interconnected, via their external ports, through a Dirac structure, then we obtain a joint PH system with energy given by the sum of the individual Hamiltonians, energy dissipation relation equal to the direct product of the individual dissipation relations, and Dirac structure given by the composition of the individual Dirac structures. 

The increase in stored energy of a PH system is less than or equal to the externally supplied power, which means that the system is passive with respect to the supply rate. PH systems may possess other conserved quantities, primarily determined by their Dirac structures. Conserved quantities which are independent of the Hamiltonian are called Casimir functions $C(x)$. They satisfy $\frac{\partial C}{\partial x}(x)^\top(J(x)-R(x)) = 0$ and can be used to construct candidate Lyapunov functions for the system \citep{ortega1999energy,ortega2008control,xu2022neural}.

Various methods have been proposed for identification of PH system models. \citet{Branford2019dirac} use energy conservation and power flows to identify linear time-varying PH systems from input-output data. They show that minimal state trajectories can be obtained from a rank-revealing factorization of an energy matrix, obtained by integrating the power balance of the effort and flow trajectories. \citet{Benner2020Identification} obtain PH realizations of strictly passive systems from frequency response data using the Loewner approach for model reduction \citep{Mayo2007realization} to enforce a PH structure on the model. \citet{Ortega2024learnability} propose a structure-preserving learning scheme for single-input single-output linear PH systems with $n$ states. The paper establishes morphisms between controllable and observable linear systems and normal form PH systems. The set of uniquely identifiable PH systems is characterized as a manifold with global Euclidean coordinates and the parameter complexity for representing such systems is shown to be $O(n)$ instead of $O(n^2)$. Thus, the paper shows interesting connections between controllable and observable canonical forms and the structure-preservation and expressive power properties of the machine learning model. 

The benefits of PH system models on learning and adaptive control are reviewed in \citet{Nageshrao2016port} with emphasis on speeding up learning via the prior knowledge of the PH structure, obtaining stability or convergence guarantees via the PH model, and interpreting the resulting control laws in the context of physical systems. Model reduction techniques that preserve PH structure as well as stabilization and optimal control of PH systems are discussed in \citet{Mehrmann_Unger_2023}. Interconnection and damping assignment (IDA) passivity-based control (PBC) \citep{vanderSchaft2014PHBook} is a widely used method for stabilizing PH systems using the system input to inject energy and reshape the total energy of the closed-loop system such that its minimum is at a desired equilibrium. \citet{plaza2022total} transform the IDA-PBC method into a supervised learning problem by designing loss functions that capture desirable properties, including prescribed damping, minimum oscillation, assignment of the desired equilibrium, Lyapunov stability, and enforcement of matching conditions ensuring that the desired closed-loop system is obtained. Further examples of applying deep learning methods for identification and control of PH systems are discussed in Sec.~\ref{exp:Hami}.


\subsubsection{Gaussian Process Models}

Nonlinear system identification can also be approached using Gaussian process (GP) regression \citep{GPBook}. The book by \citet{kocijan2016modelling} presents techniques for learning and validating GP dynamics models from measurement data, for embedding prior knowledge into the GP models, and for designing optimal control, model predictive control, and adaptive control based on the GP dynamics models. 

A GP is a stochastic process, that is, a collection of random variables indexed by time or space, such that any finite subset of the random variables has a joint Gaussian distribution. This can be viewed as a distribution $\mathcal{GP}(\mu(z),k(z,z'))$ for a function $f(z)$ with mean function $\mu(z) = \mathbb{E}[f(z)]$ and covariance function (or kernel) $k(z,z') = \operatorname{Cov}[f(z),f(z')]$.

Consider a regression problem:
\begin{equation}
    y_k = f(z_k) + \epsilon_k, \qquad \epsilon_k \sim \mathcal{N}(0,E),
\end{equation}
where the objective is to approximate the function $f(z)$ given a dataset $\mathcal{D} = \{z_k, y_k\}_k$ of Gaussian noisy measurements with zero mean and covariance $E$. The GP approach to regression imposes a GP prior distribution on $f(z)$ and computes the posterior of $f(z)$ conditioned on the data $\mathcal{D}$. The values $f(Z^*) = \begin{bmatrix} f(z_1^*)^\top & \cdots & f(z_M^*)^\top\end{bmatrix}^\top$ of the function $f(z)$ at a set of query points $Z^*$ are predicted by using that the observations $Y = \begin{bmatrix} y_1^\top & \cdots & y_N^\top\end{bmatrix}^\top$ in the dataset and the query function values $f(Z^*)$ have a joint Gaussian distribution:
\begin{equation}
    \begin{bmatrix}
        Y \\ f(Z^*)
    \end{bmatrix} \sim \mathcal{N}\left(\begin{bmatrix}
        \mu(Z) \\ \mu(Z^*)
    \end{bmatrix}, \begin{bmatrix}
        k(Z,Z) + I \otimes E & k(Z,Z^*)\\
        k(Z^*,Z) & k(Z^*,Z^*)
    \end{bmatrix}\right),
\end{equation}
where $\otimes$ is the Kronecker product and $Z$ are the locations where the training observations $Y$ were obtained from. Then, the posterior distribution of $f(Z^*)$ conditioned on the dataset $\mathcal{D}$ (of observations $Y$ and locations $Z$) is obtained as the conditional distribution from the above joint distribution. Specifically, $f(Z^*) \mid \mathcal{D}$ has a GP distribution mean function:
\begin{equation}
\mu(Z^*) + k(Z^*,Z)(k(Z,Z)+ I \otimes E)^{-1}(Y-\mu(Z))
\end{equation}
and kernel:
\begin{equation}
k(Z^*,Z^*) - k(Z^*,Z)(k(Z,Z)+ I \otimes E)^{-1}k(Z,Z^*).
\end{equation}

In system identification \citep{sarkka2021use}, GPs are used to form time series predictions (e.g., impulse response or autoregressive models) or directly represent the dynamics model and output model of a state-space representation \eqref{eq:control_system}.

\citet{Pillonetto2014Kernel} use GP regression to model the impulse response $g(t)$ in \eqref{eq:convolution} of LTI systems. Given input-output data, the impulse response is inferred via the posterior mean of a GP, with hyperparameters (kernel parameters and noise variance) estimated by maximizing the marginal likelihood. To guarantee BIBO stability of the identified model, the kernel should be absolutely integrable, $\int\int |k(t,t')| dt dt' < \infty$, which ensures that the impulse response $g(t)$, estimated by the GP mean, is almost surely integrable. The authors show that commonly used stationary kernels, such as the Gaussian (RBF) or spline kernels, fail to meet this criterion but can be modified via exponential or algebraic decay terms to ensure integrability and, hence, BIBO stability of the impulse response estimates.

A GP model can also be used to learn a state-space model \eqref{eq:control_system} directly using input-output data.  \citet{Frigola2013GPSSM,Frigola2014GPSSM} model the dynamics $f$ as a sparse GP with inducing points and the output model $h$ as a parametric density (e.g., Gaussian or Poisson), and optimize the model parameters via variational inference.  A joint variational distribution is specified over the inducing points, (latent) states, and state transitions and a particular factorization is chosen to make the Evidence Lower Bound (ELBO) computationally efficient. \citet{Eleftheriadis2017GPSSM} assume that the state posterior has a Markov-structured Gaussian distribution and introduce a bi-directional recurrent neural network to facilitate smoothing of the state posterior over the data sequences and recovery of the variational parameters. In closely related work, \citet{Doerr2018PRSSM} propose an improved factorization, which captures the true system transitions, and compute the ELBO stochastic gradients analytically via the re-parametrization trick. These methods are demonstrated to learn dynamics models of neural activity in rats’ hippocampus, of a hydraulic actuator that controls a robot arm, and of a belt-driven pulley with two electric motors. 

State-space GP models can also be used to ensure stability. \citet{Berkenkamp_SafeRoA_CDC16} introduce a residual GP model to capture uncertainty in a nominal dynamics model and estimate the region of attraction (RoA) of the closed-loop system under a fixed control policy. A Lyapunov function is used to certify a subset of the RoA by verifying that its derivative remains negative with high probability, given the GP uncertainty. The method iteratively expands the certified RoA by actively selecting new states near the boundary, updating the GP model, and recomputing the RoA without leaving the certified safe region. 
In subsequent work, \citet{Berkenkamp_SafeRL_NeurIPS2017} extended the approach to allow modifying the control policy to achieve optimal stabilization performance, captured by a cumulative cost function, while maintaining stability guarantees.

\begin{figure}[t]
\centering
\begin{subfigure}[t]{\linewidth}
    \centering
    \includegraphics[width=0.95\linewidth,trim={0 2ex 0 0},clip]{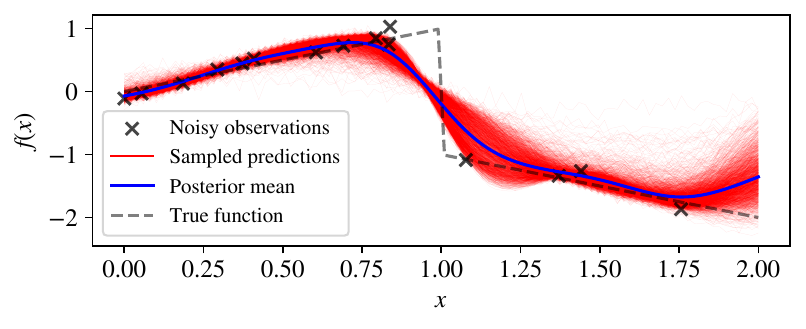}
    \caption{Unstructured GP regression.}
    \label{fig:unstructured_GP}
\end{subfigure}\\
\begin{subfigure}[t]{\linewidth}
    \centering
    \includegraphics[width=0.95\linewidth,trim={0 2ex 0 0},clip]{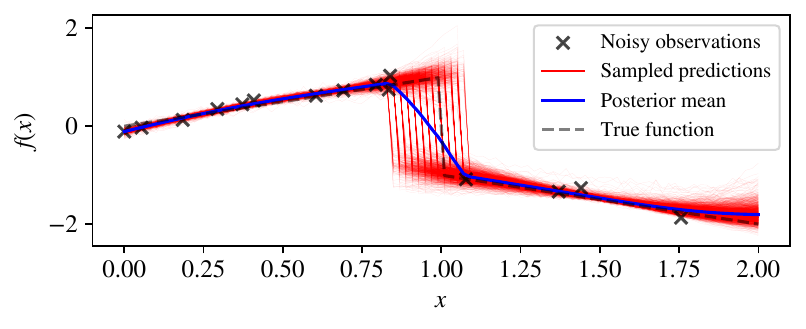}
    \caption{Structured GP regression.}
    \label{fig:structured_GP}
\end{subfigure}
    \caption{Learning a piecewise-linear dynamics function $f$ via GPJax \citep{GPJax} with structured GP regression (b) is more accurate than with unstructured GP regression (a).}
    \label{fig:GP_RoA}
\end{figure}

GP regression can be used to approximate an explicit control law for model predictive control \citep{Grancharova2007GPMPC,Binder2019GPMPC,Hewing2020GPMPC}. For example, \citet{Rose2023GPMPC} use a sampling-based scenario approach to provide bounds on the approximation errors and ensure recursive feasibility and input-to-state stability. Beyond model predictive control, recent work \citep{wang2018safe,Castaneda_GPControl_CDC21,Long_SafeControl_RAL22,Dhiman_SafeDynamicsLearning_TAC23} has extended the types of safety and stability guarantees for controllers synthesized from GP system models by considering worst-case and probabilistic formulations of control Lyapunov function and control barrier function constraints to account for the presence of model uncertainty.

In connection with Sec.~\ref{sec:phmodels}, GP regression can be used to learn Lagrangian and port-Hamiltonian system models \eqref{eq:phs} from state-input data. \citet{Evangelisti2022PCGP} introduce a physically constrained formulation of GP regression to respect the structure of Lagrangian systems and guarantee energy conservation. This requires extending the GP kernel to symmetric matrix spaces and using Cholesky decomposition to provide probabilistic guarantees that the energy GP has positive-definite quadratic form. Subsequent work by \citet{Evangelisti2024MomentumObserver} integrates the Lagrangian GP model into covariance-adaptive momentum observers to enable learning-based external force and disturbance estimation. The GP prediction error bounds are used to establish exponential stability guarantees for the observer. Regarding Hamiltonian systems, \citet{beckers2022gaussian} and \citet{beckers2023bayesian} formulate a joint GP over the Hamiltonian and its partial derivatives and use a specific matrix kernel to guarantee that the GP almost surely generates valid port-Hamiltonian trajectories. As in the deterministic case, the interconnection of two GP port-Hamiltonian systems remains a GP port-Hamiltonian system. The approach is demonstrated to learn the dynamics of an iron ball in the magnetic field of a controlled inductor. The results show that training a GP whose kernel does not capture the port-Hamiltonian system structure requires ten times more samples compared to a GP that does.

Several modern GP libraries designed for large-scale inference with GPU acceleration are available, including GPyTorch \citep{gardner2018gpytorch}, GPflow \citep{GPflow2017}, Pyro \citep{pyro}, and GPJax \citep{GPJax}. For example, GPJax \citep{GPJax} allows us to learn a dynamics function $f$ from data more accurately by incorporating prior knowledge of the system in a GP regression model in JAX, as shown in Example \ref{ex:structured_gp}. 

\begin{example}[Learning dynamics using GP regression] \label{ex:structured_gp}
Consider learning a piecewise-linear system:
\begin{equation} \label{eq:piecewise-linear-dyn}
y = f(x) = \begin{cases}
ax, & \text{if } x \leq x_{th}, \\
bx, & \text{if } x > x_{th}, \\
\end{cases}
\end{equation}
where the ground-truth parameters are $a = 1$, $b = -1$, $x_{th} =~1$. A dataset of $15$ pairs $(x,y)$ is generated from the ground-truth dynamics to train standard (unstructured) and structured GP regression models, implemented in GPJax \citep{GPJax}. While the unstructured GP model does not assume any prior knowledge of the system, the structured GP model encodes the piecewise-linearity of the dynamics $f$ by replacing the constant mean of an unstructured GP with \eqref{eq:piecewise-linear-dyn}. The unknown parameters $a$, $b$ and $x_{th}$ in \eqref{eq:piecewise-linear-dyn} are then trained to fit the data, leading to better prediction accuracy as shown in Fig.~\ref{fig:GP_RoA}.
\end{example}

\section{Deep Learning-based System Identification}
\label{sec:deep}

Deep learning for system system identification is usually formulated as a supervised learning problem. Recall that in a system identification problem, we are given a dataset $\mathcal{W} = \{(u_{i,k},y_{i,k})\}_i$ of input-output trajectories from a continuous-time dynamical system. We seek to learn a nonlinear state-space model with dynamics $f_{\theta}$ and
output model $h_{\theta}$, where the model parameters $\theta$ is obtained from
the unconstrained optimization \eqref{eq:general_sysid_optimizaiton} or the constrained system identification problem \eqref{eq:constrained_sysid_optimizaiton}. For fitting of the deep learning based system ID models, discretely sampled data from the continuous-time system is often assumed. Some methods assume access to not only the input-output trajectories, but also the state trajectories thus \(\mathcal{W}^{+} = \{(x_{i,k}, u_{i,k},y_{i,k})\}_i\), which we will discuss in context. {  The distinction between \(\mathcal{W}\) and \(\mathcal{W}^+\) is important when enforcing control-relevant properties in deep learning models. When full state trajectories are available, properties such as Lyapunov stability or dissipativity can often be imposed directly on the learned state-space model. In output-only or partially observed settings, the state trajectory used by the model is not directly measured and may instead be estimated, represented through an internal model state, or replaced by an input-output representation. Consequently, hard constraints on a learned state-space realization typically certify the realized or latent model state, and their interpretation for the physical system depends on the observer, encoder, or realization used to construct that state.}

We will start with reviewing common model choices for system identification from classic deep learning architectures (e.g., feedforward neural networks, recurrent neural networks) to physics-informed deep learning architectures (e.g., neural ordinary differential equations, recurrent equilibrium networks), and then surveying recent literature on incorporating control-relevant system properties into deep learning models.

\subsection{Deep Learning Architectures for System Identification}
\label{sec:classic_DL}

\subsubsection{Feedforward Neural Networks}
\label{sec:FNN}
One basic model for fitting the functions $\{f_{\theta}, h_{\theta}\}$ is via an $n$-layer feedforward neural network (FNN). The feedforward neural network (FNN) approach assumes access to the true system state $x_{i,k}$ and models the state-transition and output mappings as
\begin{equation}
    \begin{aligned}
    \hat{x}_{i,k+1} &= f_{\theta}(x_{i,k}, u_{i,k}) := \text{FNN}_{1}(x_{i,k}, u_{i,k})\,, \\
    \hat{y}_{i,k} &= h_{\theta}(x_{i,k}, u_{i,k}) := \text{FNN}_{2}(x_{i,k}, u_{i,k})\,.
\end{aligned}
\end{equation}
Each FNN consists of $L \in \mathbb{N}^+$ hidden layers that is defined as, 
$$\text{FNN}(\xi):= (l_L \circ l_{L-1} \circ ... \circ l_1)(\xi)\,,$$
where layers $l_i$ start with the input layer $l_0 = \xi$ and continue as 
\begin{equation*}  
\label{eq-nn2}  
l_{i+1}
(l_i,\theta_{i+1}) = \sigma(W_{i+1} l_i + b_{i+1}), \quad i=0,\ldots,L-1.
\end{equation*}
$\sigma$ is a nonlinear activation function, weights $W_{i+1}$ and biases $b_{i+1}$ are parameters in the neural network to be learned, collected into $\theta_i$, and then into $\theta = [\theta_1^{\rm T},\ldots, \theta_L^{\rm T}]^{\rm T}$. 
It has been shown that, the FNN has universal approximation capability \citep{hornik1989multilayer}, thus is able to model almost all systems with continuous $f_{\theta}, h_{\theta}$ over a compact set. The model parameters $f_{\theta}, h_{\theta}$ are optimized to minimize the error between predictions and the true state $x_{i,k+1}$ and output $y_{i,k}$.

Applications of FNN for system identification have been found in literature since the 90s \citep{kuschewski1993application, lu1998robust}. \citet{nagabandi2018neural} deploys an FNN to model the robotics dynamics, which is further used for model-based reinforcement learning. Recent applications of FNN for system identification include various robotic and mechanical systems \citep{gillespie2018learning,pan2018long,Shi2019neurallander,robinson2022physics,Roberto2021FNN,wei2022safe}. However, the formulation provided above requires direct access to the full input-state-output trajectories $\mathcal{W}^{+}$, which may not be available or observable in many real-world systems. We refer to a recent survey for more recent literature on FNN for system identification \citep{pillonetto2023deep}.

\subsubsection{Recurrent Neural Networks and Transformers} 
{ While FNNs can approximate input-output mappings, they do not generally capture temporal dependencies through the architecture itself unless past states, inputs, or outputs are included among the network inputs.} 
Recurrent Neural Networks (RNNs) address this limitation by maintaining a hidden state that evolves over time, making them particularly well-suited for modeling dynamical systems. 
Specifically, for all $k = 1, ..., T$, the RNN evolves as
\begin{subequations}
    \begin{align}
    \hat{x}_{i,k+1} & = f_{\theta}(\hat{x}_{i,k}, u_{i,k}) := \sigma(W_{hh} \hat{x}_{i,k} + W_{hx} u_{i,k} + b_h)\,,  \\
    \hat{y}_{i,k} & = h_{\theta}(\hat{x}_{i,k}, u_{i,k}) := \sigma(W_{y} \hat{x}_{i,k} + b_y)\,,
\end{align}
\end{subequations}
where $\sigma$ denotes a nonlinear activation function, and the weight matrices $W_{hx}, W_{hh}, W_y$ and biases $b_h, b_y$ are model parameters to be learned. Compared to FNNs, RNNs introduce a hidden state that evolves over time, enabling the model to infer and maintain internal representations of system dynamics from sequences of input-output data alone. This makes RNNs particularly advantageous for system identification tasks where only input-output trajectories $(u_{i,k}, y_{i,k})$ are available.

To improve the capability of RNNs to capture long-term dependency of system trajectories, features such as Long Short-Term Memory (LSTM) \citep{hochreiter1997long} and Gated Recurrent Units (GRUs) \citep{chung2014empirical} can be adopted. Deep RNNs \citep{pascanu2013construct} further improve RNN for replacing the learnable matrices with deep neural networks. 
Like FNNs, RNNs have shown to be \emph{universal function approximators} in \citet{sontag1992neural,funahashi1993approximation}. More recent theoretical developments \citep{hanson2020universal, hanson2021learning} provide quantitative error guarantees, specifying the sample size and neurons to reach a desired accuracy.

Given the advanced modeling capabilities of RNNs for dynamical systems, it is not surprising that they are popular choices for system identification and control. Early applications of RNNs and their variants for system identification can be found in \citet{delgado1995rnn,wang2006fully,Wang2017rnn}. The identified models can be used for decision making and control \citep{pan2011model}. Notable recent advances include convex RNNs \citep{chen2018optimal} for system identification, thus leading to a convex model predictive control (MPC) problem with optimality guarantee. For interested readers, we refer to \citet{Fabio2022RNN} for a recent survey of RNN models for system identification and learning-based control. 

As an emerging direction, Transformer networks~\citep{vaswani2017attention} are considered a promising alternative to RNNs for sequence modeling and system identification. 
\citet{GENEVA2022272transformer} proposes to apply the self-attention Transformers for physical dynamics modeling, which allows the model to learn longer and more complex temporal dependencies with Koopman-based embedding. 
For spatiotemporal dynamics, \citet{han2022predicting} incorporates a graph-based method for the spatial relationships and the Transformer for temporal dynamics. 
\citet{park2023simultaneous} customizes the Transformer for 
model predictive control, where the Transformers are implemented for multistep-ahead prediction. Besides modeling ODE systems, Transformers have recently emerged as a promising tool for surrogate modeling of partial differential equations (PDEs) \citep{Zijie2023transformer}. 
Because of its good performance, Transformers are widely adopted for different applications such as traffic flow \citep{xu2020spatial,REZA2022117275transformer}, fluid flow \citep{solera2024beta, hassanian2023deciphering}, mechatronic systems \citep{electronics2023transformer}, and etc.

\begin{figure*}[t]
    \centering
    \includegraphics[width=0.95\linewidth]{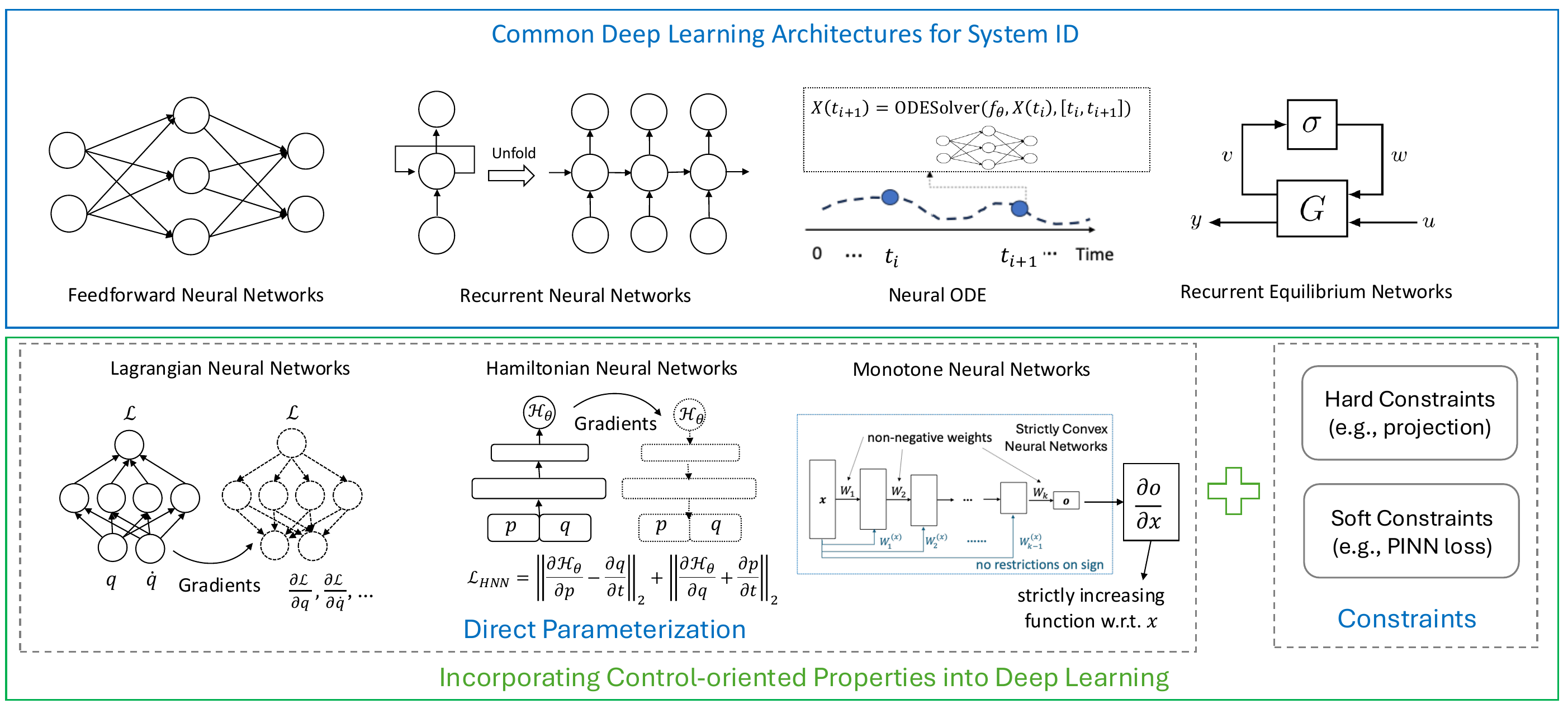}
    \caption{Overview of the deep learning architectures and control-oriented architectures for system identification.}
    \label{fig:Review_DLdiagram}
\end{figure*}

\subsubsection{Neural Ordinary Differential Equations (Neural ODEs)} 
One challenge with RNN is that as the trajectory length increases, it requires propagating information over long horizons. Thus, RNNs could lead to high computational cost and difficulties in training.  
\cite{chen2018neural} introduce neural ODEs that describe the system dynamics as $\dot{\hat{x}}_i(t) = f_{\theta}(\hat{x}_i(t), u_i(t))$ in the original continuous-time domain,
\begin{subequations}
    \begin{align}
    \hat{x}_i(t+1) &= \hat{x}_i(t) + \int_{\tau=t}^{t+1} f_{\theta}(\hat{x}_i(t), u_i(t)) d\tau\,, \label{eq:node1}\\
    \hat{y}_i(t) &= h_{\theta}(\hat{x}_i(t), u_i(t))\,, \label{eq:node2}
\end{align}
\end{subequations}
where $f_{\theta}, h_{\theta}$ are commonly modeled via FNNs. We can view Neural ODEs as introducing the structure of differential equation as a part of prior knowledge about the problem (a.k.a., inductive bias in machine learning). 
Starting with $\hat{x}_i(t)$ at time $t$, the state at time $t+1$ is obtained by integrating the neural network $f_{\theta}$ over $[t, t+1]$. 
The integral in \eqref{eq:node1} has to be solved numerically, and can be expensive since the dynamics are modeled as neural networks. 
Follow-up works \citep{Quaglino2020ODE,kelly2020learning,dupont2019augmented,bilovs2021neural} focus on retaining expressive dynamics at less computational cost. 
\citet{zhu2021neural} proposes neural delay differential equations, that can be used to learn dynamical systems with delays. \citet{chen2021ode} proposes neural event functions to include discrete events (e.g. switches) in the continuous-time dynamics modeling. To understand the representation capability of Neural ODE, \citet{ruiz2023neural} establishes universal approximation guarantees for ODEs.

For applications, \citet{Quaglino2020ODE} incorporates the spectral element methods with Neural ODEs for fast and accurate system identification, where the dynamics are interpreted as truncated series of Legendre polynomials. To improve generalization, physics-informed design is widely adopted in design of Neural ODEs. For example, 
\citet{zhong2020symplectic} incorporates the prior knowledge of Hamiltonian dynamics. \citet{hochlehnert2021learning} deploys a Central-Difference Lagrange integrator structure for modeling contact dynamics using Neural ODEs. More examples incorporating control-oriented properties are deferred to sections \ref{sec:DL_hard_constraints} and \ref{sec:PINN_systemID}. Following the benchmark result of~\citep{Rahman2022ode}, Neural ODEs have shown good performance in system identification, and have been successfully applied in structural identification \citep{lai2021structural},  power systems \citep{xiao2022feasibility}, robotics~\citep{huang2023fi,xiao2023forward} etc. Another interesting deep learning architecture sharing a similar structure as Neural ODE is the Deep Equilibrium (DEQ) model, which finds a fixed point of a dynamical system corresponding to a neural network \citep{bai2019deep}. \citet{winston2020monotone} extends DEQ by restricting the mapping to be monotone, which guarantees the existence of a unique equilibrium point.

\subsubsection{Recurrent Equilibrium Networks}
A Recurrent Equilibrium Network (REN) \citep{revay2023recurrent}
is a flexible model structure that incorporates both a recurrent structure and an equilibrium network. It was introduced to include many established models as special cases, such as FNNs and RNNs. Specifically, REN is given in the following form,
\begin{equation}
    \begin{aligned}
    \hat{x}_{k+1} &= A \hat{x}_k + B_1 w_k + B_2 u_k + b_x\,,\\
    \hat{y}_k & = C_2 \hat{x}_k + D_{21} w_k + D_{22} u_k + b_y,
    \end{aligned}
\end{equation}
in which $w_k$ is the solution of an equilibrium network DEQ
$$w_k = \sigma(D_{11} w_k + C_1 \hat{x}_k + D_{12} u_k + b_v),$$
$A, B, C, D$ are matrices of appropriate dimension, $b_x \in \mathbb{R}^n, b_y \in \mathbb{R}^p, b_v \in \mathbb{R}^q$ are bias parameters, and $\sigma$ is a nonlinear activation function. REN models can also be represented as a feedback interconnection of a linear system $G$ (composed of matrices $A, B, C, D$) and a memoryless nonlinear operator $\sigma$, as depicted in Figure \ref{fig:Review_DLdiagram}.

RENs can incorporate built-in guarantees of stability and robustness through either convex or direct parameterization to satisfy sufficient conditions for contractive models and incremental dissipativity.  
Convex parameterization introduces multipliers to add degrees of freedom, making the model parameters, stability certificate, and multipliers jointly convex. Direct parameterization maps the convex region to a set of free parameters, turning the problem into unconstrained optimization. The resulting parameters are guaranteed to be well-posed, ensuring the model generates unique state trajectories for any input and initial condition. While RENs model discrete-time dynamics, NodeRENs \citep{martinelli2023unconstrained} extend the framework to continuous-time systems by combining RENs with Neural ODEs. Like RENs, NodeRENs are contractive and dissipative by design and allow training via unconstrained optimization.

\subsubsection{Incorporating Control-oriented Properties} 
\label{sub:incor-control-properies-exp}
As discussed in Section~\ref{sec:ways_incorporating_physics}, there are three common ways of incorporating structures into system identification: \emph{direct parametrization, hard constraints, soft constraints}.
In the next subsections, we will survey deep learning based system identification literature falling under each kind. We start from the direct parametrization with three examples on Lagrangian Neural Networks, Hamiltonian Neural Networks, and Monotone Neural Networks. We then survey ways of incorporating structure as hard constraints and soft constraints, respectively. We end the discussion by comparing hard and soft stability constraints in a learning pendulum dynamics example.

\subsection{Direct Parametrization} 
\subsubsection{Lagrangian Neural Networks}
\label{exp:lagrangian}
Consider the Lagrangian dynamics of a rigid-body system expressed as
\begin{equation}
\label{eq:Lagrangian_rigid_body}
    M(q)\ddot{q} + C(q,\dot{q})\dot{q} + G(q) = \tau\,,
\end{equation}
where \( q \in \mathbb{R}^n \) represents generalized coordinates, \( M(q) \) is the inertia matrix, \( C(q,\dot{q}) \) is the Coriolis matrix, \( G(q) \) denotes conservative forces (e.g., gravity), and \( \tau \) is the non-conservative control input. To put the Lagrangian dynamics in the standard state-space representation, one could define the state vector \( x = (q, \dot{q}) \in \mathbb{R}^{2n} \), and rewrite the dynamics as 
\begin{align*}
    \dot{x}_1 &= x_2\,, \\
    \dot{x}_2 &= M^{-1}(x_1) \left( \tau - C(x_1, x_2)x_2 - G(x_1) \right)\,,
\end{align*}
so that the overall system dynamics are represented as \( \dot{x} = f(x, u) \), consistent with Eq.~\eqref{eq:control_system}. The system output is modeled as \( y = h(x, u) \), where \( y \) may represent partial observations such as positions \( q \), velocities \( \dot{q} \), or both.

Instead of directly parameterizing the system dynamics $f$ and output function $h$ as neural network, \citet{roehrl2020modeling} parametrizes the force term as a neural network \( \tau(x, u; \theta) \) with inputs \( x = (q, \dot{q}) \), control input \( u \), and parameters \( \theta \). Given a dataset of trajectories \( \mathcal{D} = \{(u_i, y_i)\}_{i=1}^N \), a one-step 4th-order Runge-Kutta method is used to integrate the dynamics and generate predicted trajectories \( \hat{\mathcal{D}} = \{(\hat{y}_i)\}_{i=1}^N \). The neural network is trained by minimizing the discrepancy between the observed and predicted outputs, i.e., \( y_i \) and \( \hat{y}_i \). Instead of modeling $\tau$ as a neural network, \citet{lutter2019deeplagrangian} models $M(q) = L(q)L(q)^\top$ and $G(q)$ as neural networks and represented $C(q,\dot{q})$ in terms of $L(q)$. The approximated inverse dynamics is compared with $\tau$ from the dataset to train the models. This approach avoids the need of an ODE solver but requires $\ddot{q}$ in the dataset. 

The Lagrangian neural networks in \citet{cranmer2020lagrangian} further extend the above methods to general Lagrangian dynamics beyond the rigid-body dynamics in \eqref{eq:Lagrangian_rigid_body}. 
\citet{cranmer2020lagrangian} also illustrates that the Lagrangian neural networks can handle systems with disconnected coordinates such as graphs.
In terms of controller design based on learned Lagrangian dynamics model, \citet{lutter2019deepunderactuated} designs an energy-based controller for a class of simple under-actuated systems such as Furuta pendulums, cartpoles, and acrobots. In \citet{gupta2019general}, the learned Lagrangian dynamics model is used to plan a trajectory using Direct Collocation trajectory Optimization and design a time-varying linear quadratic regulator controller to track the trajectory.

\subsubsection{Hamiltonian Neural Networks} 
\label{exp:Hami}
Consider the Hamiltonian dynamics of the form,
\begin{equation} 
\label{eq:hamiltonian_dynamics_Rn}
    \frac{dq}{dt} = \frac{\partial H}{\partial p}, \frac{dp}{dt} = -\frac{\partial H}{\partial q}, 
\end{equation}
where $q \in \mathbb{R}^n$ and $p\in \mathbb{R}^n$ are the generalized coordinates and momentum of a system. $H$ denotes the total energy of the system, 
and the Hamilton's equations of motion \eqref{eq:hamiltonian_dynamics_Rn} guarantees that $\frac{dH}{dt} =0$, i.e. the total energy is conserved along the trajectory. 
To express the Hamiltonian dynamics in the standard state-space form consistent with Eq.~\eqref{eq:control_system}, one can define the state vector \( x = (q, p) \in \mathbb{R}^{2n} \) and write the dynamics as \( \dot{x} = f(x) = J \nabla_x H(x) \). With control inputs, the Hamiltonian dynamics can be modified by an additional forcing term that influences the state evolution by injecting or dissipating energy.

Hamiltonian Neural Networks (HNNs) \citep{greydanus2019hamiltonian} parametrize the conserved energy function $\mathcal{H}(q, p; \theta)$ rather than directly learning the system dynamics $f$. By enforcing the Hamiltonian structure, HNNs provide a principled way to embed hard physical constraints such as energy conservation into the learned model. The training loss of HNN follows
\begin{equation*}
    c_{HNN}(\theta) = \left\Vert \frac{dq}{dt} - \frac{\partial \mathcal{H}_{\theta}}{\partial p} \right\Vert_2 + \left\Vert \frac{dp}{dt} +\frac{\partial \mathcal{H}_{\theta}}{\partial q} \right\Vert_2,
\end{equation*}
where $\theta$ are parameters of the HNN. \citet{bertalan2019learning} extends the HNN approach to handle observations of the coordinates $q$ and momentum $p$ from images. The Hamilton's dynamics are modeled using a neural network on a latent space, learned from the observations using an auto-encoder. The auto-encoder and the Hamiltonian dynamics model are trained simultaneously to minimize loss function $c_{HNN}(\theta)$. 

Hamiltonian-based models require the second-order derivation of the coordinates $q$. \citet{chen2019symplectic} explores the use of a leapfrog integrator parameterized by a recurrent neural network to get rid of the need of taking the second-order derivatives. On the same vein, \citet{zhong2020symplectic} uses neural ODEs with Hamiltonian structure to approximate the unknown dynamics. 
\citet{finzi2020simplifying} shows that it is better, in terms of accuracy and data efficiency, to learn the Lagrangian and Hamiltonian dynamics with Cartesian coordinates with explicit manifold constraints instead of generalized coordinates. 

In terms of controller design based on learned Hamiltonian dynamics model, \citet{zhong2020symplectic} designs an energy-shaping regulation controller for fully-actuated systems based on the learned dynamics. \citet{duong2021hamiltonian} considers first learning Hamiltonian dynamics on $\texttt{SE}(3)$ manifold (short for Special Euclidean group in three dimensions, describing rigid body transformations combining rotation and translation in 3D space) , that encodes Hamiltonian dynamics and the $\texttt{SE}(3)$ manifold constraints on a neural ODE network. This is suitable for learning the dynamics of single-rigid-body robots such as ground vehicles and quadrotors, where the orientation is often described by a rotation matrix on the $\texttt{SO}(3)$ manifold (short for Special Orthogonal group in three dimensions, represents all 3×3 rotation matrices that describe rotations in three-dimensional Euclidean space). An energy-based controller is then designed for trajectory tracking tasks for both fully and under-actuated systems such as quadrotors.

\begin{figure}[t]
\begin{subfigure}[t]{0.23\textwidth}
        \centering
\includegraphics[width=\textwidth, height = 0.6\textwidth]{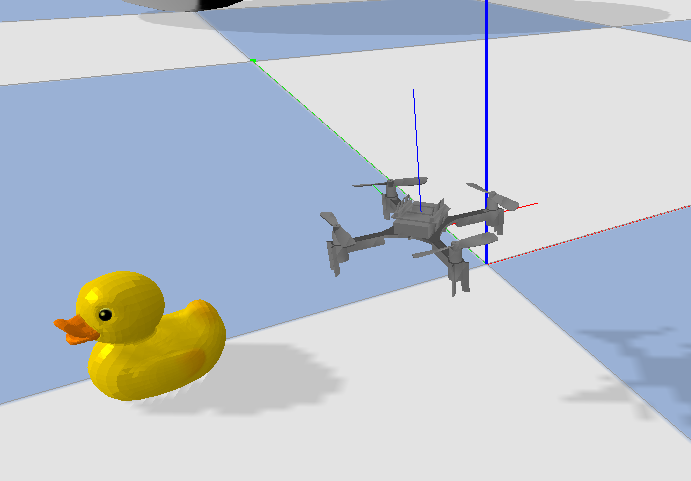}%
        \caption{Crazyflie quadrotor.}
        \label{fig:pybullet_crazyflie}
\end{subfigure}%
\hfill%
\begin{subfigure}[t]{0.23\textwidth}
        \centering
        \includegraphics[width=\textwidth, height = 0.6\textwidth]{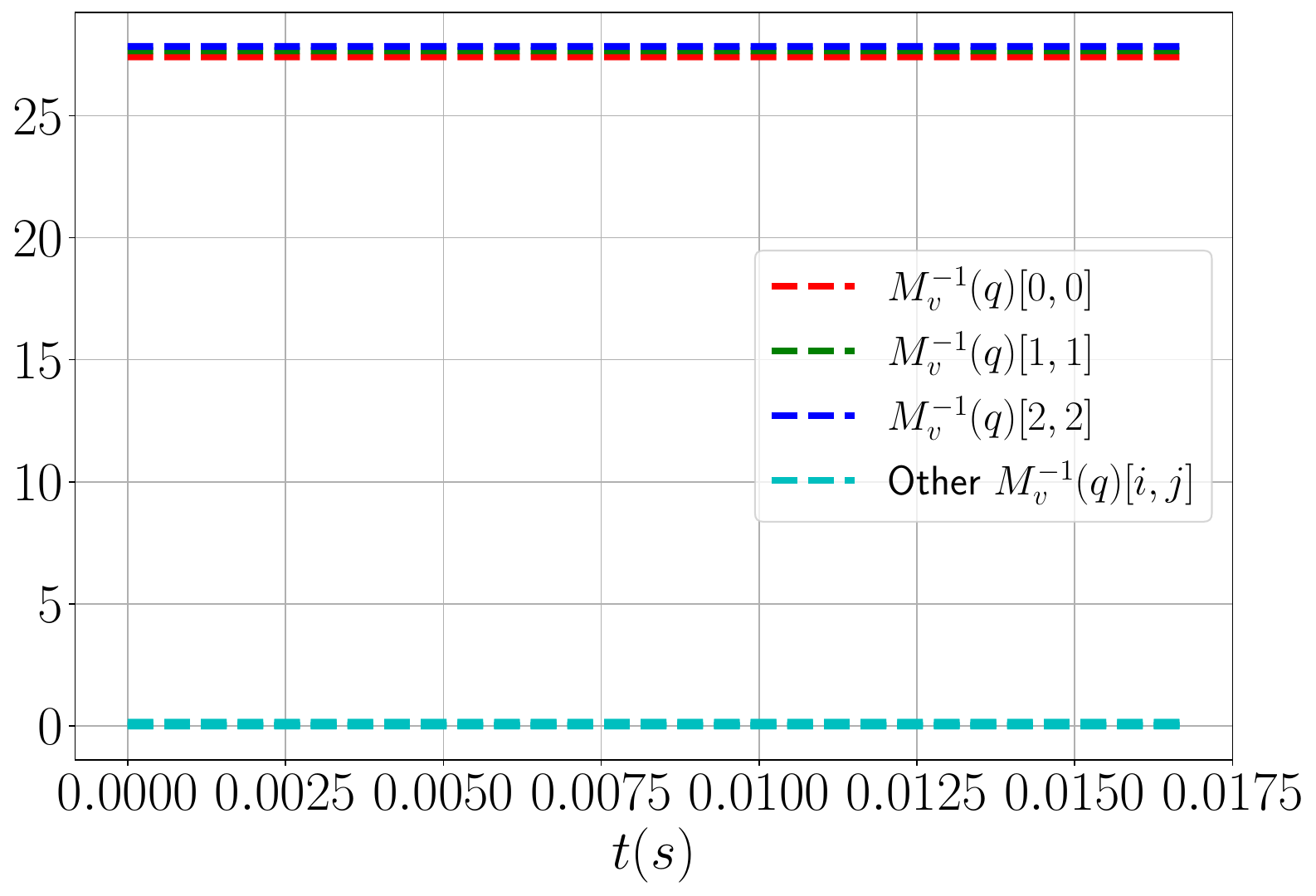}
        \caption{$M_v^{-1}(q)$'s entries}
        \label{fig:pybullet_M1_x_all}
\end{subfigure}

\begin{subfigure}[t]{0.23\textwidth}
        \centering
		\includegraphics[width=\textwidth, height = 0.6\textwidth]{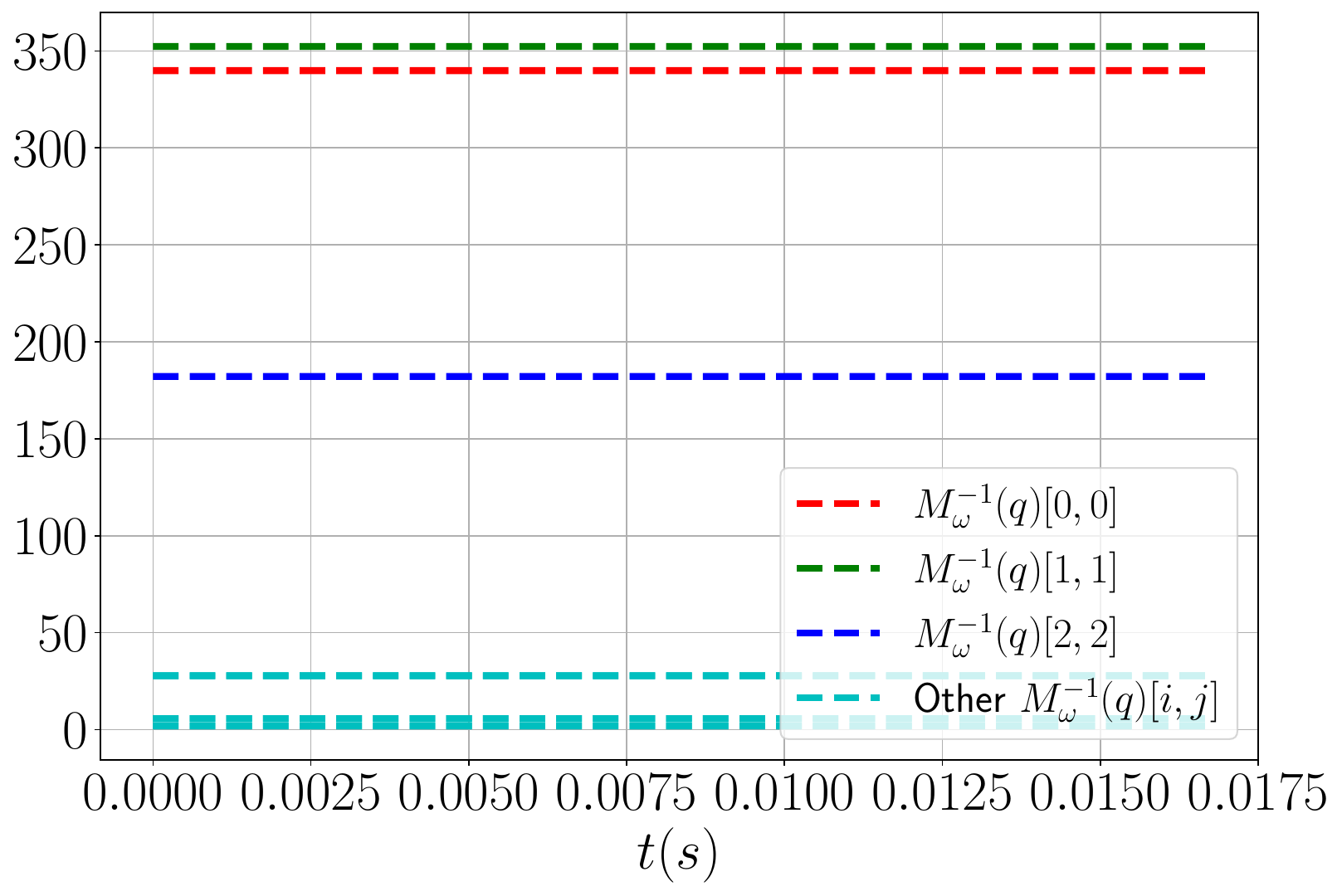}%
        \caption{$M_{\omega}^{-1}(q)$'s entries}
        \label{fig:pybullet_M2_x_all}
\end{subfigure}%
\hfill%
\begin{subfigure}[t]{0.23\textwidth}
        \centering
\includegraphics[width=\textwidth, height = 0.6\textwidth]{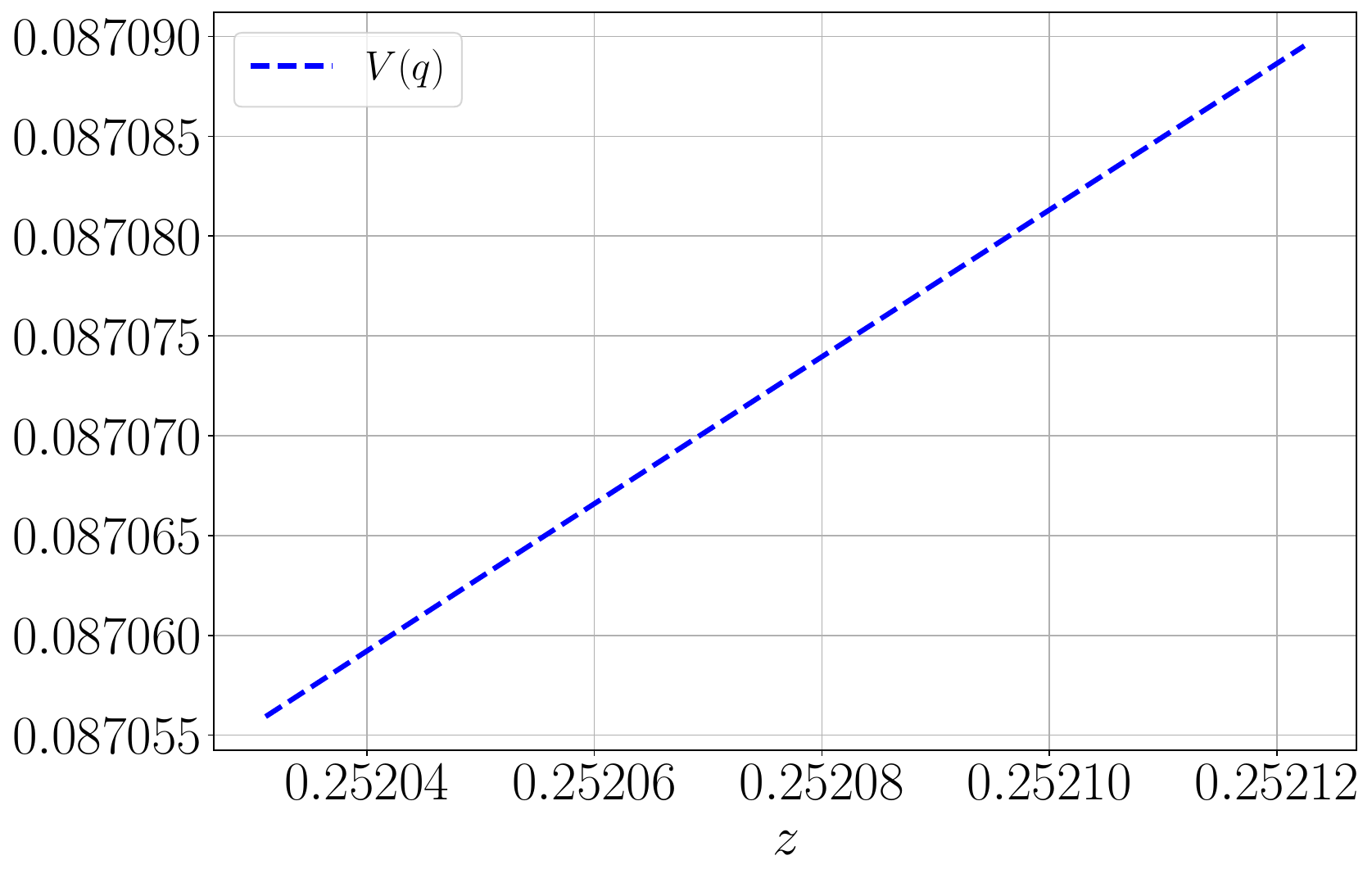}%
        \caption{$V(q)$}
        \label{fig:pybullet_Vx}
\end{subfigure}

\begin{subfigure}[t]{0.23\textwidth}
        \centering
        \includegraphics[width=\textwidth, height = 0.6\textwidth]{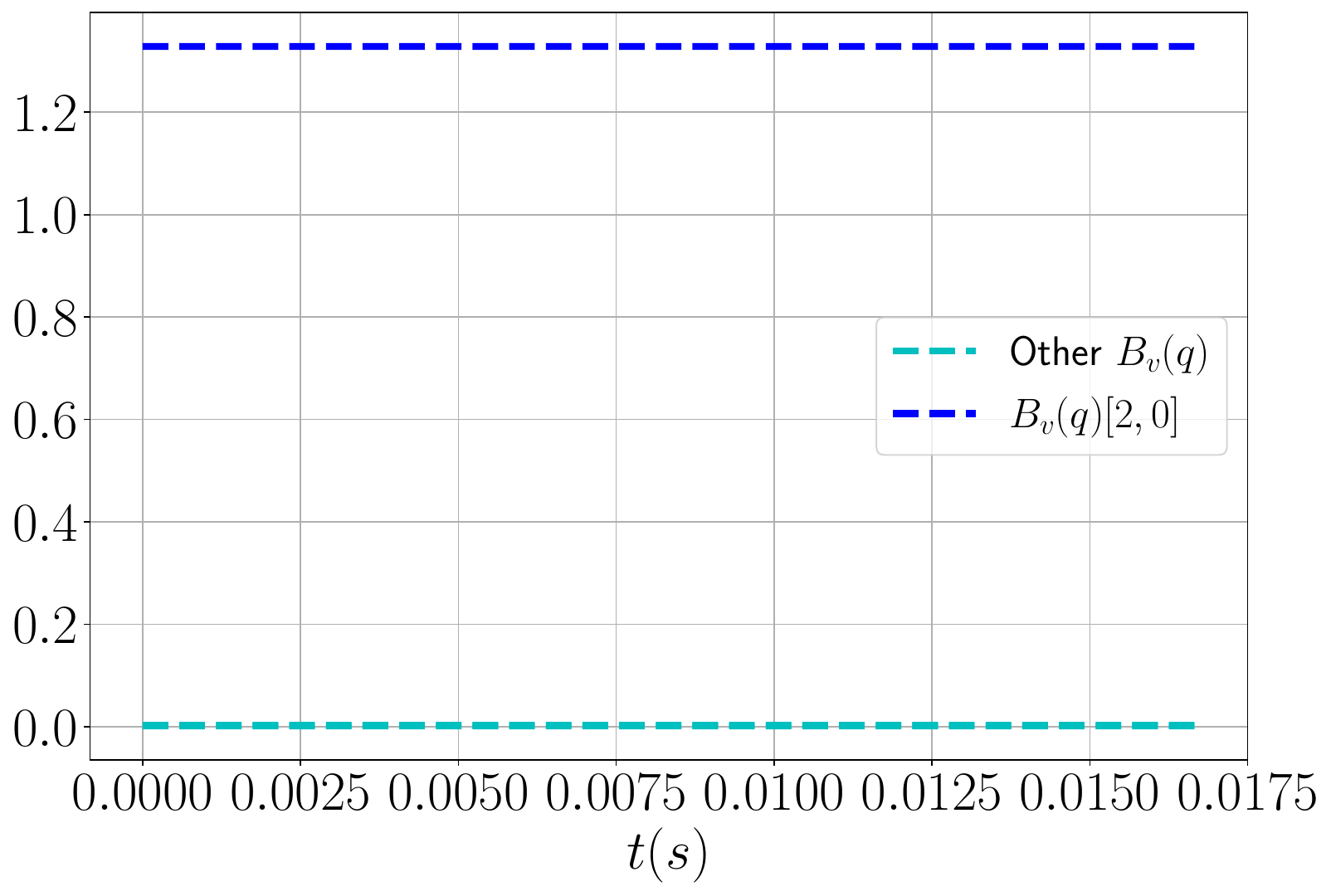}
        \caption{$B_{v}(q)$'s entries}
        \label{fig:pybullet_g_v_x_all}
\end{subfigure}%
\hfill%
\begin{subfigure}[t]{0.23\textwidth}
        \centering
        \includegraphics[width=\textwidth, height = 0.6\textwidth]{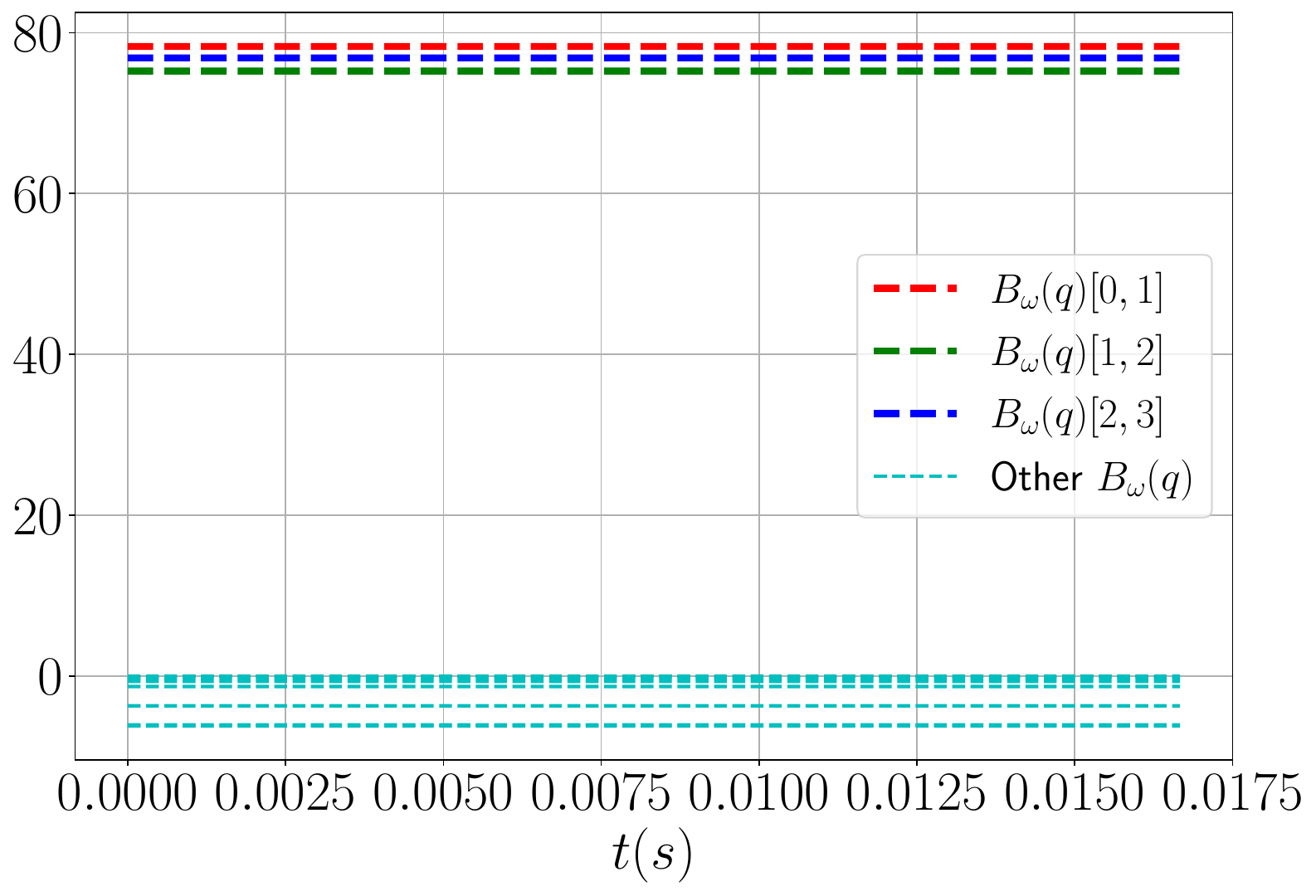}
        \caption{$B_{\omega}(q)$'s entries}
        \label{fig:pybullet_g_w_x_all}
\end{subfigure}

\begin{subfigure}[t]{0.23\textwidth}
        \centering
\includegraphics[width=\textwidth, height = 0.6\textwidth]{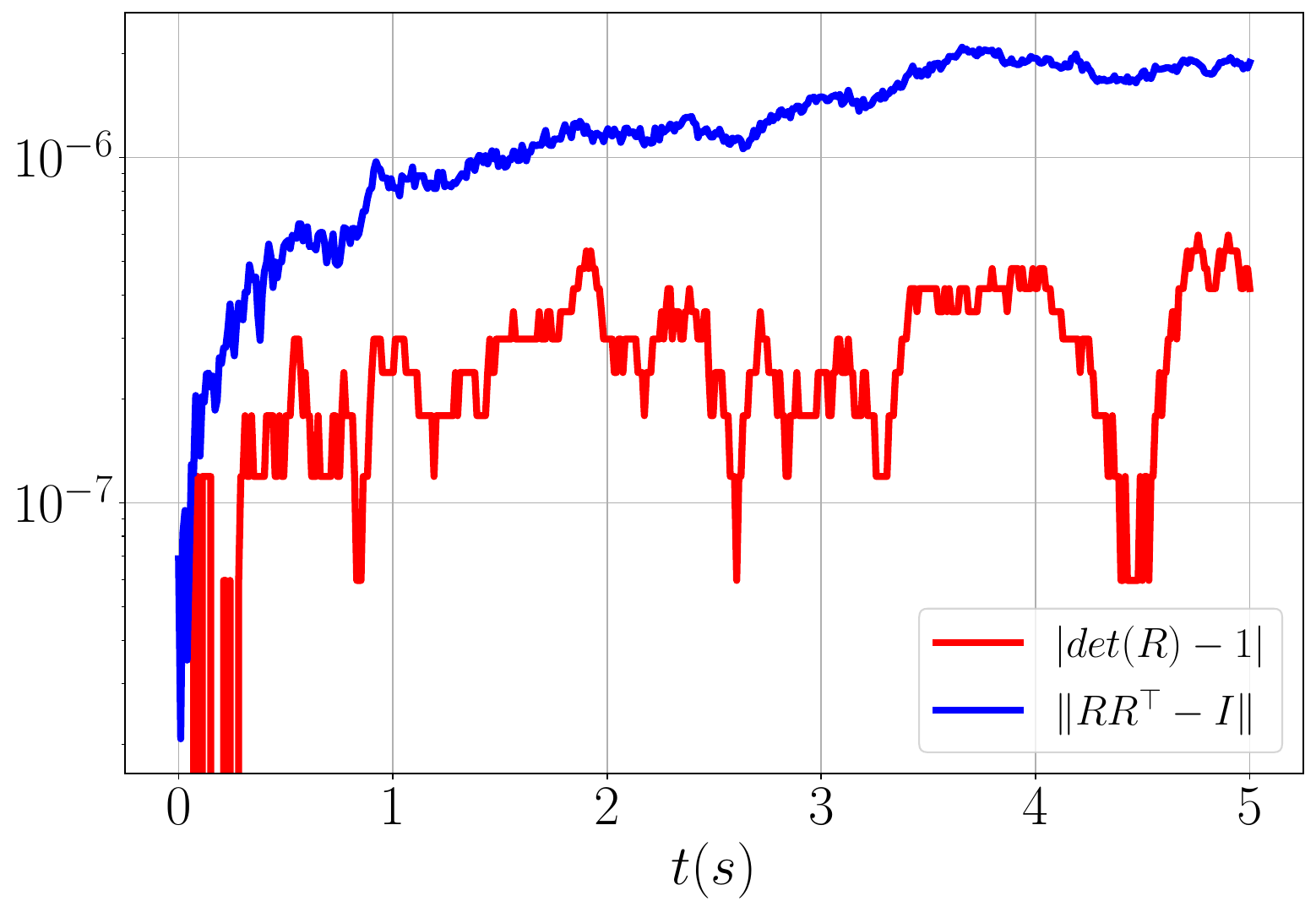}%
        \caption{SO(3) constraints.}
        \label{fig:pybullet_so3_constraints}
\end{subfigure}
\hfill%
\begin{subfigure}[t]{0.23\textwidth}
        \centering
\includegraphics[width=\textwidth, height = 0.6\textwidth]{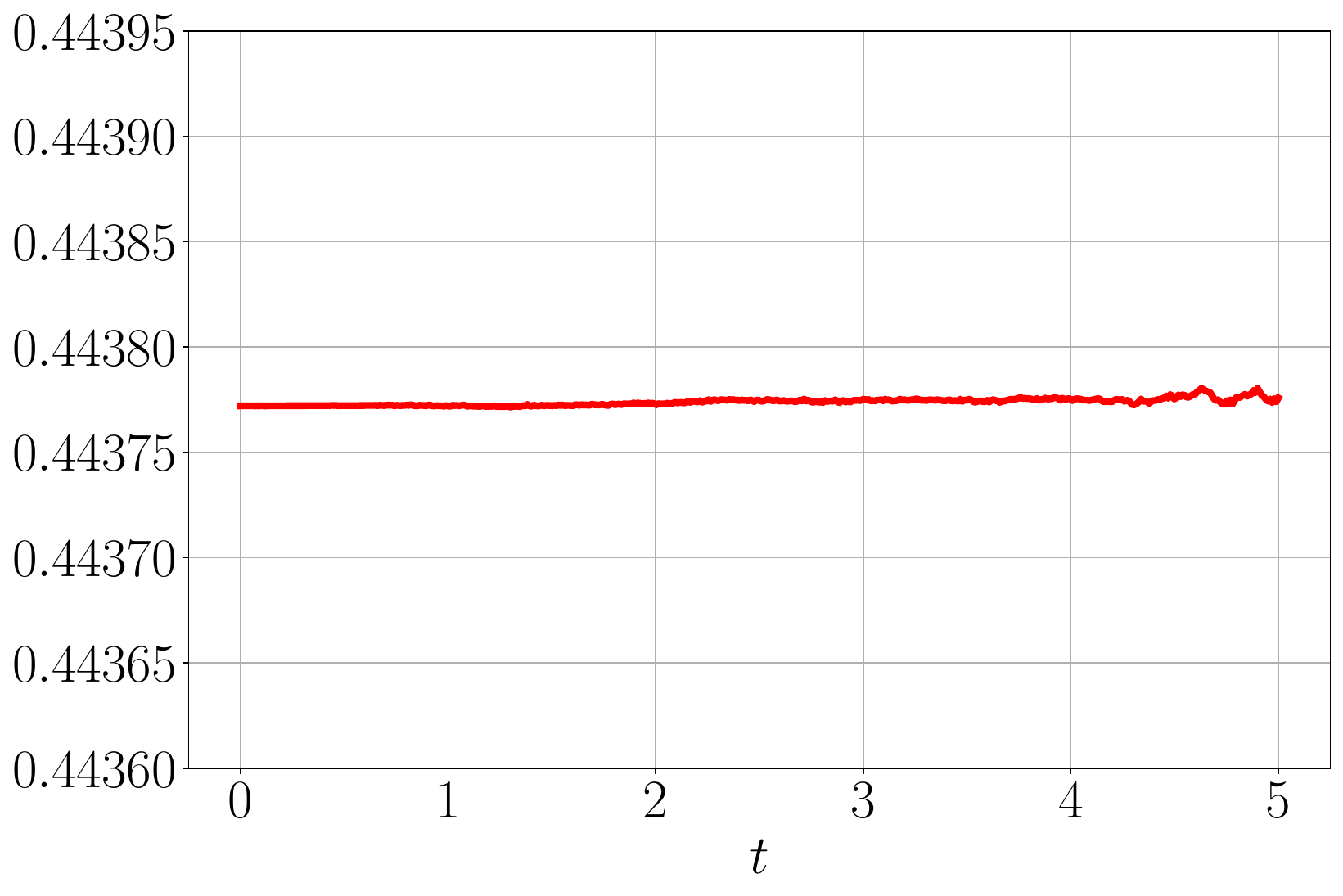}%
        \caption{Total energy.}
        \label{fig:pybullet_total_energy}
\end{subfigure}
\caption{$SE(3)$ port-Hamiltonian neural ODE network on a Crazyflie quadrotor in the PyBullet simulator \citep{gym-pybullet-drones2020}.}
\label{fig:pybullet_exp}
\end{figure}

\begin{example}[Quadrotor Dynamics Identification using Port-Hamiltonian Neural ODE Network]
\label{ex:pHNODE_quadrotor}
\begin{figure}[h]
\begin{subfigure}[t]{0.23\textwidth}
        \centering
\includegraphics[width=\textwidth]{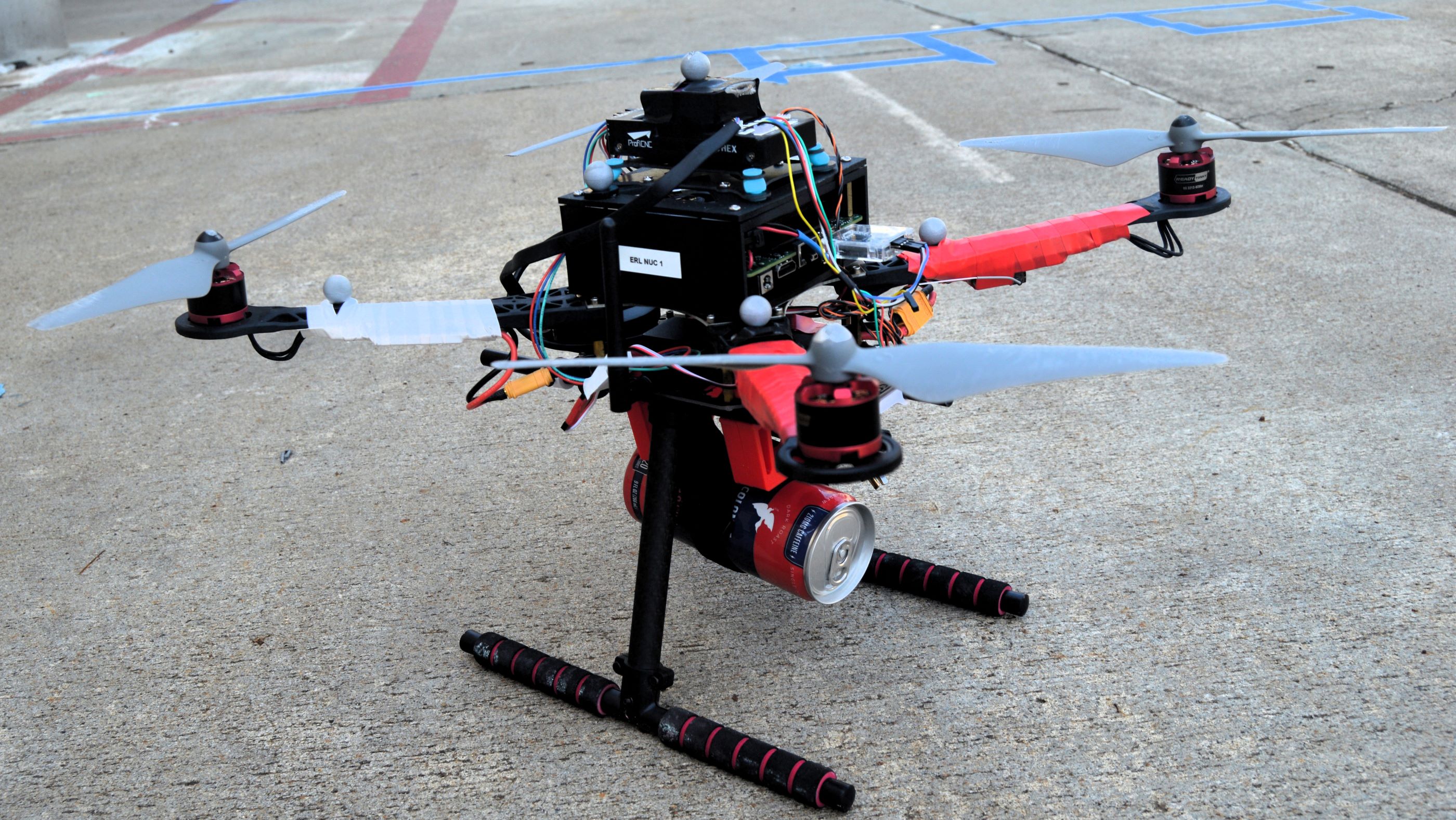}%
        \caption{An Intel NUC quadrotor.}
        \label{fig:nuc_quadrotor}
\end{subfigure}%
\hfill
\begin{subfigure}[t]{0.243\textwidth}
        \centering
        \includegraphics[width=\textwidth]{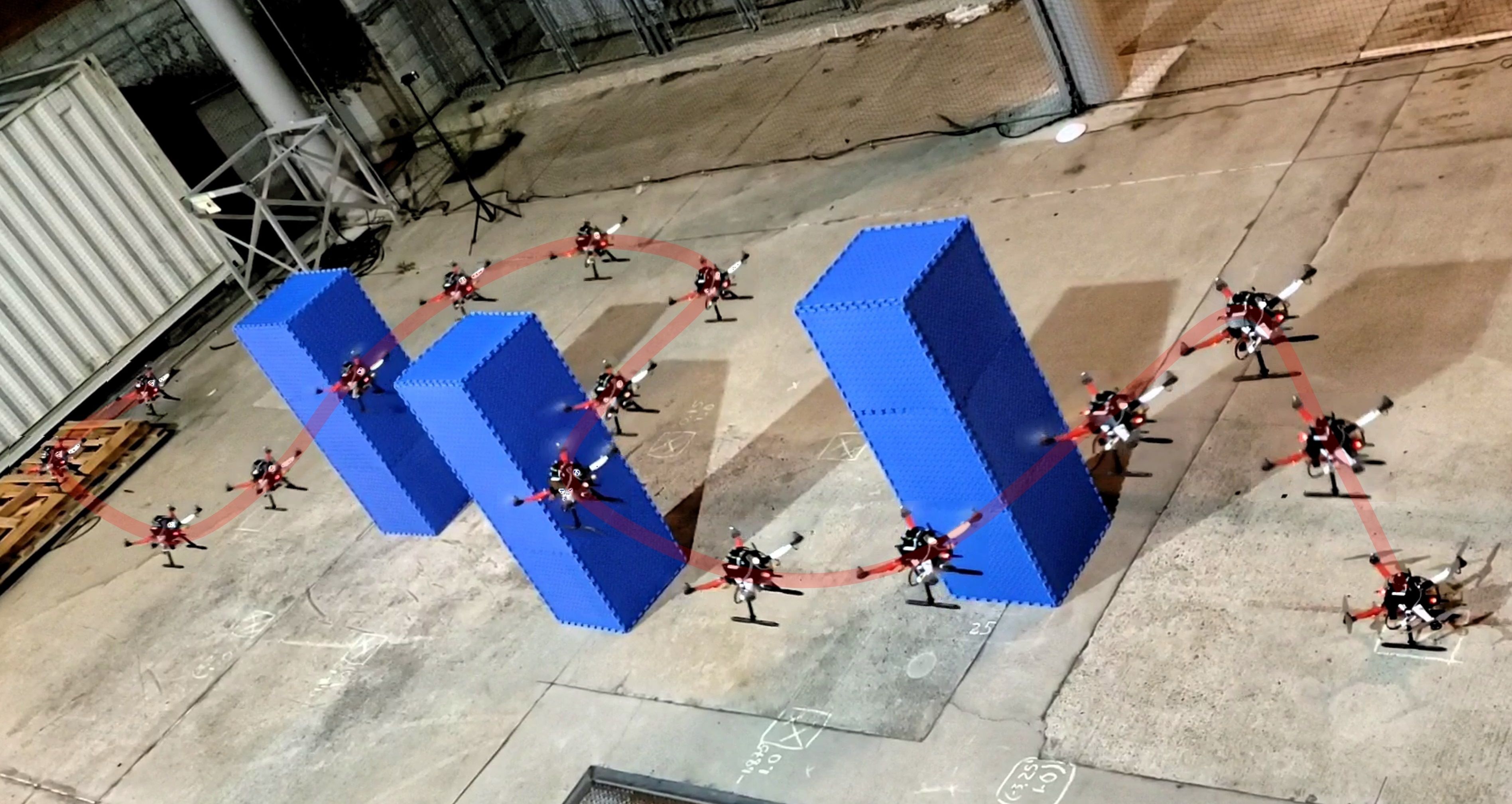}
        \caption{Trajectory tracking with learned model.}
        \label{fig:traj_tracking_nuc}
\end{subfigure}%
\caption{Real experiments with $SE(3)$ port-Hamiltonian neural ODE network \citet{Duong_PHNODE_TRO24}.}
\label{fig:real_exp_se3_ham}
\end{figure}

To illustrate a Hamiltonian neural network for system identification, we provide an example from \citet{Duong_PHNODE_TRO24} where both Hamiltonian structure and the state manifold constraints are encoded in the neural network architecture. Consider a mobile robot, e.g., a quadrotor, with its generalized coordinate $q = (z, R) \in \texttt{SE}(3)$ consisting of its position $z \in \mathbb{R}^3$ and orientation $R\in \texttt{SO}(3)$. Let $\zeta = (v, \omega) \in \bbR^6$ be the generalized velocity, consisting of the body-frame linear velocity $v \in \bbR^3$ and the body-frame angular velocity $\omega\in \mathbb{R}^3$.
The generalized momentum is defined as $p = \begin{bmatrix} {p}_{v} \\ {p}_{\omega} \end{bmatrix} = M(q)\zeta \in \mathbb{R}^6$,
where $M(q)$ is the generalized mass of the system, ${p}_{v}$ is the linear momentum, and ${p}_{\omega}$ is the angular momentum.
%
The Hamiltonian function is defined as the total energy of the system $H(q, p) = \frac{1}{2} p^\top M^{-1}(q)p + V(q)$ where $T = \frac{1}{2} p^\top M^{-1}(q)p$ is the kinetic energy and $V(q)$ is the potential energy.

The equations of motions on the $\texttt{SE}(3)$ manifold are written in port-Hamiltonian form as:
\begin{subequations} \label{eq:portham_dyn_SE3}
\begin{align}
    \;\dot{x} &=\;\;R\frac{\partial{H(q, p)}}{\partial p_{v}}, \label{eq:ham_se3_pos_dot}\\
    \dot{R_i} &=\;\;R_i \times \frac{\partial{H(q, p)}}{\partial p_{\omega}}, \quad i = 1,2,3 \label{eq:ham_se3_rot_dot}\\
    \dot{p}_{v} &=\;\;p_{v}\times \frac{\partial{H(q, p)}}{\partial p_{\omega}} - R^\top \frac{\partial{H(q, p)}}{\partial x} +  b_{v}(q)u \label{eq:ham_se3_pv_dot}\\
    \dot{p}_{\omega} &=\;\;p_{\omega} \times \frac{\partial{H(q, p)}}{\partial p_{\omega}} + p_{v}\times \frac{\partial{H(q, p)}}{\partial p_{v}} +  \label{eq:ham_se3_pw_dot}\\
    &\qquad\sum_{i = 1}^3 R_i \times \frac{\partial{H(q, p)}}{\partial R_i} + b_{\omega}(q)u, \notag 
\end{align}
\end{subequations}
where the input matrix is $B(q) = \begin{bmatrix} b_{v}(q)^\top & b_{\omega}(q)^\top \end{bmatrix}^\top$ and the notation $\times$ denotes the cross product of two vectors.

A data set $\mathcal{D} = \{t_{0:N}^{(i)}, q_n^{(i)}, \xi_n^{(i)}, u^{(i)}\}_{i=1}^D$ is collected by applying a constant control input $u^{(i)}$ to the system and sampling the state at times $t_n^{(i)}$ for $n = 0, \ldots, N$. \citet{Duong_PHNODE_TRO24} uses neural networks with parameters $\theta$ to approximate the mass $M^{-1}_{\theta}(q)$, the potential energy $V_{\theta}(q)$, the dissipation matrix $D_{\theta}(q, p)$, and the input matrix $B_{\theta}(q)$, respectively. The model is trained by minimizing a loss function combining errors in orientation, position, and generalized velocity. The orientation error is measured on \( \mathrm{SO}(3) \) via the logarithmic map, while position and velocity errors use standard Euclidean norms.

A Crazyflie quadrotor, shown in Fig. \ref{fig:pybullet_crazyflie}, simulated in the physics-based simulator PyBullet \citep{gym-pybullet-drones2020} is used to verify the approach. The control input $u = [u'\, \tau^T]^T$ includes a thrust $u' \in \mathbb{R}_{\geq 0}$ and a torque vector $\tau \in \mathbb{R}^3$ generated by the $4$ rotors. The quadrotor is controlled from a random starting point to $18$ different desired poses using a PID controller \citep{gym-pybullet-drones2020}, providing $18$ $2.5$-second trajectories. The trajectories were used to generate a dataset $\mathcal{D} = \{t_{0:N}^{(i)},\mathbf\frakq_{0:N}^{(i)}, \zeta_{0:N}^{(i)}, u^{(i)})\}_{i=1}^D$ with $N = 5$ and $D = 1080$. The $SE(3)$ port-Hamiltonian ODE network \citep{Duong_PHNODE_TRO24}  is trained for $500$ iterations. The training and test results in Fig. \ref{fig:pybullet_exp} show that the neural networks are able to approximate to the ground-truth values. Fig. \ref{fig:pybullet_so3_constraints} and \ref{fig:pybullet_total_energy} verify that when we roll out the learned dynamics for a long horizon, the law of energy conservation is satisfied via the encoded Hamiltonian structure, and the SO(3) constraint violation is negligible. The SE(3) port-Hamiltonian neural ODE networks are also trained with noisy real data collected from a quadrotor (Fig. \ref{fig:nuc_quadrotor}), and tested with trajectory tracking task (Fig. \ref{fig:traj_tracking_nuc}). 
\end{example}

\subsubsection{Monotone Neural Networks}
\label{sec:monotone_neural_networks}
Monotone systems have outputs that change in a consistent direction in response to inputs. 
For a single-input-single-output (SISO) system, there have been papers on SISO monotone neural networks \citep{sill1997monotonic,wehenkel2019unconstrained, liu2020certified,daniels2010monotone,sivaraman2020counterexample,zhang1999feedforward} for learning the dynamical system while preserving the order. 
For multi-input-multi-output (MIMO) systems, 
monotonicity has been studied under a collective manner, i.e., $f(\cdot):\mathbb{R}^n\to\mathbb{R}^n$, for all $x, x' \in \mathbb{R}^n$, $(f(x) - f(x'))^\top (x - x') \ge 0$. In this case,~\citet{cui2024structured} is inspired by the fact that gradients of convex functions are monotone, and constructs MIMO monotone neural networks (MNN) using gradients of strictly convex neural networks (termed \texttt{MNN1} in Figure \ref{fig: voltage_traj}). A strictly convex neural network is parameterized as a $L$-layer FNN with additional residual connection layers that directly pass the input to the hidden layers. With input $l_0 = x$, each layer $l_i$ evolves as,
\begin{equation}\label{eq:ICNN}
l_{i+1}=\sigma \left(W_{i+1} l_i + W_{i+1}^{(x)} x+b_{i+1}\right)\,, \quad i=0,\ldots,L-1,
\end{equation}
where ${\theta}=\left\{W_{1: L}, W_{1: L}^{(x)}, b_{1: L}\right\}$ are neural network weights and biases and $\sigma$ is the non-linear activation function. 
In order for the neural network output to be strictly convex with respect to the input, all $W_{2: L}$ are required to be \emph{positive}, $W_{1:L}^{(x)}$ could be either positive or negative, and the activation function $\sigma$ needs to be strictly convex and increasing, such as the softplus activation function~\cite{glorot2011deep}. Such neural network architecture is also known as the input convex neural network (ICNN)~\cite{amos2017input} in the literature.
\emph{Gradient} of the strictly convex ICNN output with respect to the network input, i.e., $\frac{\partial o}{\partial x}$ satisfies the monotonicity condition. An illustration of this monotone neural network design is presented in Figure \ref{fig:Review_DLdiagram}. 

\citet{wang2024monotone} introduces another MNN architecture of the form $f(x)=\mu x+m(x)$ where $\mu$ is a positive constant (termed \texttt{MNN2} in Figure \ref{fig: voltage_traj}). $m(x)$ is a feed-through neural network, with connections from each hidden layer to the input and output variables. The feed-through neural network is parameterized to guarantee monotonicity and bi-Lipschitzness via the integral quadratic constraint framework and the Carley transform. MNNs have been applied to monotone system modeling, including power system frequency control \citep{cui2024structured, feng2025freq}, power system voltage control \citep{shi2022stability,Feng2023lya}, where monotone controller policies are constructed to guarantee closed-loop stability by design.

\begin{example}[Monotone Neural Network for Power System Model Identification]
\label{ex:MNN_voltage_control}
To illustrate the idea of MNN and standard FNN for monotone system identification, we provide an example for distribution power system identification from \citet{Feng2023lya}. The distribution power network can be represented by a graph $\mathcal{G}=(\mathcal{N},\mathcal{E})$, where $\mathcal{N}:=\{1,2,3,...,n\}$ represents the node set and $\mathcal{E}$ is the edge set. Each node $i\in\mathcal{N}$ is associated with an active power injection $p_i$ and a reactive power injection $q_i$. Let $V_i$ be the complex voltage and $v_i = |V_i|^2$ be the squared voltage magnitude. For any node $j$ as a child node of node $i$, the following equations represent the branch flow model:
\begin{subequations}
\label{eq:nonlinear_powerflow}
\begin{align*}
    -p_j &= P_{ij} - r_{ij} {l_{ij}} - \sum_{k: (j, k) \in \mathcal{E}} P_{jk}, \\
    -q_j &= Q_{ij} - x_{ij} {l_{ij}} - \sum_{k: (j, k) \in \mathcal{E}} Q_{jk}, \\
    v_j &= v_i - 2(r_{ij}P_{ij} + x_{ij} Q_{ij}) + (r_{ij}^2 + x_{ij}^2) l_{ij}, (i,j)\in \mathcal{E},  
\end{align*}
\end{subequations}
where $l_{ij} = \frac{P_{ij}^2 + Q_{ij}^2}{v_i}$ is the squared current, $P_{ij}$ and $Q_{ij}$ represent the active power and reactive power flow on line $(i,j)$, and $r_{ij}$ and $x_{ij}$ are the line resistance and reactance.
With slight violation of notations, we use notations ${p}, {q}$, and ${v}$ to denote the $p_i,q_i,v_i$ stacked into a vector. The system can be approximated by the following linear equation by dropping high-order terms
$${v} = R {p} + X {q} + v_0 {1} = X{q} + {v}^{env},$$
where matrix $R={[R_{ij}]}_{n \times n}, X = {[X_{ij}]}_{n \times n}$ are given as follows,
$R_{ij}:= 2 \sum_{(h, k) \in \mathcal{P}_i \cap \mathcal{P}_j} r_{hk}, X_{ij}:= 2 \sum_{(h, k) \in \mathcal{P}_i \cap \mathcal{P}_j} x_{hk}$, and $\mathcal{P}_i \subset \mathcal{E}$ is the set of lines on the unique path from bus $0$ to bus $i$. $v^{env}$ is considered as a non-controllable constant. A fact is that for radial distribution systems, $R$ and $X$ are positive definite matrices. As a result, $v$ is monotonically increasing with respect to $q$ following the linearized branch flow model. 
Similarly, the nonlinear branch flow also exhibits local monotonicity.

All neural networks have two hidden layers, each layer has 100 hidden units. Both trained MNNs (\texttt{MNN1} and \texttt{MNN2}) achieve MSE values at the scale of $e^{-7}$ ($6.5e^{-7}$ and $3.6e^{-7}$), comparable but slightly worse compared to the FNN's performance $(2.8e^{-7})$. In addition, we observe that training monotone neural networks presents additional challenges relative to standard FNN architecture. Such challenges in training monotone neural networks have also been observed in~\citet{mikulincer2022size}. More broadly, this example illustrates a common tradeoff in control-oriented learning. Incorporating structure into the learning process can provide valuable inductive bias, improving model reliability and interpretability. However, it also limits model expressiveness and can introduce additional optimization challenges. We further discuss such tradeoffs in Section~\ref{sec:futurework_tradeoff}.
\begin{figure}[t]
    \centering
    \includegraphics[scale=0.55,trim=0cm 0.5cm 0cm 0cm]{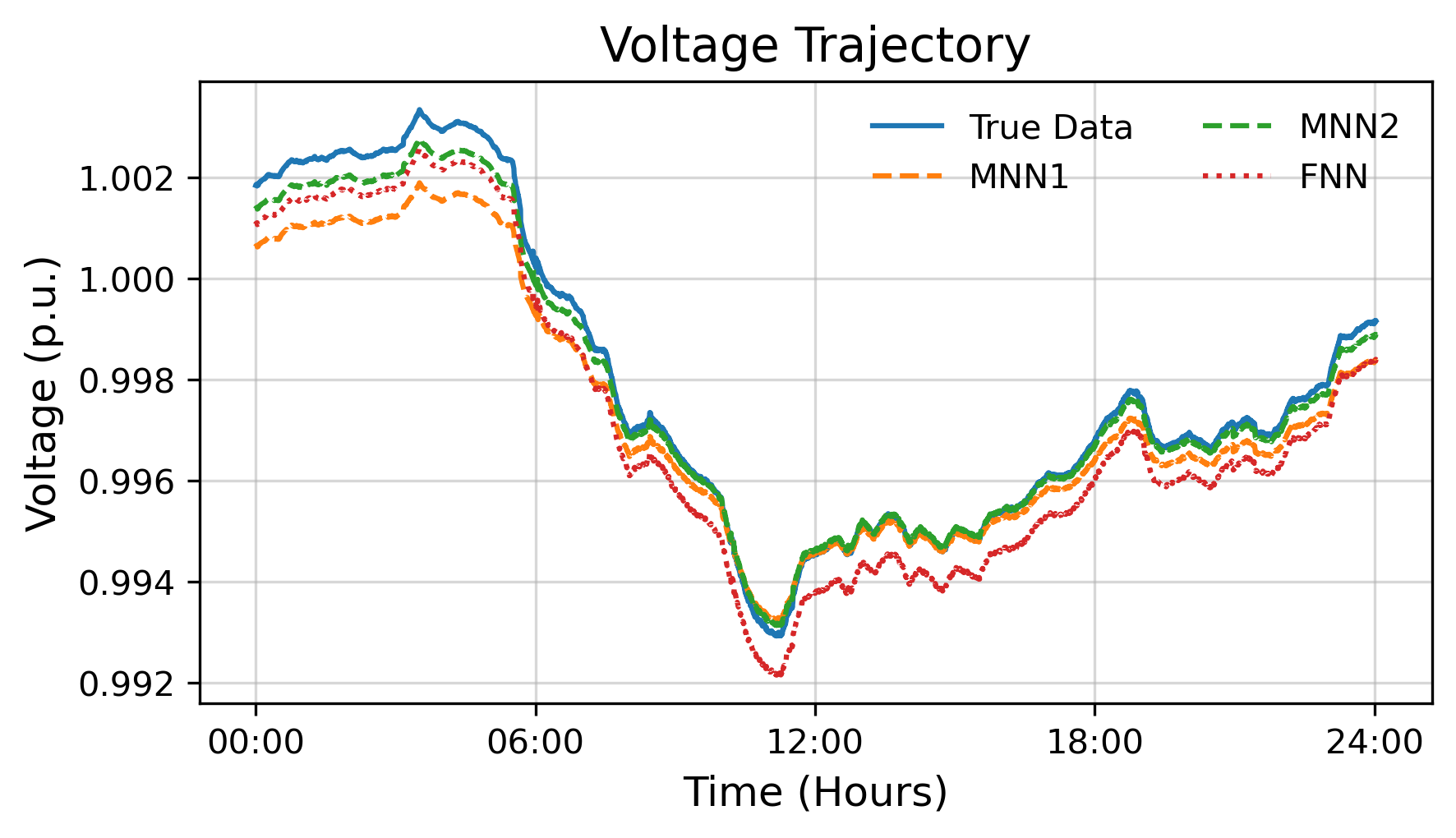}
    \caption{Voltage trajectories of the ground truth using real-world load data, the fit for standard MLP, FTN-based monotone neural network (MNN2), and gradients of input convex neural networks (MNN1) for example \ref{ex:MNN_voltage_control}. 
    }
    \label{fig: voltage_traj}
\end{figure}
\end{example}

\subsection{Hard Constraints}
\label{sec:DL_hard_constraints}
One way to introduce \emph{hard constraints} into learning-based system identification is to first solve the unconstrained system identification problem in \eqref{eq:general_sysid_optimizaiton} and project the solution to the constraint set in \eqref{eq:constrained_sysid_optimizaiton}. 
On this line, \citet{kolter2019learning} proposes to jointly learn a dynamics model 
$f$ and Lyapunov function $V$, both as neural networks, where the dynamics are inherently
constrained to be stable. Whenever the learned $f$ does not meet the Lyapunov stability constraints, it is projected to the following constraint set, 
\begin{align}
\label{eq:stable_constr}
    &\texttt{Proj}\left(\hat{f}(x), \{f: \nabla V(x)^\top f(x) \leq -\alpha V(x)\}) \right) := \nonumber\\
    & \quad \hat{f}(x) - \nabla V(x) \frac{\nabla V(x)^{\top} \hat{f}(x) + \alpha V(x)}{\|\nabla V(x) \|_2^2}.
\end{align}
\citet{lawrence2020almost} extends \citet{kolter2019learning} 
to learn discrete-time systems with disturbance $x_{k+1} = f(x_k, w_{k+1})$, where $w_k \in \mathbb{R}^d$ is a stochastic process. A Mixture density networks $p(x_{k+1}|x_k) = \sum_{i=1}^m \pi_i(x_k) \phi_i(x_{k+1}|x_k)$ is proposed to model the dynamics, where each $\phi_i$ is a kernel function (usually Gaussian), the mean $\hat{\mu}_i(x_i)$ can be parameterized as FNNs and mixing coefficients $\pi_i$ are also FNNs, which are nonnegative and sum to one. The Lyapunov stability condition is posed to the mean parameters, i.e., $V(\mu_{k+1}) \leq \beta V(\mu_k)$ with $\mu_{k+1} = \sum_{i=1}^m \pi_i(x_i) \hat{\mu}_i(x_k)$. 
\citet{takeishi2021learning} generalizes \citet{kolter2019learning} to learn dynamics models with stable invariant sets rather than an equilibrium point, by modifying the projection step to a set and ensuring invariance on the set. 
\citet{schlaginhaufen2021learning} jointly learn a stable system model (using neural ODEs) and a Lyapunov function for partially observable systems with delays. \citet{min2023data} extends this method to a system with control, by jointly learning the dynamical system, Lyapunov function, and a feedback controller all by neural networks, where the closed-loop stability is guaranteed by projection. \cite{tang2024learning} further extends this approach to chaotic systems and learning invariant set.

Dissipativity/passivity extend the concept of Lyapunov stability to
input-output dynamical systems by considering ``energy'' \citep{brogliato2007dissipative}. To incorporate dissipativity as a hard constraint into learned dynamics,  \citet{zhong2020dissipative} introduces a deep learning architecture, Dissipative SymODEN, that is built on structured Port-Hamiltonian dynamics with energy dissipation matrix for constant inputs. It learns four neural nets, representing the inverse of mass matrix, potential energy, the input matrix, and the dissipation matrix respectively, through NeuralODE. \citet{greydanus2022dissipative}, on the other hand, extends Hamiltonian Neural Networks (HNN) \citep{greydanus2019hamiltonian} to Dissipative Hamiltonian Neural Networks (D-HNN) by decomposing dynamic systems into dissipative and conserved quantities so that it accommodates the case where the total energy is not conserved. \citet{drgovna2021physics} focuses on general dissipative discrete-time autonomous dynamical systems and enforces dissipativity by formulating a feed-forward neural network as a point-wise affine map and constraining on the corresponding state-dependent affine coefficients. \citet{xu2023learning}
proposes a method that learns dissipative dynamical systems using neural models while preserving input-output dissipativity by perturbing the weights of neural networks. \citet{okamoto2024learning} leverages the Kalman-Yakubovich-Popov (KYP) lemma, which is a
necessary and sufficient condition for dissipativity to design a differentiable projection that transforms any dynamics represented by neural networks into dissipative ones and jointly learning the transformed dynamics.

\begin{example}[Learning Dissipative Dynamics as Hard Constraints]
\label{ex: learning dissipative neural}
To demonstrate the idea of learning dissipative dynamics as a hard constraint, we consider a mass-spring-damper system, whose nonlinear dynamics are as follows.
\begin{equation*}
    m\ddot{x}(t)+c\dot{x}(t)+(kx(t)+\alpha x(t)^3)=u(t),
\end{equation*}
where $m$ is the mass, $c$ is the damping factor, $k$ is the linear stiffness coefficient, and 
$\alpha$ is a constant that introduces the non-linearity in the spring. We take $m=1, c=1, k=1, \alpha=1$ and set state variables to be $x_1=x$ and $x_2=\dot{x}$. The state-space model is
\begin{equation}
\begin{bmatrix}
\dot{x}_1(t)\\
\dot{x}_2(t)
\end{bmatrix}
=\begin{bmatrix}
x_2(t)\\
-x_2(t)-x_1(t)-x_1(t)^3
\end{bmatrix}
+\begin{bmatrix}
0\\
u(t)
\end{bmatrix}
\end{equation}
We collect step response data with initial point being $[0.1,0.1]$ and introduce normal distributed noises with mean 0 and standard deviation 0.01. Following the method in \citet{xu2023learning}, we first learn a baseline model, which is a two-layer neural network with LeakyReLU activation function on the hidden layer and with 16 neurons for the hidden layer, and then perturb the weights to enforce dissipativity. At this stage, we obtain a dissipative neural dynamical system with $R=0$, $S=\frac{1}{2}I$, and $Q=0$, corresponding to the passive case. For the weight perturbation, we have the results $\lambda_1=10, \lambda_2=10,$ and $\lambda=27.822$. The distance (Frobenius norm) of weights being moved is 3.2608 and 0.267 for the first and second layer respectively (both weights have 64 entries in total). Then we retrain biases and obtain the final model. We demonstrate the ground truth, the fit for the baseline model, the model after weights perturbation, and the final model in Fig. \ref{fig: mass spring damper nonlinear} . We observe the final model closely matches the ground truth and retains closeness to the baseline model while guaranteeing incremental dissipativity.
\begin{figure}
    \centering
\includegraphics[scale=0.55,trim=0cm 0.5cm 0cm 0cm]{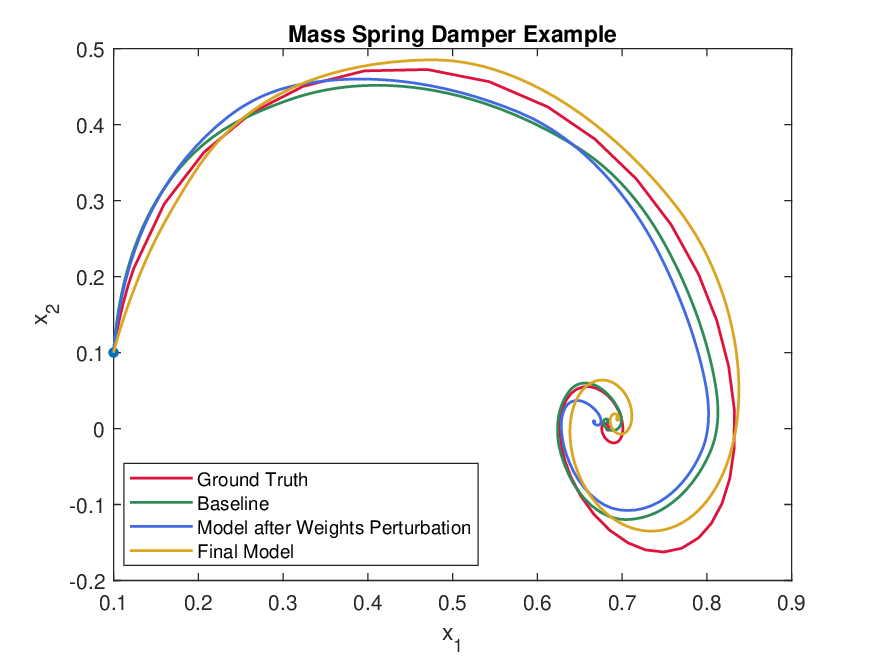}
    \caption{Trajectories of the ground truth, the fit for baseline model, model after weights perturbation and final model for Example \ref{ex: learning dissipative neural}}
    \label{fig: mass spring damper nonlinear}
\end{figure}
\end{example}

\subsection{Soft Constraints}
\label{sec:PINN_systemID}
Control-relevant properties can be incorporated into deep learning based system ID approaches as a soft penalty. There have been a large volume of literature on using soft penalty to incorporate the above properties, such as stability, dissipativity, passivity, physics and conservation laws. 

A primary example in this category is the physics-informed neural networks (PINNs). PINNs \citep{raissi2019physics} are deep learning models that integrate physical principles into the neural network architecture. 
Consider we have some prior knowledge about the dynamical system in the form of 
$P(x, u, y) = 0$.
PINNs use both a data-driven loss and a physics-informed loss for learning the system model $f_{\theta}$ and $h_{\theta}$. 
The data-driven loss term is defined as in \eqref{eq:constrained_sysid_optimizaiton}, 
    \[c_{data}(\theta; \mathcal{W}) = \sum_i \int_t \|y_i(t) - \hat{y}_i\|^2 \]
and the physics-based loss/regularization term is defined as,
    \[c_{physics}(\theta; \mathcal{W}) = \sum_i \int_t P(\hat{x}_i(t), u_i(t), y_i(t))^2. \]
Combining the data-driven and physics-based loss terms forms the total loss function $c(\theta; \mathcal{W}) = c_{data}(\theta; \mathcal{W}) + \lambda\,  c_{physics}(\theta; \mathcal{W})$, where $\lambda \, c_{physics}(\theta; \mathcal{W}):=r(\theta)$ is the soft regularization term to enforce the physics law. 
PINNs are trained to minimize $c(\theta; \mathcal{W})$ with respect to parameters \(\theta\).

{  It is also useful to distinguish PINNs from physics-guided neural networks (PGNNs). In many PINN formulations, physical laws are incorporated through residual terms in the training objective, and therefore fit naturally within the soft-constraint viewpoint. In contrast, PGNNs typically combine a physics-based model component with a trainable black-box neural component. When this combination is designed so that a control-relevant property holds by construction, PGNNs are closer in spirit to the direct-parameterization viewpoint described in Section \label{sec:ways_incorporating_physics}. For example, \citet{bolderman2024physics} propose a PGNN architecture for feedforward control in which a physics-based layer and a neural layer are combined to obtain input-to-state-stability guarantees through Lipschitz-based conditions.}

As soft constraints, PINNs can be incorporated into data-driven system identification following two general frameworks: i) By posing a certain structural prior about the system to be identified, we can turn the system identification problem into a {parameter identification} problem. A simple example is to assume that the system to be identified follows the linear time-invariant system $\dot{x} = Ax + B u$ and to identify $A, B$ matrices from the data. \citet{richards2023learning} extends this idea to learn linear time-varying system models. ii) By posing certain physics laws or/and conservation laws into system identification rather than directly assuming the model parametric forms. 
A common conservation law that can be used in dynamical system learning is the energy conservation law. A benchmark is available in \cite{zhong2021benchmarking} for the design of energy-conserving PINNs that include both Lagrangian neural networks and Hamiltonian neural networks. Until now, PINN-based modeling algorithms have been successfully incorporated in learning Lagrangian mechanics \citep{roehrl2020modeling}, robotics dynamics \citep{NICODEMUS2022pinn,liu2024physics},  building dynamics \citep{gokhale2022physics}, rotor dynamics \cite{LIU2024307}, power system dynamics \citep{Huang2023pinn}, etc.. PINNs also have achieved great success in modeling dynamical systems governed by PDEs such as fluid mechanics \citep{mowlavi2023optimal}, see a recent review by \citet{faroughi2024physics}. 

\begin{figure}
    \centering
    \includegraphics[width=\linewidth]{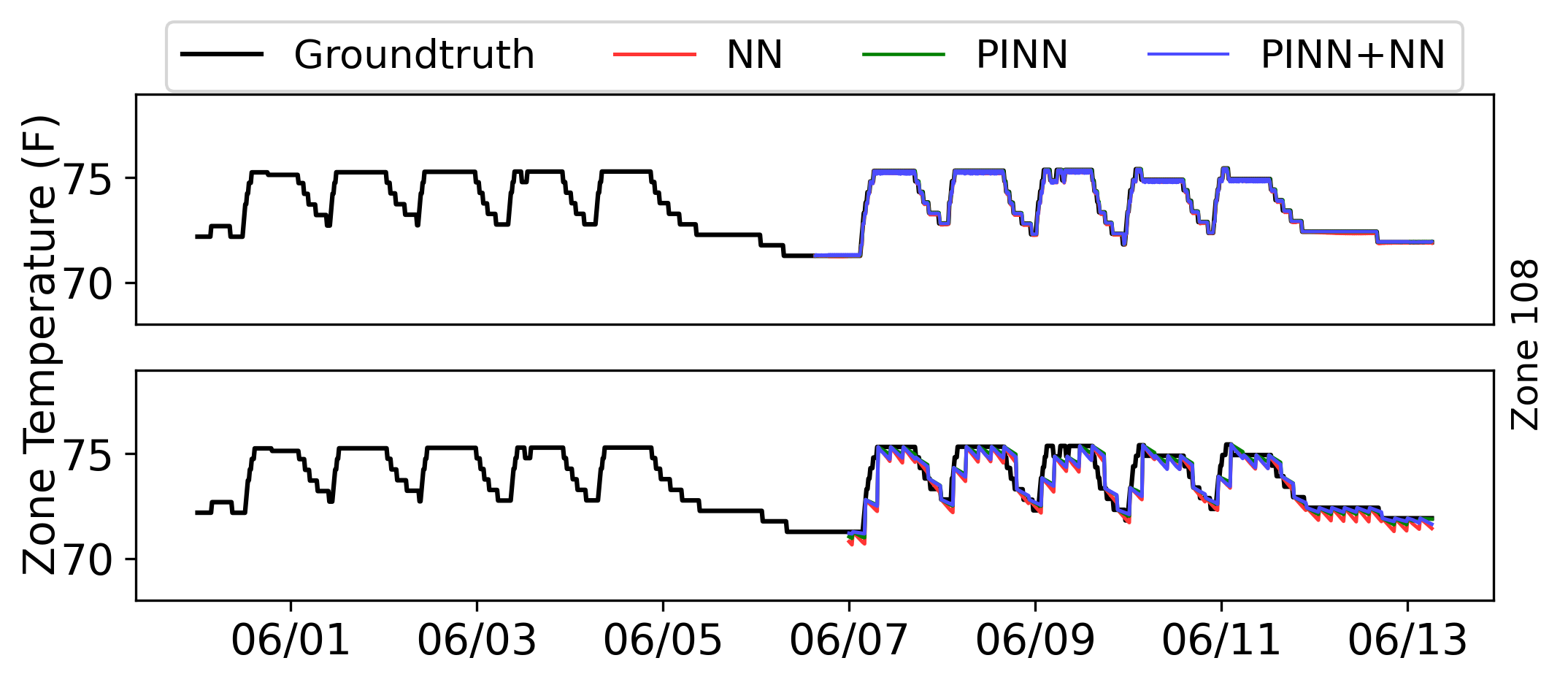}
    \caption{Temperature predictions for 15-minute (top) and 3-hour (bottom) horizons, using one week data for training and the rest for testing.}
    \label{fig:PINN_buildings}
\end{figure}

\begin{example}[PINNs for Building System Identification]
\label{ex:PINN_building}
To illustrate the idea of PINNs for system identification, we provide an example of a dynamic identification building from \citet{bian2023bear}. For PINN design, the physics knowledge is incorporated by directly posing a linear resistor-capacitor model structure, where the heat transfer between zone $i$ and zone $j$ is modeled as a resistor coefficient $R_{ij}$ and the heat generation/absorbing within zone $i$ is modeled as a capacitor coefficient $C_i$. The thermal dynamics in each zone follows,
{\small $$C_i \dot{x}_i= \frac{x_o(t)-x_i(t)}{R_{oi}} + \sum_{j\in \mathcal{N}_i}\frac{x_j(t)-x_i(t)}{R_{ji}}+ c_p u_i(t)(x_s(t)-x_i(t))+p_i\,, \text{PINN}$$}
where $x_i(t)$, $x_o(t)$, $x_s(t)$ are the indoor temperature of zone $i$, the outdoor temperature, and the temperature of supply air. $\mathcal{N}_i$ is the set of rooms that are adjacent to room $i$, $c_p$ is the specific heat capacity of the air, and $p_i$ is the internal heat gain. $u_i(t)$ denotes the supply air flow rate, which is the control action.
Parameters $\{C_i,\,R_{oi},\,R_{ji},\,c_p,\,p_i\}$ are to be identified from data through the PINN training.

In comparison, a standard NN model, as defined in Sec \ref{sec:FNN}, without prior knowledge can be learned,
$$\dot{x}_i=f_{\theta_i}(x_i(t), u_i(t),x_{\mathcal{N}_i}(t),x_o(t),x_s(t))\,, \quad \text{NN}$$
where $\theta_i$ are the NN parameters to be learned for zone $i$. 

It is also interesting to consider a combined approach that combines a PINN and a NN, 
{\small $$C_i \dot{x}_i=\frac{x_o(t)-x_i(t)}{R_{oi}} + \sum_{j\in \mathcal{N}_i}\frac{x_j(t)-x_i(t)}{R_{ji}}+f'_{\theta_i}(x_i(t), u_i(t), x_s(t))\,, \text{PINN + NN}$$}
where the learnable parameters are the combination of the NN parameters $\theta_i$ and PINN parameters $\{C_i,\,R_{oi},\,R_{ji}\}$. 

We train the three models to learn the thermal dynamics of one building zone. Two forecasting horizons are considered: 15 minutes (short-term prediction) and 3 hours (long-term prediction). The training data includes 7-day periods at a 15-minute resolution. Prediction results are presented in Figure \ref{fig:PINN_buildings}. For the 15-minute horizon, all methods achieve high accuracy with errors below 0.02 degrees Fahrenheit. For the 3-hour horizon, incorporating physics priors enables the PINN and the combined approach to outperform the standard FNN, with PINN delivering the best test performance across both horizons.
\end{example}

We end this subsection by comparing incorporating control-relevant properties via soft and hard constraints. We consider a case study of learning a stable dynamical system with Lyapunov stability constraint. 
\begin{example}[Soft v.s. Hard Constraints for Learning Stable Pendulum Dynamics]
Consider the following stable pendulum dynamics from \citet{kolter2019learning}, 
$\ddot{x}=-b\dot{x}-g\sin{x}$ 
where $x$ denotes the angular position, $\dot{x}$ is the angular velocity, and  $\ddot{x}$ is the angular acceleration.
$b$ is the damping coefficient, and $g$ is the gravity constant. 

For incorporating a stability constraint, two neural networks are learned simultaneously: a two-layer NN model $f$ for learning the dynamics, and a neural Lyapunov function $V$. The neural Lyapunov function is designed as $V(x)=\sigma(g(x)-g(0))+\epsilon \lVert x\rVert_2^2$, where $\sigma$ is a smoothed ReLU to make the Lyapunov function continuously differentiable, $g(x)$ is parameterized as an input convex neural network to avoid local optima, and the regularization term ensures strict positive definiteness. 
To jointly learn the neural dynamics $f$ and the Lyapunov function $V$, $N=10^4$ data pairs $(x_i, \dot{x}_i)$ are generated using the true dynamics. We compare two schemes for incorporating the stability constraint.

i) \emph{Hard constraint:} The loss function for jointly learning $f, V$ is defined as, 
$$c_1(f, V)=\sum_i\lVert \dot{x}_i-\hat{f}(x_i)\rVert_2,$$
where $\hat{f}$ is defined by the projection in \eqref{eq:stable_constr} to ensure hard constraint satisfaction of Lyapunov stability. 

ii) \emph{Soft penalty:} The loss function for jointly learning $f, V$ is defined as,
$$c_2(f, V)=\sum_i \sum_i\lVert \dot{x}_i-f(x_i)\rVert_2 +\sum_i\lambda\max{(0,\Delta V(x_i))},$$
where $f(x_i)$ is the estimated dynamics (without projection), $\lambda$ is a balancing coefficient, and $\Delta V(x_i)=V(f(x_i))-V(x_i)$ 
represents the difference of the Lyapunov energy decay. For stable systems, $\Delta V(x_i)<0$. Otherwise, it would incur a penalty to guide learning of $V, f$ towards constraint satisfaction.
\begin{figure}
  \centering
    \begin{subfigure}[b]{0.24\textwidth}
    \includegraphics[width=\textwidth]{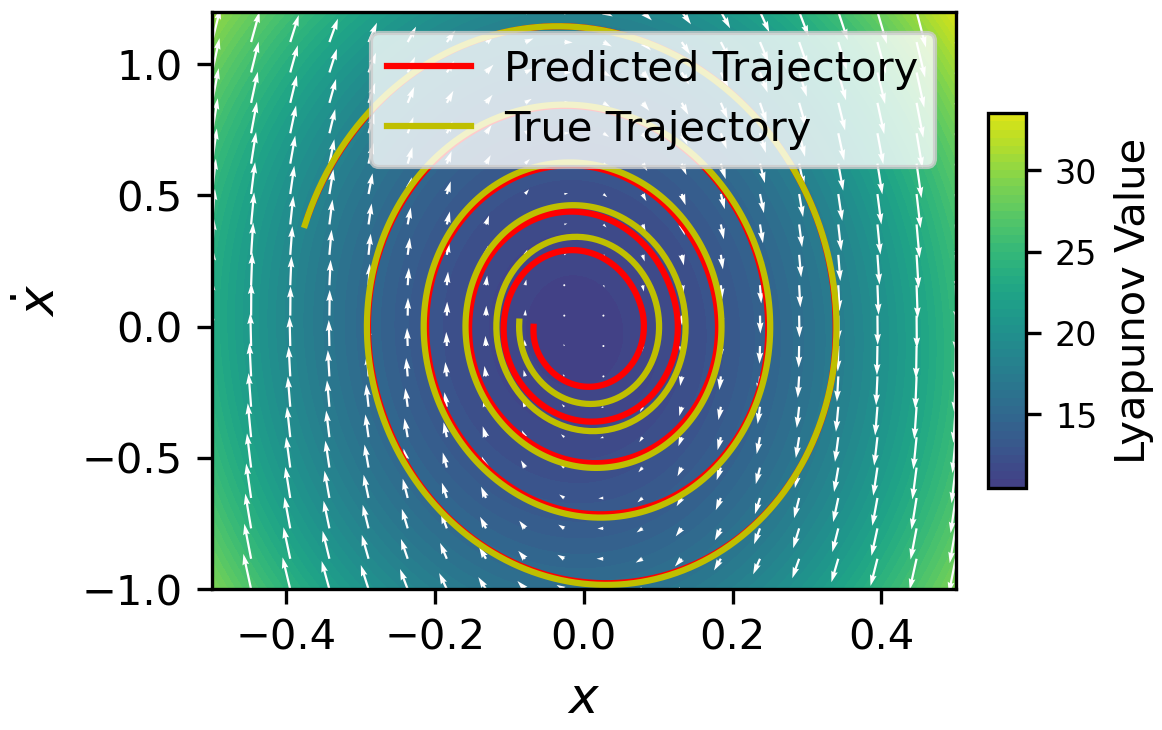}
    \caption{Stability enforced by projection.}
    \label{fig:Lyapunov_soft}
  \end{subfigure}
  \begin{subfigure}[b]{0.23\textwidth}
    \includegraphics[width=\textwidth]{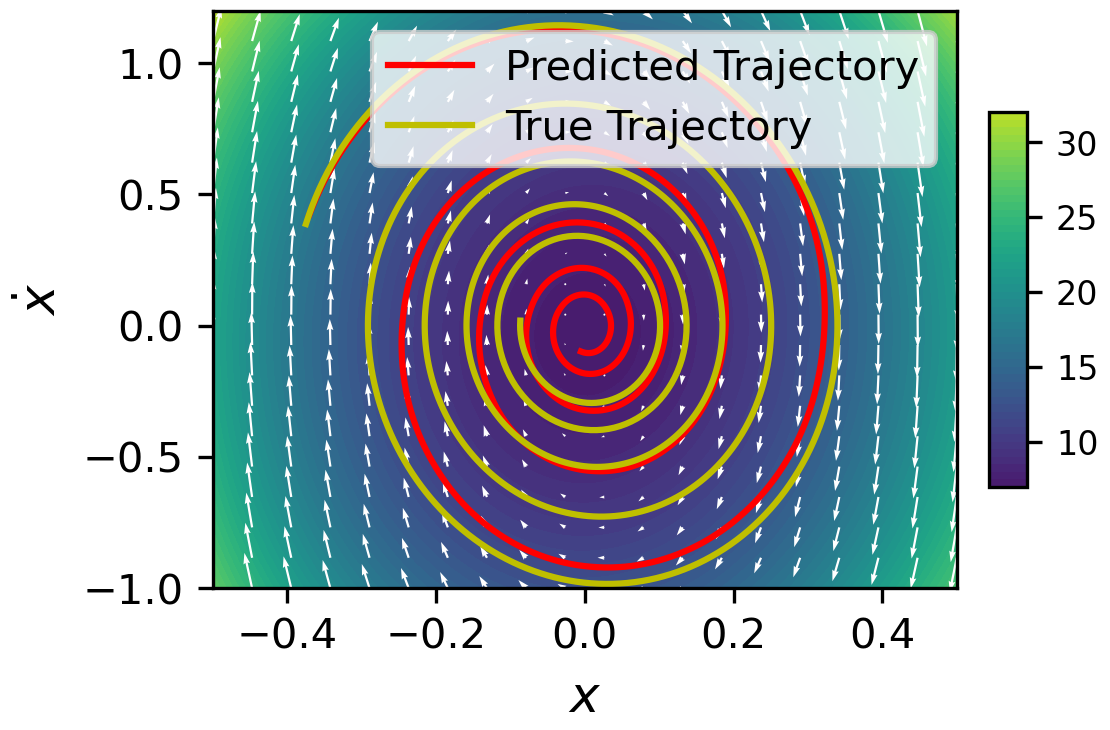}
    \caption{Stability incorporated by soft-penalty.}
    \label{fig:Lyapunov_proj}
  \end{subfigure}
  \hfill 
  \caption{Comparison of hard projection and soft penalty approach for learning stable dynamics with neural dynamics and a neural Lyapunov function. }
  \label{fig:Lyapunov}
\end{figure}
We show the trained Lyapunov function contour (after 2,000 training steps) and predicted trajectories in Figure \ref{fig:Lyapunov}. The background color represents the value of the learned Lyapunov function, with the true dynamics shown in gold and the predicted trajectory in red. Both learned trajectories exhibit stable behavior, while the trajectory with stability enforced by projection achieves higher accuracy. This might be due to the projection provides a stronger prior compared to regulation terms, and the soft penalty adds complexity to the optimization landscape. 
\end{example}

\subsection{Discussions}
Despite the remarkable flexibility of deep learning models for system identification, several fundamental challenges and significant opportunities for future research remain. 

One major difficulty lies in imposing hard constraints—such as stability, passivity, or physics knowledge—during training, as these constraints are often nonconvex and make the learning process difficult. This calls for the development of advanced optimization algorithms that can account for such constraints and guide learning toward meaningful and feasible solutions. 

In parallel, there is a growing need to design specialized neural network architectures to achieve a balance between structural constraints and model expressiveness. On the one hand, incorporating structural priors or leveraging physics-inspired parameterizations can help improve generalization and model interpretability. However, as illustrated in Example \ref{ex:MNN_voltage_control}, these constraints may also limit model expressiveness and performance, underscoring the trade-off between enforcing physical properties and maintaining flexibility in function approximation. Finally, a promising direction lies in the data-driven \emph{discovery} of system properties. The learned models not only fit the observed trajectories but also uncover underlying system properties—such as conservation laws, stability, controllability, and causal dependencies—that are often difficult to characterize without an exact system model. This aspects will be further discussed in the next section on direct identification and verification of control-relevant properties.

\section{Direct Identification and Verification of Control-Relevant Properties}
\label{sec:direct}
A recently emerging area within control arises from behavioral system theory \citep{behavioral_system_theory}, where a representation-free perspective is adopted, and a dynamical system is described through a collection of its input-output trajectories, rather than a transfer function or state-space model. 
Control-relevant properties like dissipativity can also be extended to this setting, which allows a verification of these properties directly from measured data; this problem has been studied under the broad umbrella of `data-driven system analysis'.

Motivated by these developments, the viewpoint in this section is different compared to the previous sections: 
we are not explicitly interested in identifying a transfer function or state space model for the system. Instead, we aim to verify from a behavioral systems perspective, that is based on measured system trajectories, whether the underlying unknown system satisfies a certain control-theoretic property, or to check whether the set of all systems that could generate the observed data satisfies a control-relevant property like dissipativity. 
In particular, we are interested in verification methods with rigorous guarantees regarding the verified system property that can be translated to closed-loop settings.

\subsection{Control-relevant Properties in the Behavioral Framework}\label{Sec_Behav}
In this section, we review the direct verification of dissipativity \citep{SA_Maupong, OneShot} and more general input-output properties \citep{IQC_Anne} for LTI systems from data based on Willems' fundamental lemma \citep{Willems} as described in Section \ref{sec: sys_modeling}. Since the fundamental lemma characterizes finite horizon trajectories of systems, \cite{SA_Maupong} introduce the notion of $L$-dissipativity.
\begin{definition}[$L$-dissipativity]\label{LDissiDef}
The discrete-time LTI system \eqref{LTI_DT} is $L$-dissipative with respect to the supply rate $s(u,y)=\begin{bmatrix}u\\y\end{bmatrix}^\top \Pi\begin{bmatrix}u\\y\end{bmatrix}$ if
\begin{equation}\label{L_dissi_cond}
	\sum_{k=0}^{L-1}\begin{bmatrix}u_k\\y_k\end{bmatrix}^\top  \Pi\begin{bmatrix}u_k\\y_k\end{bmatrix}\geq 0,
\end{equation}
for all input-output trajectories $\{u_k,y_k\}_{k=0}^{L-1}$ of \eqref{LTI_DT} with initial condition $x_0=0$.
\end{definition} 
{ Note that the dissipativity in Definition~\ref{LDissiDef} is given in an input-output setting and thus does not explicitly include a storage function. Its connection to the classical dissipativity definition in Section 2.3.2 can be understood as follows. In discrete time, classical dissipativity requires \(\sum_{k=0}^{L-1}s(u_k,y_k)\geq V(x_L)-V(x_0)\) for a nonnegative storage function \(V\). For trajectories with \(x_0=0\) and \(V(0)=0\), this condition implies \(\sum_{k=0}^{L-1}s(u_k,y_k)\geq 0\), since \(V(x_L)\geq 0\). Thus, (75) can be viewed as a finite-horizon, zero-initial-state input-output counterpart of the classical dissipativity inequality. The converse, namely when such input-output inequalities imply the existence of an appropriate storage function, requires additional conditions; for a detailed treatment of this relationship between state-space and input-output dissipativity notions, we refer { the reader} to \cite{Hill1980DissipativeDS}.} 

The key idea to verify $L$-dissipativity from a single measured trajectory is to parameterize all trajectories of length $L$ of LTI system \eqref{LTI_DT} with $x_0=0$, as required in condition~\eqref{L_dissi_cond}, by the fundamental lemma \eqref{WFL}. By the resulting condition on the parameter $\alpha$ together with the application of Finsler's lemma, \citet[(Theorem 2)]{OneShot}  provides a computationally appealing linear matrix inequality (LMI) to check whether an LTI system is $L$-dissipative based on a single measured trajectory. Extensions of this result include data-driven dissipativity verification for LPV systems \citep{LPV_dissi} using the fundamental lemma for LPV systems \citep{LPV_FundamentalLemma}. Furthermore, for linear systems, \cite{IQC_Anne} investigates the determination of optimal input-output properties characterized by time domain IQCs.  

Inspired by the fundamental lemma, \cite{PersisLinear} establishes a direct data-driven closed-loop characterization of an LTI system, which builds the basis for a data-driven state-feedback design. Furthermore, this result has led to direct data-driven state-feedback design for polynomial systems \citep{DePersis1}, for Lur'e-type systems \citep{Persis_quad_constr}, for LPV systems \citep{LPV_direct}, and for nonlinear systems via LPV embedding \citep{LPV_NL} and via nonlinearity cancellation \citep{NonlinearityCancell}. Moreover, the fundamental lemma is suitable for data-driven simulation and output matching control \citep{Markovsky}, data-driven predictive control with guarantees for LTI systems \citep{deepc,DDMPC}, LPV systems \citep{LPV_PC}, feedback linearizable systems \citep{FeedbackLin}, bilinear systems \citep{yuan2022data}, nonlinear systems \citep{BerberichNL}, and system level synthesis \citep{SLS}. 

We conclude that the application of Willems' fundamental lemma enables a simple and direct verification of dissipativity and related system properties from a single input-output trajectory. However, the analysis via the fundamental lemma calls for noise-free measurements, is restricted to finite horizon, and an extension to general nonlinear systems is unknown. {  Recent work has  begun to investigate nonlinear extensions of this viewpoint; for example, \citet{lazar2025product} uses a product Hilbert space framework to develop a generalized Koopman operator and a nonlinear analogue of the fundamental lemma.}

\subsection{Set-Membership Approaches}
\label{Sec_SetMem}
Set-membership approaches \citep{Set_memerbship_Lin} aim to determine the set of all systems that could generate the observed data. Since this feasible system set contains the underlying system, analyzing all systems within this set yields guarantees for the true system even if it cannot be identified due to, e.g., noisy data. Set theoretic methods for control, set membership based identification, and broadly, identification under uncertainty and identification for robust control, have a long history (see books such as \cite{blanchini2008set,milanese2013bounding,chen2003control} and surveys such as \cite{makila1995worst,milanese1991optimal}). 
\emph{Given the wide scope of set membership approaches, we do not consider general set membership based system identification and control here; rather, we focus our attention on set membership approaches for direct identification of control-relevant properties like dissipativity, stability, and monotonicity.}

For the ease of presentation, we focus on unknown LTI systems and refer afterwards to possible extensions in various directions. Consider the discrete-time LTI system
\begin{equation}\label{SS_LTI}
    x_{k+1}=A^*x_k+B^*u_k,
\end{equation}
with state $x_k\in\mathbb{R}^{n_x}$ and input $u_k\in\mathbb{R}^{n_u}$. To infer on the unknown system matrices $A^*$ and $B^*$, we assume input-state samples $\{x_k,u_k\}_{k=0}^{N-1}$ are given with $x_{k+1}=A^*x_k+B^*u_k+d_k,k=0,\dots,N-2$. The unknown disturbance vector $d_k$ can incorporate process noise, nonlinear dynamics, or effects from inexact state measurements. We suppose that the disturbance is bounded $||d_k||_2\leq\epsilon,k=0,\dots,N-2$, with known bound $\epsilon>0$. Note that the samples can also stem from multiple trajectories.

Under the given data and the pointwise-in-time noise characterization, we can infer on the set of system matrices 
\begin{equation}\label{Set_mem}
    \Sigma=\left\{\begin{bmatrix}A & B\end{bmatrix}: \bigg|\bigg|x_{k+1}-\begin{bmatrix}A & B\end{bmatrix}\begin{bmatrix}x_k\\u_k\end{bmatrix}\bigg|\bigg|_2\leq\epsilon,i=0,\dots,N-2\right\}
\end{equation}
consistent with the data. Note that the true matrices $\begin{bmatrix}A^* & B^*\end{bmatrix}$ are contained within $\Sigma$ as they satisfy $\bigg|\bigg|x_{k+1}-\begin{bmatrix}A^* & B^*\end{bmatrix}\begin{bmatrix}x_k\\u_k\end{bmatrix}\bigg|\bigg|_2\leq\epsilon,i=0,\dots,N-2$. While the set membership \eqref{Set_mem} could directly be exploited to verify dissipativity following \cite{Berberich} or \cite{MartinPoly}, we will consider here an ellipsoidal outer approximation of \eqref{Set_mem} given by
\begin{equation}\label{Set_mem2}
    \tilde{\Sigma}=\left\{\begin{bmatrix}A & B\end{bmatrix}: \begin{bmatrix}I\\ \begin{bmatrix}A & B\end{bmatrix}\end{bmatrix}^\top  \Delta\begin{bmatrix}I\\ \begin{bmatrix}A & B\end{bmatrix}\end{bmatrix}\preceq0\right\}\supseteq \Sigma,
\end{equation}
as proposed in \citet[Proposition 1]{MartinIQC}. The computation of matrix $\Delta$ boils down to an SDP with an LMI constraint because $\Sigma$ corresponds to an intersection of $N-1$ quadratic constraints \citep{Boyd}. Thereby, $\tilde{\Sigma}$ is described by a single quadratic matrix inequality. However, this adds an additional conservatism due to a non-tight description of $\Sigma$. Since the set of feasible system matrices is characterized by the quadratic uncertainty description \eqref{Set_mem2}, we can analyze the set of linear systems with system matrices $(A,B)$ where 
\begin{equation}\label{FSS}
    x_{k+1}=Ax_k+Bu_k,\quad \begin{bmatrix}A & B \end{bmatrix}\in\tilde{\Sigma},
\end{equation}
using modern robust control techniques. For that purpose, we equivalently write \eqref{FSS} as a linear fractional representation \citep{SchererLMI}
\begin{equation}\label{LFT}
\begin{bmatrix}
	x_{k+1}\\q_k
\end{bmatrix}=\begin{bmatrix}0 & 0 & I \\ \begin{bmatrix}I\\0\end{bmatrix} & \begin{bmatrix}0\\I\end{bmatrix} & 0 
\end{bmatrix}\begin{bmatrix}
x_k\\ u_k\\ w_k
\end{bmatrix}
\end{equation}
with uncertainty $w_k = \begin{bmatrix}A & B\end{bmatrix} q_k$ satisfying the quadratic constraint
\begin{equation*}
\begin{bmatrix}q_k\\w_k\end{bmatrix}^\top  \Delta\begin{bmatrix}q_k\\w_k\end{bmatrix}\preceq0.
\end{equation*}
Now, all systems of \eqref{FSS} are $(Q,S,R)$-dissipative for the storage function $V(x)=x^\top  Xx,X\succeq0$, on $\mathcal{X}\times\mathcal{U}=\mathbb{R}^{n_x+n_u}$ if
\begin{equation}\label{DissiIneq_S}
    w^\top  Xw-x^\top  Xx-\begin{bmatrix}y\\u\end{bmatrix}^\top  \begin{bmatrix}Q & S\\ S^\top   & R\end{bmatrix}\begin{bmatrix}y\\u\end{bmatrix}\preceq0,\quad \forall w:\begin{bmatrix}q\\w\end{bmatrix}^\top  \Delta\begin{bmatrix}q\\w\end{bmatrix}\preceq0.
\end{equation}
The infinitely many constraints \eqref{DissiIneq_S} are implied by 
\begin{equation*}
    w^\top  Xw-x^\top  Xx-\begin{bmatrix}y\\u\end{bmatrix}^\top  \begin{bmatrix}Q & S\\ S^\top   & R\end{bmatrix}\begin{bmatrix}y\\u\end{bmatrix}-\alpha\begin{bmatrix}q\\w\end{bmatrix}^\top  \Delta\begin{bmatrix}q\\w\end{bmatrix}\preceq0,
\end{equation*}
for any $\alpha\geq0$. This technique is well-know in robust control as the S-procedure \citep{Boyd}. With $q=\begin{bmatrix}x\\u\end{bmatrix}$ and $y=Cx+Du$, we can write equivalently
\begin{align}\label{LMI_condidtion}
    \begin{bmatrix}x\\u\\w\end{bmatrix}^\top  \underbrace{\phi^\top  
	\begin{bmatrix}\begin{array}{cc|cc|c}
	-X & 0 & 0 & 0 & 0\\
    0 & X & 0 & 0 & 0 \\\hline
    0 & 0 & -Q & -S & 0\\
    0 & 0 & -S^T & -R & 0\\\hline
    0 & 0 & 0 & 0 & -\alpha\Delta
	\end{array}	\end{bmatrix}	
	\phi}_{=\Omega}    
    \begin{bmatrix}x\\u\\w\end{bmatrix}\preceq0,
\end{align}
with
\begin{equation*}
    \phi = \begin{bmatrix}\hspace{0.15cm}\begin{matrix} I & 0 & 0\\0 & 0 & I  \\\hline
	C & D &0 \\ 0 & I & 0  \\\hline
	I & 0 & 0 \\ 0 & I & 0\\ 0 & 0 & I 
	\end{matrix} \end{bmatrix}.
\end{equation*}
Therefore, if there exist a matrix $X\succeq0$ and a scalar $\alpha\geq0$ such that $\Omega\preceq0$, then all systems within the set membership \eqref{Set_mem}, including the true system \eqref{SS_LTI}, are dissipative. Note that $\Omega$ is linear with respect to $X$ and $\alpha$. Thus, suitable values can be efficiently found by standard LMI solvers \citep{YALMIP}. Similar results can be be found in \cite{AnneDissi} exploiting the LMI-based robust control framework of \cite{SchererLMI}. Moreover, \cite{WaardeDissi} exploits a matrix S-lemma to verify dissipativity for the feasible system set \eqref{FSS}.

In contrast to the behavioral approach in Section~\ref{Sec_Behav}, the set-membership approach enables dissipativity determination from noisy data and over an arbitrary time horizon, as required for the application of feedback laws \citep{Khalil}. 
Set membership approaches have also been extended to various model classes beyond LTI systems and control-relevant properties like stability and monotonicity as discussed in the sequel.

 \textit{Polynomial dynamics:} For systems with polynomial dynamics, common in applications such as fluid mechanics \citep{Polynomial_sys} and robotics \citep{Poly_Robotic}, the dynamics can be expressed as $x_{k+1}=Fz(x_k,u_k)$ with unknown coefficient matrix $F$ and a known vector of monomials $z(x_k,u_k)$. Dissipativity can then be verified using sum-of-squares relaxations, leading to an SDP formulation \citep{MartinPoly}. More generally, input-output properties such as integral quadratic constraints (IQCs) can also be verified using this representation \citep{MartinIQC}.
 
 \textit{General nonlinear dynamics:} While linear and polynomial systems directly lead to a set membership suitable for an analysis by SDPs, general nonlinear systems yield non-convex optimization problems, even in the model-based case. To circumvent the non-convexity and to provide guarantees for the verification of system properties, the nonlinear system behavior can be embedded into a system representation suitable for an analysis using SDPs as summarized in the survey \cite{TMSurvey}. For example, dissipativity has been verified using Taylor-based approximations \citep{MartinTP2}, piecewise polynomial models \citep{MartinTP1}, or embedding into linear parameter-varying (LPV) representations \citep{LPV_set_membership,LPV_NL}. 
 \textit{General control-relevant properties:} Set-membership based identification has also been extended to learn systems that are known to satisfy control-relevant properties. For example, works such as \cite{cerone2011enforcing,lauricella2020set,lauricella2020thesis,tobenkin2013stable} develop set membership approaches explicitly incorporating a priori information on
system stability in an error-in-variable (EIV)
framework. Set-membership based identification for monotone dynamical systems has also been considered in the literature \citep{makdesi2023data}. In fact, exploiting a priori knowledge on monotonicity or dissipativity of the system has also been shown to improve the accuracy and data efficiency of set-membership approaches \citep{ramdani2006set,Berberich}.

Summarizing, the set-membership perspective combined with modern robust control techniques establishes a flexible framework for verifying system properties for linear and nonlinear systems from noisy data with guarantees. Related to this approach, the data-informativity framework \citep{Inform_Survey} provides conditions on when the collected samples are informative to infer system properties as controllability or stabilizability. We now present an example to demonstrate direct data-driven verification of dissipativity from noisy data using set-membership approaches.
\begin{example}
 We consider a mass-spring-damper system $m\ddot{y}(t)+d\dot{y}(t)+Ky(t)=u(t)$ with position of the mass $y(t)$, external force $u(t)$, mass $m=1\,\mathrm{kg}$, damping coefficient $d=1\,\frac{\mathrm{kg}}{\mathrm{s}}$, and stiffness $K=1\,\frac{\mathrm{kg}}{\mathrm{s}^2}$. The continuous-time system is discretized in time with time step $T=0.5\,\mathrm{s}$ and the Euler method, which yields the discrete-time state-space representation
\begin{equation*}\label{MSD}
\begin{aligned}
    x_{k+1}&=\begin{bmatrix}1 & T\\ -\frac{KT}{m} & -\frac{dT}{m}+1\end{bmatrix}x_k+\begin{bmatrix}0\\T\end{bmatrix}u_k,\quad t\in\mathbb{N},\\
    y_k&=\begin{bmatrix}1 & 0\end{bmatrix}x_k.
\end{aligned}
\end{equation*}
We assume that the system order $n_x=2$ is known but the system matrices are completely unknown. To infer on the unknown system dynamics, the mass-spring-damper system is simulated over $N-1$ time steps for initial condition zero, input $u_k=\cos(0.05k)$, and randomly drawn process noise $||d_k||_2\leq\epsilon$ with various but known $\epsilon>0$. Thereby, noisy state-input data $\{x_k,u_k\}_{k=0}^{N-1}$ are available to derive the set membership \eqref{Set_mem}.

\begin{figure}
	\centering
	\includegraphics[width=0.75\linewidth]{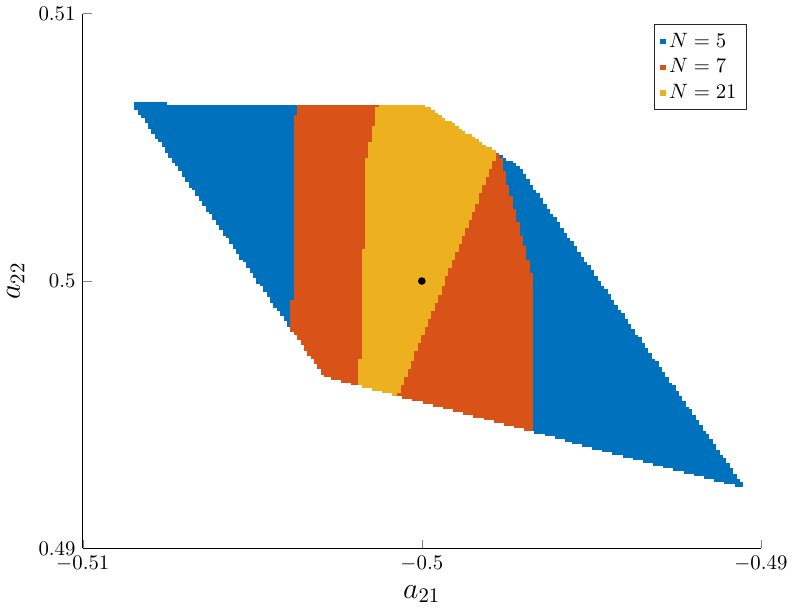}
    \includegraphics[width=0.75\linewidth]{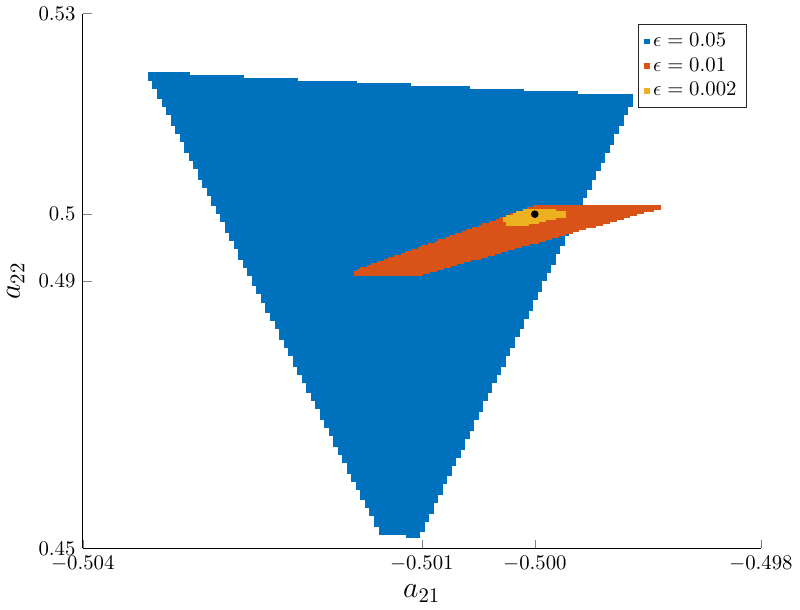}
	\caption{Projected set membership \eqref{ProjSetMem} for $\epsilon=0.005$ and different data length $N$ (top) and \eqref{ProjSetMem} for $N=21$ and different noise bounds $\epsilon$ (bottom). True coefficients are depicted by a black dot.}
	\label{Fig.Set_Mem}
\end{figure}

Figure~\ref{Fig.Set_Mem} illustrates the projected set membership 
\begin{equation}\label{ProjSetMem}
\begin{aligned}
    &\left\{a_{21}, a_{22}\in\mathbb{R}: \bigg|\bigg|x_{k+1}-\begin{bmatrix}1 & T\\ a_{21} & a_{22} \end{bmatrix}x_k-\begin{bmatrix}0\\ T \end{bmatrix}u_k\bigg|\bigg|_2\leq\epsilon,\right.\\ &\left.\hspace{4.5cm}\phantom{\begin{bmatrix}0\\ T \end{bmatrix}}k=0,\dots,N-2\right\}.
\end{aligned}
\end{equation}
As expected, the true coefficients $a_{21}^*=-0.5$ and $a_{22}^*=0.5$ are always elements of the set membership. The top figure illustrates that \eqref{ProjSetMem} decreases for an increasing amount of samples $N$. In particular, additional samples never increase the set, i.e., the set membership for $N=21$ is a subset of the set membership for $N=5$ and $N=7$. Based on \cite{MartinIQC}, \eqref{Set_mem} even converges to a singleton containing only the true coefficient matrices for $N\rightarrow\infty$ and persistently exciting data. The bottom figure illustrates \eqref{ProjSetMem} when the simulated trajectory is affected by differently large noise. Since the process noise $d_k$ influences the trajectory, the set memberships are not necessarily subsets of each other.  

For dissipativity verification, we first apply Proposition~1 of \cite{MartinIQC} to obtain ellipsoidal outer approximations of \eqref{Set_mem} while minimizing the volume of the ellipses. Subsequently, we search for a matrix $X\succeq0$ and a scalar $\alpha\geq0$ such that $\Omega\preceq0$ from \eqref{LMI_condidtion}. In this numerical example, we examine on the one hand the $\mathcal{L}_2$-gain from the input $u$ to the position $y$. Thus, we consider $(Q,S,R)$-dissipativity with $Q=-1, S=0$, and $R=\gamma^2,\gamma>0$, where $\gamma$ corresponds to an upper bound on the $\mathcal L_2$-gain. Hence, we additionally minimize over $\gamma>0$. Moreover, we investigate input-feedforward passivity, which corresponds to the maximal $\rho$ such that the system is $(Q,S,R)$-dissipativity with $Q=0, S=0.5$, and $R=\rho$. The resulting SDPs with LMI constraint are solved in MATLAB using YALMIP \citep{YALMIP} and the SDP solver MOSEK \citep{mosek}.

Figure~\ref{Fig.L2_gain} and Figure~\ref{Fig.Passivity} compare the inference on the $\mathcal L_2$-gain and input-feedforward passivity parameter, respectively, obtained from the true system and from data for various data lengths $N$ and different large noise $\epsilon$.  
\begin{figure}
	\centering
	\includegraphics[width=0.75\linewidth]{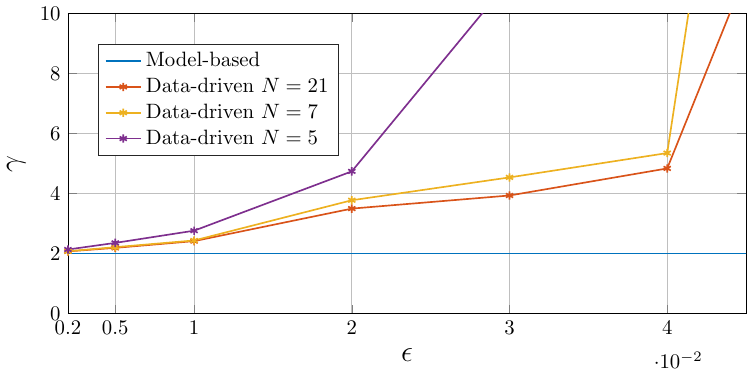}
	\caption{Data-driven inference on the $\mathcal L_2$-gain.}
	\label{Fig.L2_gain}
\end{figure}
\begin{figure}
	\centering
	\includegraphics[width=0.75\linewidth]{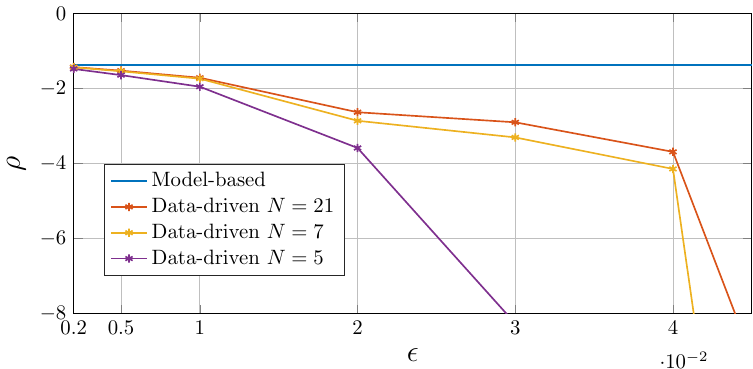}
	\caption{Data-driven inference on the  input-feedforward passivity parameter.}
	\label{Fig.Passivity}
\end{figure}
Since the obtained inferences are guaranteed system properties, we only obtain larger values than the true $\mathcal L_2$-gain and smaller values than the true input-feedforward passivity parameter. As expected, the determined bounds are more accurate for a larger set of data and smaller noise. Indeed, the data-driven inferences correspond almost to the true $\mathcal L_2$-gain and passivity parameter for very small noise and even for a small amount of samples. Furthermore, since the data-driven inferences do not significantly improve from $N=7$ compared to $N=21$, we can conclude that the data $\{x_k,u_k\}_{k=8}^{21}$ is not informative for the $(Q,S,R)$-dissipativity verification. To further improve the obtained bounds, one could consider a more exiting input or experiments with different initial conditions.
\end{example}

\subsection{Online Schemes}\label{Sec_Iter_Verification}
The behavioral and the set-membership approach suppose given trajectories, that is, both are offline methods according to \cite{DDC_Survey}. In contrast, 
online methods strive to improve the data-driven inference by iteratively performing experiments if the system or a simulation is available for further experiments. Thus, online schemes provide in each iteration an input trajectory that can be applied to refine the data-driven inference on the system property.

For discrete-time LTI SISO systems with $u_k=y_k=0$ for $k\leq0$, a finite output sequence can be computed by
\begin{align*}
	\underbrace{\begin{bmatrix}y_1\\y_2\\ y_3\\ \vdots \\y_L	\end{bmatrix}}_{=y}=\underbrace{\begin{bmatrix}g_0 & 0 & 0 & \cdots & 0\\ g_1 & g_0 & 0 & \cdots & 0\\ g_2 & g_1 & g_0 & \cdots & 0\\ \vdots & \vdots  &  &\ddots & \vdots\\ g_{L-1} & g_{L-2} & \cdots & g_1 & g_0\end{bmatrix}}_{=G}\underbrace{\begin{bmatrix}u_1\\u_2\\ u_3\\ \vdots\\u_L	\end{bmatrix}}_{=u},
\end{align*}
with impulse response $g_0,\dots,g_{L-1}$. Thus, the $\mathcal L_2$-gain over a finite time horizon $L$ is given by
\begin{equation}\label{L2_gain_iter}
    \min_{u\in\mathbb{R}^L\backslash\{0\}}\frac{y^\top  y}{u^\top  u}=\min_{u\in\mathbb{R}^L\backslash\{0\}}\frac{u^\top  G^\top  Gu}{u^\top u}=\min_{u\in\mathbb{R}^L\backslash\{0\}}J(u).
\end{equation}
To solve this optimization problem for unknown $G$, \cite{IterL2} shows that the gradient of $J(u)$ can be obtained by evaluating the system response for two specific input sequences. Thus, a power iteration can be applied to solve \eqref{L2_gain_iter} without knowledge of $G$. In \cite{Koch_Iter}, 
this idea is unified as a framework of iterative sampling schemes for determining the operator gain, passivity, and conic relations of LTI systems. Such approaches have also been extended to finding lower order surrogate models \cite{MartinLTI} for linear systems while establishing finite horizon system properties.  Works such as \cite{wu2013experimentally,welikala2022line,tanemura2019closed} further focus on estimating stability properties such as $\mathcal{L}_2$ gain and various passivity indices (e.g., input feedforward and output feedback), and lower bounds on gain and phase margins \citep{isoshima2023data} from data. There are also schemes to improve the efficiency,  convergence speed, number of required experiments, and the robustness of data-driven estimation of properties such as passivity and dissipativity \citep{tanemura2018efficient,iijima2020reduction}.

To handle noisy data and nonlinear systems, and to obtain finite data guarantees, \cite{Anne_GP} solves, among others, the $\mathcal L_2$-gain optimization problem \eqref{L2_gain_iter} by Gaussian process optimization. By modelling $J(u)$ as Gaussian process, the worst-case input with respect to the system property is identified and subsequently applied to the system. Then the Gaussian process model can be updated by the new input-output sample. To achieve an iterative scheme for nonlinear systems, noisy data, and deterministic guarantees, \cite{MartinIterative} considers the Lipschitz approximation of \cite{Kinky} for the input-output mapping of a dynamical system. This approximation can intrinsically be used to obtain an upper bound on the $\mathcal L_2$-gain. Subsequently, the worst-case input can be identified and applied in order to refine the model and the inference on the  $\mathcal L_2$-gain. 
Summarizing, the presented schemes provide an iterative experiment design to refine data-driven inference of control-relevant properties. However, only finite horizon guarantees can be provided and the number of samples for a meaningful inference is larger than that required for the behavioral and set-membership approach. 

\subsection{Input Design}
We conclude this section by making some remarks on the challenges associated with the input design within the behavioral framework. In this setting, \cite{ExpDesign} and \cite{ExpDesign_FeebackLin} present ideas for the design of inputs to ensure persistently exciting input-output sequences for locally identifying nonlinear systems and for certain classes of nonlinear systems, respectively. While these works determine one experiment to identify the system, multiple experiments are typically necessary for learning an uncertain nonlinear dynamics. In Section~\ref{Sec_Iter_Verification}, we also reviewed schemes for verifying control-relevant properties by iteratively performing experiments. These schemes provide an input sequence in each iteration step  to refine the data-driven inference. However, these iterations typically converge rather slowly, requiring a large amount of experiments. Thus, the application of these iterative schemes is time consuming and the suggested input sequences may not obey any safety conditions, limiting their practical application. To reduce the size of the set membership \eqref{Set_mem} by additional experiments, one could consider state-input pairs which generate informative constraints on \eqref{Set_mem}. Here the challenge is to identify these pairs and to steer the system safely to the corresponding state. Moreover, while this procedure would reduce the conservatism of the set membership, it is not guaranteed to refine the inference on the system property. { While these considerations are specific to behavioral and direct data-driven representations, where the input sequence must not only excite the system but also support finite-horizon trajectory representations or direct inference of input-output properties, we discuss broader experiment-design questions pertaining to control-oriented identification in Section \ref{sec:expt_design}.}



\section{Future Directions}\label{sec: future}
As we have outlined earlier, there is an impressive array of work that has now appeared, and continues to be done, utilizing the recent developments from learning to extend and complement classical system identification algorithms. Nevertheless, there remain rich problems that are unsolved in this area. In the following, rather than focusing on specific challenges on the various directions surveyed in the paper, we outline some major open directions that have not been explored much in the literature so far, yet represent important problems of great practical and theoretical interest. 

\subsection{Networked System Identification}
Another major open question is the identification of networked systems, where imposing interconnection structure on the learned system matrices in addition to control-oriented and physics properties is a challenge. Many systems of practical interest are networked sub-systems that are dynamically coupled. Thus, for instance, it will be of interest when learning a model for a transportation network to preserve the structure of arterial roads and traffic intersections. Similarly, for a chemical reaction network, it may be of interest to learn the reaction rates while preserving the structure of the reactants and products. 

A simple model of these interconnections for a linear system can be imposing certain elements of the system matrices $A$ and $B$ to be zero. Unfortunately, even for simple linear systems, identifying the system matrices typically yields a full matrix with scant regard to the desired structure. It is not clear that just setting the specified elements in the learned system matrices to be zero is necessarily the best solution. There has been some initial work~\cite{fattahi2019learning} on imposing a sparsity pattern on the learned system matrices; however, that work requires special assumptions on the matrices which may be difficult to check. Similarly, for special interconnection structures (i.e., those satisfying the so-called partially nested information pattern), some works~\cite{ye2022sample,ye2022regret} have shown that imposing sparsity constraints may not be important for learning optimal control laws; however, once again, it is not clear how generalizable these results are. The general problem of \emph{imposing interconnection structure} during networked system identification remains open and an important avenue for future work.

Further, several networked systems like chemical reaction networks, power grids, and traffic flow networks satisfy control-relevant properties like monotonicity that can be exploited for scalable control synthesis \cite{sivaranjani2017distributed}. There have been some recent developments in this direction in \cite{revay2021distributed}, where an identification for large-scale networked systems with monotonicity guarantees is proposed. However, network identification preserving both the graph structure and various control-relevant properties is still an open problem.

\subsection{Network Identification} 
A related problem is the identification of network structure itself from the data. In other words, consider an interconnected system as before, but now the interconnection structure is unknown and needs to be estimated from the data along with the parameters of the nodes describing the dynamical subsystems and edges describing the interconnections. This problem naturally arises in a variety of settings, with a concrete example being the identification of the structure of infrastructure systems such as the distribution grid in power systems or transportation network flows. There are classical fields such as network tomography \citep{vardi1996network,coates2002internet} that address such a problem, although typically under simplifications such as absent or known node dynamics. From the domain of data science, a powerful idea that has been used in fields such as identification of gene regulatory networks is that if the observations from different nodes could be treated purely as a data series, one could associate statistical closeness among the series (as measured using quantities such as mutual information) with a weighted edge between the nodes \citep{villaverde2014mider}. Extensions such as considering directed mutual information have been proposed to include directed edges that represent which node affects the other one. 

Recent developments in causal inference \citep{pearl2010causal} have also aimed to consider similar problems of identifying causal networks whereby random variables representing data series at nodes affect each other causally. Granger causality \citep{granger2001investigating} based methods have emerged as particularly powerful when the nodes represent random processes, as in the general framework considered in this paper. However, Granger causality based approaches necessitate presence of delays in the dynamic dependencies. New techniques have been proposed for the case when the nodes represent linear time-invariant subsystems \citep{veedu2023causal}. However, it is fair to say that the general problem when the node dynamics can be non-linear, edges may have dynamics, and partial information about the network structure may be available, remains largely open.

\subsection{Switched {and Time-varying} Systems}
Our discussion so far has assumed that both the node dynamics and the interconnection structure are time-invariant. For the general case when these quantities can be arbitrarily time-varying, it is difficult to collect enough data to perform any identification. However, for the case when there is some additional structure on the time-variation, it may be possible to obtain useful identification algorithms. A popular and practically important case is when the node dynamics can be modeled as switched (or more generally, hybrid) system. Switched systems, where the system switches between a set of individual modes that may be linear or non-linear, have proven to be immensely useful in modeling many situations of practical interest such as power grid operation~\citep{feng2025freq,ochoa2023control}, robotics~\citep{tomlin2002conflict,grizzle2014models}, etc..
There is a large theory for modeling, analysis, and control of such systems available, with further classification into classes such as state-dependent or state-independent switching, deterministic or stochastic switching, autonomous or controlled switching, and so on.

The easiest extension of the methods we considered in this paper to switched system identification is if the active mode is known at any time and each mode is identified separately. Even in this case, if there is some additional information available about how the dynamics in each mode are related (e.g., the dynamics of temperature increase when a heater is on v/s decrease when a heater is off have many common parameters), one may envisage utilizing data from one mode to identify the dynamics of other modes. More generally, if the mode is unknown, then the identification problem is even more difficult since it involves simultaneous identification of a discrete variable (i.e., which mode is active) and the continuous dynamics. Relevant classical fields here include change point detection~\citep{fu2024simultaneous} and hidden Markov models~\citep{smyth1994hidden} and many numerical algorithms have been proposed to identify the dynamics of the switched systems~\citep{lauer2018hybrid}. However, the fundamental difficulty of the optimization problem here constrains the performance and analysis of these algorithms except under some specific conditions. For instance, if the switching signal is not being controlled, and some of the modes of the system are unstable, the problem largely remains open~\citep{shi2022finite}. 

If additionally, as we argued in the paper, we aim to preserve physics-relevant properties during identification, an additional step that needs to be solved is to define the properties of interest. As an example, extension of properties such as dissipativity and passivity to switched systems requires care since the modes that are not active may gain energy~\citep{zhao2008dissipativity}. The core intuition is the same as in classic results on stability of switched autonomous systems -- it is easy to construct examples where a system that switches among individually stable modes is unstable. There is existing work on defining dissipativity to switched systems keeping this subtlety in mind~\citep{lavaei2022dissipativity,mccourt2012stability,li2016dissipativity,zhao2007dissipativity,haddad2006impulsive,agarwal2016dissipativity,wang2013passivity,xia2016passivity,sivaranjani2018conic,zhao2016feedback}. However, these works tend to require multiple storage functions and become computationally complex. There is little, if any work, on utilizing these definitions during identification of the system and further using them for control. Given the generality and importance of switched and time-varying systems in practical applications, control-informed system identification for these systems is an important direction for future work.


\subsection{Experiment Design}\label{sec:expt_design}
Experiment design, broadly pertaining to the  design of input sequences, data filtering, and sampling approaches are central to system identification. The design of input sequences for identification is particularly critical. Ideally, the input sequence must yield rich and meaningful data that excite all modes. On the other hand, the collection of data samples for system identification may be expensive, especially in online settings. Typically, the goal is to balance these tradeoffs. 

This problem has been extensively studied in classical system identification, including closed-loop identification for control (see works such as \citep{goodwin1977dynamic,hjalmarsson2005experiment,gevers1986optimal} for a survey of this area) and the persistent excitation conditions~\citep{narendra1987persistent,willems2005note,markovsky2023persistency}. Input design for ML-based system identification is especially critical, since learning algorithms (such as neural network training) often require massive amounts of data, which is difficult to acquire in control applications. However, this area has received limited consideration in learning-based identification. A key open question is how to optimize input design while simultaneously ensuring that the learned model satisfies control-relevant properties. Imposing such properties places nontrivial structural constraints on the feasible excitation signals, thus altering the underlying geometry of the optimal experiment design problem. Further, imposing control-theoretic constraints may require balancing information gain against performance degradation or violations of control-relevant properties during data collection. Therefore, developing principled methods for experiment design while imposing control-relevant properties, particularly in online or data-limited regimes, is a critical direction.

\subsection{Machine Learning vs Data-driven System Representations}
We discussed the emerging paradigm of data-driven representations for control in Section \ref{sec:direct}, including approaches to provide guaranteed inferences on control-theoretic properties even in presence of noise. However, the system representations for nonlinear systems are tailored towards system analysis and control design using LMI techniques. Therefore, the nonlinearity of the system can not precisely be incorporated into a data-driven system representation, thus yielding conservative results that may only hold locally. On the other hand, models from machine learning are flexible and can explain complex nonlinear phenomena. However, the obtained system representations are usually strongly nonlinear, and are hence difficult to exploit for system analysis and control. Thus, closing the gap between data-driven system representations tailored for control and machine learning approximations can be an interesting direction for future research. Thereby, the benefits of both worlds could be fused into a new class of surrogate models. Initial works in this direction include, for instance, \cite{Persis_kernel} and \cite{Fiedler}. { {Nonlinear extensions of behavioral representations, including recent connections to generalized Koopman operators, provide another promising direction in this context \citep{lazar2025product}.}} Related to merging these two worlds, it is important to note that data-driven control lacks standardized datasets and benchmarks as is common in the machine learning community, which makes meaningful comparison of different frameworks challenging. Developing such benchmarks and meaningful control-relevant use cases will be important to enable rapid advances in this area.

\subsection{Tradeoffs in Control-Oriented System Identification}
\label{sec:futurework_tradeoff}
A central challenge in both classical and learning-based system identification is the tradeoff between control-relevant constraints and model expressivity. On one hand, physics-based priors and control-relevant constraints provide useful inductive biases that can reduce sample complexity, yield tractable formulations for learning-based identification, aid identifiability in problems with scarce data, and enable safe and interpretable closed-loop control. { {Physics-guided neural network architectures with built-in stability guarantees provide one example of this tradeoff, since the imposed architectural structure can certify properties such as input-to-state stability while restricting the admissible model class \citep{bolderman2024physics}.}} On the other hand, hard constraints or rigid model parameterizations can introduce model bias that limits the ability to capture complex real-world dynamics and generalize to unseen data, especially when the true system deviates from nominal operating conditions or exhibits non-ideal behavior, potentially undermining the very advantages that make data-driven models attractive in control applications. This merits a re-examination of classical considerations such as the closely related bias-variance tradeoff and the impact of structure and constraints on generalizability in the context of learning-based and data-driven models \citep{pillonetto2025deep}. A key direction for future research is to establish universal approximation theorems for structured model classes that satisfy control-theoretic constraints. Such results would formalize the representational capacity of control-relevant models and guide the design of models and architectures that are both expressive and guaranteed to satisfy physics or control properties of interest.

\section*{Acknowledgments}
The work of Y. Shi and J. Feng was supported by NSF grant ECCS-2442689, DOE grant DE-SC0025495, a Schmidt Sciences AI2050 Early Career Fellowship, and the Michael R Anastasio LANL-UC Early Career Faculty Fellowship. The work of J. Feng was also supported by the UC-National Laboratory In-Residence Graduate Fellowship L24GF7923. The work of Y. Xu, S. Sivaranjani, and V. Gupta was supported in part by the  the Air Force Office of Scientific Research under
Grant FA9550-23-1-0492. The work of V. Gupta was partially supported under ARO grant W911NF2310111 and ONR grant N000142312604. Y. Xu was also partially supported by the Purdue University College of Engineering Seed Funding for High Impact Papers, Books, and Monographs. The work of N. Atanasov and T. Duong was supported by NSF grants CCF-2112665 (TILOS) and CCF-2402689 (ExpandAI). The work of F. Allgöwer was funded by the Deutsche Forschungsgemeinschaft (DFG, German Research Foundation) under Germany’s Excellence Strategy – EXC 2075 – 390740016 and within grant AL 316/15-1 – 468094890. T. Martin thanks the Graduate Academy of the SC SimTech for its support.



\bibliographystyle{elsarticle-harv} 
\bibliography{bib/reference,bib/DL_SysID,bib/atanasov,bib/thai,bib/TM_Bib,bib/siva}






\end{document}